\documentclass[12pt]{article}
\usepackage{amsmath, latexsym, amssymb, amscd, tensor, slashed, mathrsfs}
\numberwithin{equation}{section}

\newcommand{\LL}{\mathfrak L}
\newcommand{\R}{\mathbb R}
\newcommand{\C}{\mathbb C}

\newcommand{\p}{\partial}

\newcommand{\Ta}{T_a}

\newcommand{\Bold}[1]{{\boldsymbol{\mathit{#1}}}}

\newcommand{\ti}{\mathrm{t}}

\newcommand{\x}{\mathrm{x}}
\newcommand{\s}{\mathrm{s}}

\newcommand{\zz}{\bar{z}}

\newcommand{\y}{\mathrm{y}}

\newcommand{\z}{b}

\newcommand{\QED}{\hspace{.2in}\square\newline}
\newcommand{\qED}{\hspace{.2in}\boxminus\newline}

\newtheorem{theorem}{Theorem}[section]
\newtheorem{corollary}[theorem]{Corollary}
\newtheorem{proposition}[theorem]{Proposition}
\newtheorem{definition}[theorem]{Definition}

\newtheorem{lemma}[theorem]{Lemma}
\newtheorem{example}[theorem]{Example}
\newtheorem{application}[theorem]{Application}

\newtheorem{remark}[theorem]{Remark}

\begin{document}

\title{Mellin-type Functional Integrals \\ with Applications to Quantum Field Theory and Number Theory}
\author{J. LaChapelle}
\maketitle

\begin{abstract}
Conventional functional/path integrals used in physics are most often defined and understood, either explicitly or implicitly, as the infinite-dimensional analog of Fourier transform. In this paper, the infinite-dimensional analog of Mellin transform is defined and developed. The associated functional integrals are useful tools for probing non-commutative function spaces in general and $C^\ast$-algebras in particular. Functional Mellin transforms are used to define the functional analogs of resolvents, complex powers, traces, logarithms, and determinants. Several aspects of these objects are examined and applied to various constructs in mathematical physics. As substantial applications, we construct Mellin-based QFT generating functionals for bosonic and fermionic degrees of freedom, explore connections between functional complex powers and scattering amplitudes, interpret renormalization from a functional Mellin perspective, define a parameter-dependent entropy that formally justifies the replica trick, and explore $L$-functions associated with functional traces and determinants.
\end{abstract}



\section{Introduction}
\subsection{Motivation}
Functional integration had its beginnings in the study of stochastic
processes --- particularly the Wiener process --- and was therefore
deeply rooted in probability theory \cite{DY/YU}--\cite{FRI1}. Later on, mathematically non-rigorous functional integrals in the form of Feynman integrals \cite{FE/HI} were introduced and found to be useful tools in quantum theory and partial differential equations. They have been extensively developed and utilized in the mathematical physics literature, which is by now quite mature. \cite{GROS}--\cite{AL}

One consequence of this heritage is that functional integration
methods in mathematical physics borrow heavily from probability
constructs and are mostly confined to transformations or
expansions/perturbations around quadratic-type functional integrals.
These archetypical functional integrals are distinguished by the fact that the characteristic function of a Gaussian probability distribution is again a Gaussian, and the probability analogy is
used to carry this notion over to the context of quantum physics. This allows functional integrals to be interpreted as a functional Fourier transform between dual Banach spaces, and this forms the basis of the vast majority of applications in physics.

But, while the probability analogy can be inspirational and fruitful, it can also be
restrictive. On one hand, probability theory is a useful complement
to intuition, and it is easy to imagine that functional integrals
based on probability distributions other than Gaussian would be useful in
mathematical physics. On the other hand, expressing probability
distributions through their characteristic functions can lead to an
emphasis on Fourier transform (which expresses Pontryagin duality between locally compact abelian topological groups) as a guiding principle, and this encourages customary expansions around Gaussian backgrounds in the restricted setting of \emph{abelian} topological groups.

However, Fourier transform is not the only game in town. It is not hard to see the functional analog of the Mellin transform
  also could be a useful tool. And, similar to the functional Fourier transform, functional Mellin encodes a duality --- but
an algebraic rather than group duality.  Likewise, for functional Mellin transforms the probability analogy remains a profitable guide; except this time in the context of non-commutative Banach algebras. It is significant that functional Mellin provides a means to represent and probe Banach algebras in the setting of \emph{non-abelian} topological groups which lies beyond the power of Fourier. The purpose of this paper is to construct and develop this tool with an eye towards primarily physics applications.

\subsection{Outline of paper}
The functional integral framework we will use to construct functional Mellin transforms is based on topological groups, so we start with a brief exposition of some pertinent results concerning locally compact topological groups and their associated integral operators on Banach algebras. Our proposed definition of functional integrals is then briefly reviewed, but we refer to \cite{LA1} for details.

The remainder of the paper  concentrates on elaborating the functional analog of the finite-dimensional Mellin
transform. We use the functional Mellin transform to define functional
analogs of resolvent, trace, log and determinant. In the functional context, Mellin is not just a simple
transformation of Fourier, so developing and investigating
the infinite dimensional analog of Mellin transform is worthwhile.

With certain restrictions, functional Mellin represents ${C}^\ast$-algebras and is therefore an effective tool for quantum physics. Notably, one can construct and multiply operators with complex powers. We explore some properties of such operators in several examples. In particular, in $\S$5 we examine a certain fractional-power operator whose trace is given by a sum of beta functions that reproduces the four-point tree-level tachyon string scattering amplitude\cite{POL} on the disk up to normalization and momentum conservation. Moreover, the algebraic product of two such operators, which is realized through a convolution integral, yields the closed tachyon tree-level amplitude on the sphere; thereby echoing the KLT relation\cite{KLT}. This is \emph{not} to claim that functional Mellin has any direct connection with string theory or contains any stringy physics. Rather, it supports our primary thesis; that functional Mellin is well suited to represent the $C^\ast$-algebras of interesting quantum theories and to construct interesting objects related to number theory.

To further support this claim, \S\ref{compare} compares Fourier and Mellin representations of bosonic and fermionic $n$-point generating functions in QFT. Our treatment is not comprehensive as it does not address gauge symmetry issues. Nevertheless, the Mellin representations of functional resolvent, trace, log, and determinant are profitable in this context. In particular, we show the functional log justifies the replica trick, we observe connections between functional Mellin and scattering processes in QFT and string theory, and we show the regularization and scaling aspects of renormalization in QFT are intrinsic to functional Mellin. Further, in Appendix \ref{sec. key theorem} we prove a theorem generalizing the matrix relation
$\mathrm{exp}\,\mathrm{tr}\,M=\det \mathrm{exp}\,M$. Roughly stated, it says the Mellin transform and exponential map commute under appropriate conditions. Finally, in \S \ref{number theory} we defined certain classes of $L$-functions (typically formulated in number theory) in terms of functional traces and determinants, and we investigate some of their properties.

There are further reasons --- beyond representing $C^\ast$-algebras --- to expect that functional Mellin transforms will be useful in physics and applied mathematics. To give just a few: Crossed products, which are effective tools for $C^\ast$-algebraic quantization \cite{LANDS1}--\cite{LANDS3}, are closely related to functional Mellin transforms. (Appendix \ref{relation to crossed products} looks closer at this relationship.) The Mellin transform features prominently in the world-line formalism \cite{STR,SCH} of perturbative QFT and celestial holography \cite{LD}. Properties of the Mellin transform allow efficient analytic treatment of harmonic integrals, asymptotic analysis of harmonic
sums, and Fuchsian type partial differential equations
\cite{FL}--\cite{ZEM}.  Finally, functional integrals based
on the gamma probability distribution, which are
particular classes of functional Mellin transforms, show up in the study of
constrained function spaces \cite{LA3}, in the combinatorics of Hurwitz numbers and Hodge integrals \cite{GGR}, and prime $k$-tuples \cite{LA1}.

A word about notation: Elements of an algebra of functionals mapping a topological group into a target Banach algebra are denoted by uppercase roman (e.g. $\mathrm{A}$ and $\mathrm{E}^{-\mathrm{A}}$) and sometimes referred to as ``observables''. Elements of the target Banach algebra are denoted by uppercase italic (e.g. $A$ and $e^{-A}$)  and sometimes referred to as ``operators''.

\section{Functional integration scheme}\label{scheme}
In this section, we briefly recall some relevant features of the functional integral scheme proposed in \cite{LA1}.

Our functional integrals are based on the data $\left(G,\mathfrak{B},G_\Lambda\right)$
where $G$ is a Hausdorff topological group, $\mathfrak{B}$ is a
Banach space that may have additional  {associative} algebraic structure, and
$G_\Lambda:=\{G_{\lambda},\lambda\in\Lambda\}$ is a {countable} family of
locally compact topological groups indexed by surjective homomorphisms $\lambda:G\rightarrow G_{\lambda}$.

\begin{definition}
A Hausdorff topological group $G$ is a group endowed with a topology
such that; (i) multiplication $G\times G\rightarrow G$ by
$(g,h)\mapsto gh$ and inversion $G\rightarrow G$ by $g\mapsto
g^{-1}$ are continuous maps, and (ii) $\{e\}$ is closed.
\end{definition}
Topological groups come equipped with one-parameter subgroups.
\begin{definition}\emph{(\cite[ch.~5]{HM})} A
one-parameter subgroup ${\phi}:\R\rightarrow G$ of a topological group is the unique extension of a continuous
homomorphism $f\in
\mathrm{Hom}_C(I\subseteq\R,G)$ such that $f(t+s)=f(t)f(s)$ and $f(0)=e\in G$. Let $\mathfrak{L}(G)$ denote the set
of all one-parameter subgroups $\mathrm{Hom}_C(\R,G)$ endowed with the uniform convergence topology on compact sets in
$\R$. The exponential function is defined by
\begin{eqnarray}
\mathrm{exp}_G:\mathfrak{L}(G)&\rightarrow& G \notag\\
\phi&\mapsto&\mathrm{exp}_G\,\phi=\phi(1)\;.
\end{eqnarray}
\end{definition}

Locally compact topological groups posses a crucial structure:
\begin{definition}$G$ is locally compact if every $g\in G$ has a neighborhood
basis comprised of compact sets.
\end{definition}

\begin{theorem}
If $G$ is locally compact, then there exists a unique (up to
positive scalar multiplication) Haar measure. If $G$ is compact,
then it is unimodular.
\end{theorem}
Thanks to this theorem, the set $G_\Lambda$ gives us access to Banach-valued integration.
\begin{proposition}\emph{(\cite[prop.~B.34]{W})}\label{banach integration}
Let $G_\lambda$ be a locally compact topological group, $\mu$ its associated
Haar measure, and $\mathfrak{B}$ a Banach space possibly with an
algebraic structure. Then the set of integrable
functions $L^1(G_\lambda,\mathfrak{B})\ni f$, consisting of equivalence classes of measurable functions equal almost everywhere
with norm $\|f\|_1:=\int_{G_\lambda}\|f(g_\lambda)\|d\mu(g_\lambda)\leq\|f\|_\infty\,\mu(\mathrm{supp}\,f)<\infty$, is a
Banach space. Moreover, $f\mapsto\int_{G_\lambda}f(g_\lambda)\;d\mu(g_\lambda)$ is a linear map
such that
\begin{equation}
\|\int_{G_\lambda}f(g_\lambda)\;d\mu(g_\lambda)\|\leq\|f\|_\infty\,\mu(\mathrm{supp}\,f)
\end{equation}
for all $f\in L^1(G_\lambda,\mathfrak{B})$,
\begin{equation}
\varphi\left(\int_{G_\lambda}f(g_\lambda)\;d\mu(g_\lambda)\right)
=\int_{G_\lambda}\varphi(f(g_\lambda))\;d\mu(g_\lambda)
\end{equation}
for all $\varphi\in\mathfrak{B}'$, and
\begin{equation}
L_B\left(\int_{G_\lambda}f(g_\lambda)\;d\mu(g_\lambda)\right) =\int_{G_\lambda}L_B(f(g_\lambda))\;d\mu(g_\lambda)
\end{equation}
for bounded linear maps
$L_B:\mathfrak{B}_1\rightarrow\mathfrak{B}_2$. Moreover, Fubini's
theorem holds for all equivalence classes $f\in L^1(G_1\times G_2,\mathfrak{B})$.
\end{proposition}

\begin{corollary}\emph{(\cite[lemma.~1.92]{W})}\label{H-representation}
Let $\mathfrak{B}^\ast$ be a $\ast$-algebra and
$\pi:\mathfrak{B}^\ast\rightarrow L_B(\mathcal{H})$ a
representation with
$L_B(\mathcal{H})$ the algebra of bounded linear operators on Hilbert space
$\mathcal{H}$. Then
\begin{equation}
\left\langle \pi\left(\int_{G_\lambda}f(g_\lambda)\;d\mu(g_\lambda)\right)v|w\right\rangle
=\int_{G_\lambda}\left\langle \pi\left(f(g_\lambda)\right)v|w\right\rangle\; d\mu(g_\lambda)\;,
\end{equation}
\begin{equation}
\left(\int_{G_\lambda}f(g_\lambda)\;d\mu(g_\lambda)\right)^\ast =\int_{G_\lambda}f(g_\lambda)^\ast\; d\mu(g_\lambda)\;,
\end{equation}
and
\begin{equation}
a\int_{G_\lambda}f(g_\lambda)\;d\mu(g_\lambda)b=\int_{G_\lambda}af(g_\lambda)b\;d\mu(g_\lambda)
\end{equation}
where $v,w\in\mathcal{H}$ and $a,b\in M(\mathfrak{B}^\ast)$ with
$M(\mathfrak{B}^\ast)$ the multiplier algebra of
$\mathfrak{B}^\ast$.
\end{corollary}

It is well known (see e.g. \cite[appx. B]{W}) that  $L^1(G_\lambda,\mathfrak{B}^\ast)$ is a Banach $\ast$-algebra when equipped with: i) the $\|\cdot\|_1$ norm,
ii) the convolution
\begin{equation}
f_1\ast f_2 (g_\lambda):=\int_{G_\lambda}f_1(h_\lambda)f_2(h_\lambda^{-1}g_\lambda)d\mu(h_\lambda)\;,
\end{equation}
and iii) the involution
\begin{equation}
f^\ast (g_\lambda):={f(g_\lambda^{-1})}^\ast\Delta(g_\lambda^{-1})
\end{equation}
where $\Delta$ is the modular function on $G_\lambda$.

The data $\left(G,\mathfrak{B},G_\Lambda\right)$ together with its associated Banach-valued integration motivates our definition of functional integral.
\begin{definition}\label{int-def}
Let $\overline{\mathbf{F}}(G)$ represent a space of functionals\footnote{In this paper the term `functional' will refer to a map from a topological group to a Banach algebra. One could just as well use the term `function' since a functional is a particular type of function. We adopt `functional' to emphasize and remind of the functional integral context.}  $\mathrm{F}:G\rightarrow\mathfrak{B}$, and denote the restriction of $\,\mathrm{F}$ to $G_\lambda$ by $f:=\mathrm{F}|_{G_\lambda}$. Let $\nu$ be a left Haar measure on $G_\lambda$.

A family of integral operators
$\mathrm{int}_\Lambda:\overline{\mathbf{F}}(G)\rightarrow \mathfrak{B}$ is
defined by
\begin{equation}\label{FI}
\mathrm{int}_\lambda(\mathrm{F})\equiv\int_G\mathrm{F}(g)\mathcal{D}_\lambda
g:=\int_{G_\lambda}f(g_\lambda)\;d\nu(g_\lambda)
\end{equation}
such that $f\in L^1(G_\lambda,\mathfrak{B})$ for all
$\lambda\in\Lambda$. We say $\mathrm{F}$ is integrable with respect to the integrator family
$\mathcal{D}_\Lambda g$, and $\mathbf{F}(G)\subseteq\overline{\mathbf{F}}(G)$ is the space of integrable functionals (with respect to $\Lambda$).

In addition, if $\mathfrak{B}$ is an algebra, define the functional $\ast$-convolution and $\star$-convolution by
\begin{equation}
\left(\mathrm{F}_1\ast \mathrm{F}_2\right)_{{\lambda}}(g)
:=\int_G\mathrm{F}_1(\tilde{g})\mathrm{F}_2(\tilde{g}^{-1}g)
\mathcal{D}_{{\lambda}}\tilde{g}
\end{equation}
and
\begin{equation}\label{star convolution}
\left(\mathrm{F}_1\star \mathrm{F}_2\right)_{{\lambda}}(g)
:=\int_G\mathrm{F}_1(\tilde{g}g)\mathrm{F}_2(\tilde{g}\tilde{g})
\mathcal{D}_{{\lambda}}\tilde{g}
\end{equation}
for each $\lambda\in\Lambda$.
\end{definition}

\begin{proposition}
$\mathbf{F}(G)$  equipped with the $\ast$-convolution is a Banach algebra when completed with respect to the norm
$\|\mathrm{F}\|_{\mathbf{F}}:=\mathrm{sup}_\lambda\|\mathrm{int}_\lambda(\mathrm{F})\|$.
\end{proposition}
\emph{Proof}: For any given
$\lambda$, the integral operator is linear and bounded according to
\begin{equation}
\|\mathrm{int}_\lambda(\mathrm{F})\|\leq\int_{G_\lambda}\|f(g_\lambda)\|
\;d\nu(g_\lambda)=\|f\|_{1,\lambda}<\infty\;.
\end{equation}
Linearity is obvious. To see it is bounded, use Cauchy-Schwarz  along with Proposition \ref{banach integration}.

Since $\|f\|_{1,\lambda}$ is a norm on $L^1(G_\lambda,\mathfrak{B})$, it follows that $\|\mathrm{F}\|_{\mathbf{F}}:=\mathrm{sup}_\lambda\|\mathrm{int}_\lambda\mathrm{F}\|$ is a norm on $\mathbf{F}(G)$. Being $\mathbf{F}(G)$ a normed space, its completion (which will be denoted by the same symbol) is a Banach space. The $\ast$-convolution then implies
\begin{eqnarray}\label{product}
\mathrm{int}_\lambda(\mathrm{F}_1\ast \mathrm{F}_2)
&=&\int_G(\mathrm{F}_1\ast \mathrm{F}_2)(g)\mathcal{D}_\lambda g\notag\\
&=&\int_{G_\lambda\times G_\lambda}f_1(\tilde{g}_\lambda)
f_2(\tilde{g}_\lambda^{-1}g_\lambda)
\;d\nu(\tilde{g}_\lambda,g_\lambda)\notag\\
&=&\int_{G_\lambda\times G_\lambda}f_1(\tilde{g}_\lambda)
f_2(g_\lambda)
\;d\nu(\tilde{g}_\lambda,g_\lambda)\notag\\
&=&\int_{G_\lambda}\int_{G_\lambda}f_1(\tilde{g}_\lambda)
f_2(g_\lambda)
\;d\nu(\tilde{g}_\lambda)d\nu(g_\lambda)\notag\\
&=&\mathrm{int}_\lambda(\mathrm{F}_1)\,\mathrm{int}_\lambda(\mathrm{F}_2)
\end{eqnarray}
where the third line follows from left-invariance of the Haar measure and the fourth line follows from Fubini. A similar computation (using left-invariance and Fubini) establishes associativity $(\mathrm{F}_1\ast \mathrm{F}_2)\ast \mathrm{F}_3=\mathrm{F}_1\ast (\mathrm{F}_2\ast \mathrm{F}_3)$. Finally, given that $\mathfrak{B}$ is Banach, eq. (\ref{product}) implies $\|\mathrm{F}_1\ast \mathrm{F}_2\|_{\mathbf{F}}\leq \|\mathrm{F}_1\|_{\mathbf{F}}\,\|\mathrm{F}_2\|_{\mathbf{F}}$.
$\QED$

If $\mathfrak{B}$ has an involutive structure, then $\mathbf{F}(G)$ inherits this structure:
\begin{proposition}\emph{(\cite[prop.~2.21]{LA1})}
Suppose $\mathfrak{B}$ is a Banach $\ast$-algebra. Then the integral operator $\mathrm{int}_\lambda$ is a $\ast$-homomorphism, and $\mathbf{F}(G)$  is a Banach
$\ast$-algebra when endowed with a suitable topology and
involution given by $\mathrm{F}^\ast (g):={\mathrm{F}(g^{-1})}^\ast\Delta(g^{-1})$ and completed with respect to
the norm $\|\cdot\|_{\mathbf{F}}$.
\end{proposition}

\begin{corollary}
If $\mathfrak{B}$ is a $C^\ast$-algebra, then $\mathbf{F}(G)$ is $C^\ast$-algebra when completed w.r.t. the norm $\|\cdot\|_{\mathbf{F}}$.
\end{corollary}

\emph{Proof}:
First,
\begin{eqnarray}\label{star}
\mathrm{int}_\lambda(\mathrm{F}^\ast)
&=&\int_G\mathrm{F}^\ast(g)\mathcal{D}_\lambda g\notag\\
&=&\int_{G_\lambda}f^\ast(g_\lambda)\;d\nu(g_\lambda)\notag\\
&=&\int_{G_\lambda}f(g_\lambda^{-1})^\ast\Delta(g_\lambda^{-1})
\;d\nu(g_\lambda)\notag\\
&=&\int_{G_\lambda}f(g_\lambda)^\ast \;d\nu(g_\lambda)\notag\\
&=&\left(\int_{G_\lambda}f(g_\lambda)\;d\nu(g_\lambda)\right)^\ast\notag\\
&=&\mathrm{int}_\lambda(\mathrm{F})^\ast\;,
\end{eqnarray}
where the fourth line follows by virtue of the Haar measure. Together with (\ref{product}), this shows that the integral operators are $\ast$-homomorphisms. In particular, $\mathrm{Id}^\ast=\mathrm{Id}$.

It remains to verify the $\ast$-algebra axioms. The $\ast$-operation is continuous for a suitable choice of topology, and linearity is obvious.  Next,
\begin{equation}
(\mathrm{F}^\ast)^\ast(g):={\mathrm{F}^\ast(g^{-1})}^\ast\Delta(g^{-1})=(\mathrm{F}(g)^\ast)^\ast\Delta(g)\Delta(g^{-1})=\mathrm{F}(g)
\end{equation}
and
\begin{eqnarray}
\left(\mathrm{F}_1^\ast\ast \mathrm{F}_2^\ast\right)_{{\lambda}}(g)
&:=&\int_{G_\lambda}f^\ast_1(\tilde{g}_\lambda)
f^\ast_2(\tilde{g}_\lambda^{-1}g_\lambda)
\;d\nu(\tilde{g}_\lambda)\notag\\
&=&\int_{G_\lambda}\left(f_2(g_\lambda^{-1}\tilde{g}_\lambda)
\Delta(g_\lambda^{-1}\tilde{g}_\lambda)f_1(\tilde{g}^{-1}_\lambda)
\Delta(\tilde{g}_\lambda^{-1})\right)^\ast
\;d\nu(\tilde{g}_\lambda)\notag\\
&=&\left(\int_{G_\lambda}f_2(g_\lambda^{-1}\tilde{g}_\lambda)
 f_1(\tilde{g}^{-1}_\lambda)
\Delta(g_\lambda^{-1})\;d\nu(\tilde{g}_\lambda)
\right)^\ast\notag\\
&=&\left((\mathrm{F}_2\ast \mathrm{F}_1)_{{\lambda}}(g^{-1})\right)^\ast\Delta(g^{-1})\notag\\
&=&\left(\mathrm{F}_2\ast \mathrm{F}_1\right)^\ast_{{\lambda}}(g)
\end{eqnarray}
where we used the definition of involution, left-invariance of the Haar measure, and the fact that the modular function $\Delta$ is a homomorphism. For the norm, $\mathfrak{B}$ a $\ast$-algebra and (\ref{star}) imply $\|\mathrm{int}_\lambda(\mathrm{F}^\ast)\|=\|\mathrm{int}_\lambda(\mathrm{F})^\ast\|
=\|\mathrm{int}_\lambda(\mathrm{F})\|$ which implies $\|\mathrm{F}^\ast\|_{\mathbf{F}}=\|\mathrm{F}\|_{\mathbf{F}}$. Conclude that $\mathbf{F}(G)$ is a $\ast$-algebra.

Lastly, if $\mathfrak{B}$ is a $C^\ast$-algebra, the corollary follows from (\ref{product}) and (\ref{star}) since now
\begin{equation}
\|\mathrm{int}_\lambda(\mathrm{F}\ast\mathrm{F}^\ast)\|
=\|\mathrm{int}_\lambda(\mathrm{F})\,\mathrm{int}_\lambda(\mathrm{F})^\ast\|
=\|\mathrm{int}_\lambda(\mathrm{F})\|\|\mathrm{int}_\lambda(\mathrm{F})^\ast\|
=\|\mathrm{int}_\lambda(\mathrm{F})\|^2
\end{equation}
implies $\|\mathrm{F}\ast\mathrm{F}^\ast\|_{\mathbf{F}}=\|\mathrm{F}\|^2_{\mathbf{F}}$.
$\QED$

\section{Functional Mellin transform}\label{main def.}
\subsection{Motivation}
One might question the utility of the functional integration scheme: After all, it eventually just boils down to a space of functions on a countable family of locally compact groups that inherits the structure of some target Banach space. But, as stressed in \cite{LA1}, the accompanying organization and structure carries motivational value.

To further reveal its value, consider integrals of the type
$\int_{G_\lambda}\pi(f(g_\lambda))U(g_\lambda)d\nu(g_\lambda)$ where $\mathfrak{B}^\ast$ is a $C^\ast$-algebra, $\pi:\mathfrak{B}^\ast\rightarrow L_B(\mathcal{H})$ is a non-degenerate representation,  and $U:G_\lambda\rightarrow U (\mathcal{H})$ is a unitary representation furnished
by some Hilbert space $\mathcal{H}$. This integral represents a Fourier transform if $G_\lambda$ is abelian, but generically $G_\lambda$ will be non-abelian. Unfortunately, as it stands the integral is not well-defined
because $\pi(f(g_\lambda))U(g_\lambda)$ is not a continuous function in general. However, if the multiplier algebra of $\mathfrak{B}^\ast$ (viewed as a Hilbert $\mathfrak{B}^\ast$-module) is equipped with the strict topology, continuity is restored since then
$\pi(f(g_\lambda))U(g_\lambda)$ \emph{is} continuous for $f(g_\lambda)\in M_s(\mathfrak{B}^\ast)$ where
$M_s(\mathfrak{B}^\ast)$ denotes $M(\mathfrak{B}^\ast)$ endowed with the strict topology (\cite[\S
1.5]{W}). Adopting the strict topology and restricting to $f(g_\lambda)\in M_s(\mathfrak{B}^\ast)$ maintains Corollary \ref{H-representation} in the form;
\begin{proposition}\label{matrix elements}\emph{(\cite[lemma~1.101]{W})}
For $f\in C_C(G_\lambda,M_s(\mathfrak{B}^\ast))$ (i.e. $f:G_\lambda\rightarrow M_s(\mathfrak{B}^\ast)$ is continuous and compactly supported), and
$\overline{\pi}:M(\mathfrak{B}^\ast)\rightarrow L_B(\mathcal{H})$ a non-degenerate representation,
there exists a linear map $f\mapsto\int_{G_\lambda}f(g_\lambda)d\nu(g_\lambda)$ from
$C_C(G_\lambda,M_s(\mathfrak{B}^\ast))$ to $M(\mathfrak{B}^\ast)$ such that
\begin{equation}
\left\langle\overline{\pi}\left(\int_{G_\lambda}f(g_\lambda)d\nu(g_\lambda)\right)v|w\right\rangle
=\int_{G_\lambda}\left\langle\overline{\pi}\left(f(g_\lambda)\right)v|w\right\rangle
d\nu(g)\;,
\end{equation}
and
\begin{equation}
\overline{l}\left(\int_{G_\lambda}f(g_\lambda)d\nu(g_\lambda)\right)
=\int_{G_\lambda}\overline{l}\left(f(g_\lambda)\right)d\nu(g_\lambda)
\end{equation}
for $\overline{l}:M(\mathfrak{B}^\ast_1)\rightarrow
M(\mathfrak{B}^\ast_2)$ a non-degenerate homomorphism.
\end{proposition}
Under these conditions, the integral $\int_{G_\lambda}\overline{\pi}(f(g_\lambda))U(g_\lambda)d\nu(g_\lambda)$ embodies the crossed-product approach to algebraic quantization \cite{LANDS3}:
\begin{proposition}\emph{(\cite[ch.~2.3]{W})}
\begin{equation}
{\overline{\pi}}\rtimes U(f):=\int_{G_\lambda}{\overline{\pi}}\left(f(g_\lambda)\right)U(g_\lambda)d\nu(g_\lambda)
\end{equation}
defines a $\ast$-representation of $C_C(G_\lambda,\mathfrak{B}^\ast)$ on
$\mathcal{H}$.
\end{proposition}

As our ultimate goal is to apply functional integration in the context of quantum physics, this connection to crossed products is suggestive. However, to our knowledge, virtually all crossed-product quantizations are built from some `dynamical system' (see appx. D). That is, they begin with a classical system, form a commutative algebra of functions, and then build a non-commutative $C^\ast$-algebra via the crossed product. This is the time-honored method of classical$\rightarrow$quantum quantization.

But the functional integral context inspires a different approach. We have seen that certain algebras have units that comprise a topological group.\cite{LA1} This suggests beginning with an abstract algebra $\mathfrak{A}$ whose group of units is isomorphic to the topological group $G$ that partially defines a functional integral. In particular, for quantum physics we should start with an abstract non-commutative $C^\ast$-algebra. Then it is enough to specify $G\rightarrow G_\Lambda$ and its relevant representations in order to construct a concrete realization of $\mathfrak{A}$ expressed through the algebra $\mathfrak{B}^\ast$. This understanding together with the crossed-product construction viewed from the functional integral perspective motivates the introduction of functional Mellin transforms.

\subsection{The definition}
In order to define the functional Mellin transform, specialize the functional integral data to $(G^\C,\mathfrak{C}^\ast,G^\C_\Lambda)$ where
 $\mathfrak{C}^\ast$ is a unital $C^\ast$-algebra and $G^\C$ a complex topological Lie group isomorphic to the group (or a subgroup) of units $A_{\mathfrak{A}}$ of a complex CCIA $\mathfrak{A}$.
\begin{definition}
A complete continuous inverse algebra(\emph{CCIA}) is a Mackey complete, unital topological algebra $\mathfrak{A}$ whose group of  units(invertible elements) $A_{\mathfrak{A}}$ is open and group inversion is continuous.
\end{definition}
\begin{definition}\emph{(\cite[\S~1.3]{GLO})}
A locally convex space $S_\diamond$ is Mackey complete iff the integral $\int_a^b\gamma(\ti)\;d\ti$ exists for any smooth curve $\gamma:(\alpha,\beta)\rightarrow S_\diamond$ with $\alpha<a<b<\beta$.
\end{definition}
Having a CCIA is enough to construct a functional calculus on $\mathfrak{A}$, and one can construct complex-analytic exponential and logarithm maps;
\begin{definition}\emph{(\cite[defs.~4.7, 5.1]{GLO})}
Suppose $(z\mathbf{1}-\mathfrak{a})\in A_{\mathfrak{A}}$. Let $B_1(\textbf{\emph{1}})$ be the unit ball about the identity element $\textbf{\emph{1}}\in\mathfrak{A}$. The exponential
$\mathrm{exp}_{\mathfrak{A}}:\mathfrak{A}\rightarrow \mathfrak{A}$ and
logarithm
$\log_{\mathfrak{A}}:B_1(\textbf{\emph{1}})\rightarrow\mathfrak{A}$
are defined by
\begin{eqnarray}\label{log def}
&&\mathrm{exp}_{\mathfrak{A}}\mathfrak{a}
:=\frac{1}{2\pi i}\int_\Gamma e^z(z\mathbf{1}-\mathfrak{a})^{-1}\;dz
\;\;\;\;\;\;\;\;\forall\,\mathfrak{a}\in\mathfrak{A},
\;\notag\\
&&\log_{\mathfrak{A}}\mathfrak{a}
:=\frac{1}{2\pi i}\int_\Gamma \log z (z\mathbf{1}-\mathfrak{a})^{-1}\;dz\;\;\; \;\;\;\;\;
\forall\,\mathfrak{a}\in B_1(\textbf{\emph{1}})
\end{eqnarray}
for $\Gamma$ a partially smooth contour in $\C$ enclosing the spectrum $\sigma(\mathfrak{a})$.
\end{definition}
Then, as an open subset of $\mathfrak{A}$, the units $A_{\mathfrak{A}}$ inherit a manifold structure, and
the complex-analytic structure of a CCIA is enough to endow $A_{\mathfrak{A}}$ with a Lie group structure.
\begin{definition}
A topological group $G$ is a Lie group if there exists a neighborhood $U$ of $\{e\}$
such that, for every subgroup $H$, if $H\subseteq U$ then $H=\{e\}$.
\end{definition}
\begin{theorem}\emph{(\cite[prop.~3.2, prop.~3.4, \S~4]{GLO})}
Let $A_{\mathfrak{A}}$ be the set of units of a \textbf{complex} \emph{CCIA} $\mathfrak{A}$. Then group inversion $\mathrm{inv}:A_{\mathfrak{A}}\rightarrow A_{\mathfrak{A}}$ is complex-analytic and, hence, $A_{\mathfrak{A}}$ is a complex-analytic Lie group  with exponential map $\mathrm{exp}_{A_{\mathfrak{A}}}
=\mathrm{exp}_{\mathfrak{A}}|_{A_{\mathfrak{A}}}:T_e(A_{\mathfrak{A}})\rightarrow A_{\mathfrak{A}}$ by $\mathfrak{a}\mapsto \mathrm{exp}_{\mathfrak{A}}(t\mathfrak{a})=\phi_{\mathfrak{a}}(t)$ such that $d\phi_{\mathfrak{a}}(d/dt)=\mathfrak{a}\in T_e(A_{\mathfrak{A}})$ and $t\in\R$. If $\mathfrak{A}$ is \textbf{real} \emph{CCIA}, then inversion is real-analytic  and $A_{\mathfrak{A}}$ is a real-analytic Lie group.
\end{theorem}

Moreover, since $\mathfrak{A}$ is Mackey complete, $A_{\mathfrak{A}}$ becomes a \textrm{BCH}-Lie group:
\begin{definition}\emph{(\cite[def.~5.5]{GLO})}
A \emph{\textrm{BCH}}-Lie group $G_{[\,,\,]}$ is a complex-analytic Lie group such that: i) there exists an open $0$-neighborhood $U\subset T_e(G_{[\,,\,]})$  with $V:=\mathrm{exp}_{G_{[\,,\,]}}(U)$ open in $G_{[\,,\,]}$ such that the map $\varphi:=\mathrm{exp}_{G_{[\,,\,]}}|_U^V:U\rightarrow V$ is a diffeomorphism; and ii) there exists a $(0,0)$-neighborhood $W\subseteq U\times U$ such that $\mathrm{exp}_{G_{[\,,\,]}}\mathfrak{a}\cdot\mathrm{exp}_{G_{[\,,\,]}}\mathfrak{b}\subseteq V$ and $\varphi^{-1}(\varphi(\mathfrak{a})\varphi(\mathfrak{b}))$ is the \emph{BCH} series for all $\mathfrak{a},\mathfrak{b}\in W$.
\end{definition}
\begin{proposition}\label{BCH}\emph{(\cite[th.~5.6]{GLO})}
If $\mathfrak{A}$ is \emph{CCIA}, then the group of units $A_{\mathfrak{A}}$ is a \emph{BCH}-Lie group.
\end{proposition}

Given that CCIA $\mathfrak{A}$ yields a complex-analytic Lie group $A_{\mathfrak{A}}$ and $G$ is isomorphic to $A_{\mathfrak{A}}$ by assumption, we can construct the complex Lie group $G^\C$. Recall the definition of the exponential function on a topological Lie group $G$: it associates a $g^1\in G$ with a one-parameter subgroup $\phi_{\mathfrak{g}}\in\mathfrak{L}(G)$ by
$\mathrm{exp}_G\,\phi_{\mathfrak{g}}=\phi_{\mathfrak{g}}(1)=:g^1$ where $\mathfrak{g}=d\phi_{\mathfrak{g}}(d/dt)$. Then,
by the definition of one-parameter subgroups,
$\phi_{\mathfrak{g}}(t)=\mathrm{exp}_G(t\mathfrak{g})=:g^t$ with $t\in\R$, and the short-hand notation $g^t$ can be formally interpreted as the $t$-th power of $g$ in the sense that $g^t=\mathrm{exp}_G(t\,\mathrm{log}_G(g^1))$. Evidently the $t$-th power is characterized by two fiducial points $\phi_{\mathfrak{g}}(0)=e$ and $\phi_{\mathfrak{g}}(1)=g^1=g$.

To extend one-parameter subgroups to complex $G^\C$ follow \cite[pg. 23]{LEE}. Let $\mathcal{L}(\C)$ denote the Lie algebra of $\C$ and consider the morphism $\gamma':\mathcal{L}(\C)\cong\C\rightarrow \mathfrak{G}^\C$ by $z\mapsto z\mathfrak{g}$. The real analytic group $\C_\R$ underlying $\C$ is simply connected so there is a unique, real analytic homomorphism $\widetilde\phi_{\mathfrak{g}}:\C\rightarrow G^\C$ defined by $\widetilde\phi_{\mathfrak{g}}(z):=(\exp_G\circ\gamma')(z)=\exp_G(z\mathfrak{g})$. Now, since $d\widetilde{\phi}_\mathfrak{g}=\gamma'$ is a morphism of complex Lie algebras, $\widetilde{\phi}_\mathfrak{g}$ is in fact complex analytic and $d\widetilde{\phi}_\mathfrak{g}(1)=\mathfrak{g}$. Hence $\widetilde{\phi}_\mathfrak{g}$ is a one-parameter subgroup of $G^\C$. Denote the set of complex one-parameter subgroups by $\mathfrak{L}(G^\C):=\mathrm{Hom}_C(\C,G^\C)$.  This allows the definition of the exponential map $\exp_{G^\C}:\mathfrak{L}(G^\C)\rightarrow G^\C$ by $\widetilde{\phi}_{\mathfrak{g}}\mapsto\exp_{G^\C}\,\widetilde{\phi}_{\mathfrak{g}}=\widetilde{\phi}_{\mathfrak{g}}(1) $. Note that $\widetilde{\phi}_{\mathfrak{g}}(0)=e$ and
$\widetilde{\phi}_{\mathfrak{g}}(1)=g^1=g$. Formally interpret $g^z=\mathrm{exp}_{G^\C}(z\,\mathrm{log}_{G^\C}(g^1))$  as a complex power of $g$.

\begin{proposition}\emph{(\cite[prop.~5.40]{HM} ; \cite[th.~1.15]{LEE})}
The exponential $\exp_{G^\C}:\mathfrak{L}(G^\C)\rightarrow G^\C$ is a complex analytic map such that $\exp_{G^\C}=\exp_{G}$.
\end{proposition}

\emph{Proof}:
First, $\mathfrak{L}(G)$ is homeomorphic to $\mathfrak{G}$; which can be identified with the Lie algebra of $G^\C$ over $\R$ so $\exp_{G^\C}\mathfrak{g}=\widetilde{\phi}_{\mathfrak{g}}(1)=\exp_{G}\mathfrak{g}$ for all $\mathfrak{g}\in\mathfrak{G}^\C$. Therefore $\exp_{G^\C}=\exp_{G}$. Now let $U\subset\mathfrak{G}^\C$ be an open connected neighborhood of $0\in\mathfrak{G}^\C$ where $\exp_{G}$ is invertible. The inverse $\exp_{G}^{-1}$ provides a local chart of $V:=\exp_{G}(U)\subset G^\C$ at the identity, and $\exp_{G}$ commutes with the almost complex structures on $U$ and $V$. Hence, $\exp_{G}$ is complex analytic in $U$. To extend the analyticity to all of $\mathfrak{G}^\C$, pick any $\mathfrak{g}\in\mathfrak{G}^\C$ and note that $n^{-1}\mathfrak{g}\in U$ for some $n\in\mathbb{N}_+$. Then $\exp_{G}(n^{-1}\mathfrak{g})\in V$ is complex analytic. Therefore $\left(\exp_{G}(n^{-1}\mathfrak{g})\right)^n=\exp_{G}\mathfrak{g}$ implies $\exp_{G}$ (and hence $\exp_{G^\C}$) is complex analytic at any $\mathfrak{g}\in\mathfrak{G}^\C\cong\mathfrak{L}(G^\C)$.
$\QED$

We are ready to define the functional Mellin transform given data $(G^\C,\mathfrak{C}^\ast,G^\C_\Lambda)$ where $G$ is isomorphic to the group of units of a complex CCIA.
\begin{definition}\label{Mellin def.}
Given a topological group $G^\C$, let $\rho:G^\C\rightarrow
\mathfrak{C}^\ast$  be a strictly-continuous injective representation, and let
$\pi:\mathfrak{C}^\ast\rightarrow L_B(\mathcal{H})$ be a representation. Consider the space of integrable, equivariant
functionals ${\widetilde{\mathbf{F}}}(G^\C)\subseteq{\mathbf{F}}(G^\C)$ where $\mathrm{F}\in \mathrm{Mor}_C(G^\C,\mathfrak{C}^\ast)$ is
equivariant under right-translations by $G^\C$ according to $\mathrm{F}(gh)=\mathrm{F}(g)\rho(h)$.\footnote{This prescription is for left-invariant Haar measures. For right-invariant Haar measures impose equivariance under left-translations.} Let $\mathfrak{C}^\ast$ be a unital $C^\ast$-algebra whose involution induces an involution on $\widetilde{\mathbf{F}}(G^\C)$ given by $\mathrm{F}^\ast(g g^\alpha):=\rho(g^{-\alpha})^\ast \mathrm{F}(g^{-1})^\ast
\Delta(g^{-1})$ with $\alpha\in\C$ and $\Delta(g^{-1})$ defined by restriction to $G^\C_\Lambda$\emph{\cite{LA1}}. Then the functional Mellin
transform
$\mathcal{M}_\lambda:\widetilde{\mathbf{F}}(G^\C) \rightarrow \mathfrak{C}^\ast$ is
defined by
\begin{equation}
\mathcal{M}_\lambda\left[\mathrm{F};\alpha\right]
:=\int_{G^\C}\mathrm{F}(gg^{\alpha})\;\mathcal{D}_\lambda g=\int_{G^\C}\mathrm{F}(g)\rho(g^{\alpha})\;\mathcal{D}_\lambda g
\end{equation}
such that  $g^\alpha:=\exp_G(\alpha\log_{G^\C}g)$, and $\pi\left(\mathrm{F}(gg^{\alpha})\right)=\pi(\mathrm{F}(g)\rho(g^{\alpha}))\in
L_B(\mathcal{H})$ where the space of bounded linear operators $L_B(\mathcal{H})$ is given the strict topology. Denote the space of Mellin integrable functionals by $\mathbf{F}_{\mathbb{S}}(G^\C)$.\footnote{The class of
functional Mellin transforms defined here includes the integrated form of a covariant representation of a dynamical system\cite{W} as a special case. To relate
the integrated form to functional Mellin transforms, require $\pi\circ\rho$ to be a strongly continuous unitary
representation $U:G^\C_{\lambda}\rightarrow L_B(\mathcal{H})$. Then $\pi(f(g_\lambda
h_\lambda))=\pi(f(g_\lambda)\rho(h_\lambda)) =\pi(f(g_\lambda))U(h_\lambda)$ and the integrated form $\pi\rtimes
U(f)$ is equivalent to $\pi(\mathcal{M}_\lambda\left[\mathrm{F} ;1\right])$; although our definitions of $\ast$-convolution and involution are different. See Appendix \ref{relation to crossed products} for further discussion.}  Unless specified otherwise, we take the standard branch for $\mathrm{log}_{G^\C}$ when $\pi(\rho(g))\in L_B(\mathcal{H})$.\emph{\cite{TAO}}
\end{definition}

Since we don't have a definition of Mellin transform for generic locally compact topological groups, we have first defined the \emph{functional} Mellin transform. According to Definition \ref{int-def} then, a sufficient condition for the functional Mellin transform to exist is for $f(g)\rho(g^{\alpha})$ to be integrable precisely when $\alpha\in\mathbb{S}$. This extends the usual definition of Mellin transform to the case of Banach-valued integrals over locally compact topological groups:
\begin{definition}
Given a locally compact topological group $G_\lambda^\C$, let $\rho:G_{\lambda}^\C\rightarrow \mathfrak{C}^\ast$  be a strictly-continuous injective
representation, and consider equivariant functions ${f}\in \mathrm{Mor}_C(G_{\lambda}^\C,\mathfrak{C}^\ast)$ by $g^{1+\alpha}\mapsto f(g)\rho(g^\alpha)$  such that ${f}\in C_C(G_{\lambda}^\C,\mathfrak{C}^\ast)$  for all $\alpha\in
\mathbb{S}\subset\C$.\footnote{Proposition \ref{banach integration} and  Definition \ref{int-def} were stated for $f\in
L^1(G_\lambda,\mathfrak{B})$. In order to quote \cite{W} precisely and thereby avoid introducing technical
difficulties, we restrict here to $f\in C_C(G_\lambda,\mathfrak{B}^\ast)$ where $C_C(G_\lambda,\mathfrak{B}^\ast)$
denotes the set of continuous, compactly-supported functions $f:G_\lambda\rightarrow \mathfrak{B}^\ast$ and $\mathfrak{B}^\ast$ is a
$C^\ast$-algebra. But we point out that $C_C(G_\lambda,\mathfrak{B}^\ast)$ is dense in
$L^1(G_\lambda,\mathfrak{B}^\ast)$ since $G_\lambda$ is locally compact.\cite{W}} Then the functional $\mathrm{F}$ is Mellin integrable if
\begin{equation}\label{Mellin}
  \left|\mathcal{M}_\lambda\left[\mathrm{F};\alpha\right]\right|
  \leq\int_{G^\C_{\lambda}}|f(g_\lambda)\rho(g_\lambda^\alpha)|
  \,d\nu(g_\lambda)<\infty\,,\;\;\;\;\alpha\in\mathbb{S}\;.
\end{equation}
We say the Mellin transform $\mathcal{M}_\lambda\left[\mathrm{F};\alpha\right]$ exists in the fundamental region $\mathbb{S}$ (which may be empty). To emphasize  the fundamental region depends on $\lambda$, we  sometimes write $\mathbb{S}_\lambda$.

Identifying the tangent space $T_eG^\C$ with $\mathfrak{L}(G^\C)$, the Mellin integral can be explicitly formulated on $\mathfrak{L}(G_\lambda^\C)$;
\begin{equation}\label{mellin integral}
  \int_{G_{\lambda}^\C}f(g_\lambda)\rho(g_\lambda^\alpha)
  \,d\nu(g_\lambda)=\int_{\mathfrak{L}(G_{\lambda}^\C)}f(\exp_{G_{\lambda}}\mathfrak{g})
  \rho(\exp_{G_{\lambda}}(\alpha\mathfrak{g}))\,
  |\det\,d_{\mathfrak{g}}\exp_{G_{\lambda}}\mathfrak{g}|\;d\mathfrak{g}\;.
\end{equation}
This is a multiple integral depending on the dimension of $G_{\lambda}^\C$. The notation $\rho(g_\lambda^\alpha)$ is understood as
\begin{equation}\label{rho}
\rho(g_\lambda^\alpha)=\exp_{G_{\lambda}}(\alpha\rho'(\mathfrak{g}))=\exp_{G_{\lambda}}\left(\alpha\sum_{i=1}^n c_i\,\rho'(\mathfrak{e}_i)\right)
=:\exp_{G_{\lambda}}\left(\sum_{i=1}^n\alpha_i\,\rho'(\mathfrak{e}_i)\right)
\end{equation}
where $\mathrm{dim}_\C(G_{\lambda}^\C)=n$ and $\alpha$ is a multi-index relative to the basis $\{\mathfrak{e}_i\}$.

\end{definition}

\begin{example}\label{elementary kernel}
Consider a massless free particle  on $\R^d$. The degrees of freedom are encoded by a continuous map $x:\R\rightarrow \R^d$. In this context, $\mathcal{H}=L^2(\R^d)$, $\mathfrak{C}^\ast=L_B(L^2(\R^d))$,  and the observable $\mathrm{F}=\mathrm{E}^{-\Delta}$ for position-to-position boundary conditions gives rise to the heat kernel $\langle x_a'|\mathrm{F}(g)|x_a \rangle\equiv e^{-i\mathrm{S}_{x_a,x'_a}(g)}$.  Position-to-position free-particle evolution dictates the topological localization $\lambda_{\R_+}:G^\C\rightarrow\R_+$ where $\R_+$ is the multiplicative group of positive reals, and the action functional localizes to
 \begin{equation}
  \mathrm{S}_{x_a,x'_a}(g)\stackrel{\lambda_{\R_+}}{\rightarrow}
  {S}_{x_a,x'_a}(\tau)=\pi|\mathrm{x}_{a'}-\mathrm{x}_a|^2/\tau+\frac{d}{2}\log \tau
\end{equation}
where $\tau\equiv g_\lambda\in\R_+$. For the standard Haar measure \emph{(sH)} on $\R_+$, then\footnote{We will often use the $\lambda$ subscript in $\mathcal{M}_\lambda$ to indicate both the localization $\lambda:G^\C\rightarrow G^\C_\lambda$ and the chosen normalization of the Haar measure $\nu(g_\lambda)$.} $\mathcal{M}_{\R_+,\mathrm{sH}}\left[\mathrm{E}^{-{\mathrm{S}_{x_a,x'_a}}};1\right]$  is the elementary kernel of the Laplacian $\Delta$ on $\R^d$ obeying position-to-position boundary conditions. Explicitly, the position-to-position elementary kernel of the Laplacian on $\R^d$ is given by
\begin{eqnarray}
  {K}(\mathrm{x}_{a'},\mathrm{x}_{a})
  &=&\langle\mathrm{x}_{a'}|\mathcal{M}_{\R_+,\mathrm{sH}}\left[\mathrm{E}^{-{\Delta}};1\right]|\mathrm{x}_{a}\rangle\notag\\
  &=&\mathcal{M}_{\R_+,\mathrm{sH}}\left[\langle\mathrm{x}_{a'}|\mathrm{E}^{-{\Delta}}|\mathrm{x}_{a}\rangle;1\right]\notag\\
  &=&\mathcal{M}_{\R_+,\mathrm{sH}}\left[\mathrm{E}^{-{\mathrm{S}_{x_a,x'_a}}};1\right]\notag\\
  &=&\int_{\R_+}\exp\left\{\frac{-\pi|\mathrm{x}_{a'}-\mathrm{x}_a|^2}{\tau}\right\}\tau^{1-d/2}\;d\log \tau\notag\\
  &=&\left\{\begin{array}{ll}
              -2\log(\pi|\mathrm{x}_{a'}-\mathrm{x}_a|) & \;\;\;\;d=2 \\
              \pi^{1-n/2}\Gamma(n/2-1)|\mathrm{x}_{a'}-\mathrm{x}_a|^{2-d} & \;\;\;\;d> 2
            \end{array}\right.\;.
\end{eqnarray}
Notice the same result obtains for $\tau\rightarrow\tau^{-1}$ as a consequence of the Haar measure.
\end{example}

\subsection{Algebraic properties}
The functional Mellin transform inherits important properties from $\mathrm{int}_\Lambda$ that follow from equivariance, Proposition \ref{banach integration}, and Definition
\ref{int-def}. First note that if $\alpha=0\in\mathbb{S}$ then functional Mellin reduces $\mathcal{M}_\lambda\rightarrow \mathrm{int}_\lambda$, so we will not consider this case any longer.

It's easy to show that $(\mathrm{F}_1\ast\mathrm{F}_2)(gg^\alpha)=(\mathrm{F}_1\ast\mathrm{F}_2)(g)\rho(g^\alpha)$ so $\widetilde{\mathbf{F}}(G^\C)$ is a subalgebra of $\mathbf{F}(G^\C)$. Define a norm on $\mathbf{F}_{\mathbb{S}}(G^\C)$ by $\|\mathrm{F}\|_{\mathbb{S}}
:=\mathrm{sup}_{\alpha,\lambda}\|\mathcal{M}_\lambda\left[\mathrm{F};\alpha\right]\|$ with $\alpha\in\mathbb{S}_\lambda$. Complete $\mathbf{F}_{\mathbb{S}}(G^\C)$ with respect to $\|\cdot\|_{\mathbb{S}}$ (or some other suitably defined norm). Then $\mathbf{F}_{\mathbb{S}}(G^\C)$ is Banach because $\mathfrak{C}^\ast$ is Banach.

\begin{lemma}\label{pi}
If $\mathrm{F}\in \mathbf{F}_{\mathbb{S}}(G^\C)$, then
\begin{equation}
\mathcal{M}^\pi_\lambda\left[\mathrm{F};\alpha\right]
:=\pi(\mathcal{M}_\lambda\left[\mathrm{F};\alpha\right])
=\mathcal{M}_\lambda\left[\pi\circ\mathrm{F};\alpha\right]\;.
\end{equation}
\end{lemma}
\emph{Proof}: By definition, $\mathrm{F}\in \mathbf{F}_{\mathbb{S}}(G^\C)$ implies $f$ is Mellin integrable for some $\mathbb{S}_\lambda$. So
\begin{eqnarray}
\pi(\mathcal{M}_\lambda\left[\mathrm{F};\alpha\right])
&=&\pi\left(\int_{G^\C}\mathrm{F}(gg^{\alpha})\mathcal{D}_\lambda g\right)\notag\\
&=&\pi\left(\int_{G_{\lambda}^\C}f(g_\lambda g_\lambda^{\alpha})
\;d\nu(g_\lambda)\right)\,,\;\;\;\;\;\;\alpha\in\mathbb{S}_\lambda\notag\\
&=&\int_{G_{\lambda}^\C}\pi\left(f(g_\lambda g_\lambda^{\alpha})\right)
\;d\nu(g_\lambda)\,,\;\;\;\;\;\;\alpha\in\mathbb{S}_\lambda
\end{eqnarray}
and the third line follows from Proposition \ref{banach integration}.
$\QED$

Crucially, under certain conditions $\mathcal{M}_\lambda$ is a $\ast$-homomorphism:

\begin{lemma}\label{commutative}
If $\mathfrak{C}^\ast$ is commutative, then
\begin{equation}
\mathcal{M}_{\lambda}\left[\left(\mathrm{F}_1\ast
\mathrm{F}_2\right);\alpha\right]
= \mathcal{M}_{\lambda}\left[
\mathrm{F}_1;\alpha\right]\mathcal{M}_{\lambda}\left[ \mathrm{F}_2;\alpha\right]
\end{equation}
and
\begin{equation}
\mathcal{M}_{\lambda}\left[\left(\mathrm{F}_1\star
\mathrm{F}_2\right);\alpha\right]= \mathcal{M}_{\lambda}\left[
\mathrm{F}_1;\alpha\right]\mathcal{M}_{\lambda}\left[ \mathrm{F}_2;1-\alpha\right]\;.
\end{equation}
\end{lemma}
\emph{Proof}: For $\mathfrak{C}^\ast$ commutative, $[\rho({g}),\rho({h})]=0$ for all ${g},{h}\in{G}^\C$ and $[\rho'(\mathfrak{g}),\rho'(\mathfrak{h})]=0$ for all $\mathfrak{g},\mathfrak{h}\in\mathfrak{G}^\C$. Use Proposition \ref{BCH}  to write $\exp_G\mathfrak{g}\,\exp_G\mathfrak{h}
=\exp_G\left(\mathfrak{g}+\mathfrak{h}+\mathrm{C}([\mathfrak{g},\mathfrak{h}])\right)$ where $\mathrm{C}([\mathfrak{g},\mathfrak{h}])$ represents the commutator terms in the BCH series. Then
\begin{eqnarray}
\rho((gh)^\alpha)
&=&\rho((\exp_G\mathfrak{g}\,\exp_G\mathfrak{h})^\alpha)\notag\\
&=&e^{\alpha\,\rho'(\mathfrak{g}+\mathfrak{h}+\mathrm{C}([\mathfrak{g},\mathfrak{h}]))}\notag\\
&=&e^{\alpha\,\rho'(\mathfrak{g})}e^{\alpha\,\rho'(\mathfrak{h})}
=\rho(h^\alpha)\rho(g^\alpha)\;.
\end{eqnarray}
We used (\ref{rho}) in line two. Line three follows from  $[\rho({g}),\rho({h})]=0$ and $\mathrm{C}([\mathfrak{g},\mathfrak{h}])=-\mathrm{C}([-\mathfrak{g},-\mathfrak{h}])$ implies $\rho'(\mathrm{C}([\mathfrak{g},\mathfrak{h}]))=0$ and then using $[\rho'(\mathfrak{g}),\rho'(\mathfrak{h})]=0$.

Now put $g\rightarrow \tilde{g}g$ in the functional Mellin transform of the $\ast$-convolution. The
integral is invariant under this transformation by virtue of the
left-invariant Haar measure on $G^\C_\lambda$;
\begin{eqnarray}
\mathcal{M}_{\lambda}\left[(\mathrm{F}_1\ast \mathrm{F}_2);\alpha\right]&=&\int_{G^\C\times G^\C} \mathrm{F}_1(\tilde{g})\mathrm{F}_2(g(\tilde{g} g)^{\alpha})
\mathcal{D}_{\lambda}{\tilde{g}}\mathcal{D}_{\lambda} g\notag\\
&=&\int_{G^\C_\lambda\times G^\C_\lambda}
f_1(\tilde{g}_{\lambda})\rho(\tilde{g}_{\lambda}^{\alpha})
f_2({g}_{\lambda})\rho(g_{\lambda}^{\alpha})
\;d\nu_{{\lambda}}(\tilde{g}_{\lambda})d\nu_{\lambda} (g_{\lambda}) \notag\\
&=&\int_{G^\C\times G^\C}
\mathrm{F}_1(\widetilde{g}\widetilde{g}^{\alpha})\,\mathrm{F}_2({g}g^{\alpha})\,\mathcal{D}_{\lambda}
\widetilde{g} \mathcal{D}_{\lambda}{{g}}\notag\\
&=&\mathcal{M}_{\lambda}\left[\mathrm{F}_1;\alpha\right]
\mathcal{M}_{\lambda}\left[\mathrm{F}_2;\alpha\right]\;.
\end{eqnarray}
The second line uses commutativity of $\mathfrak{C}^\ast$, and the last equality uses Fubini which follows from Proposition \ref{banach
integration} and Definition \ref{int-def}. The $\star$-convolution proof follows similarly. $\QED$

\begin{lemma}\label{noncommutative}
If $\mathfrak{C}^\ast$ is non-commutative, but $G^\C$ is abelian, $\rho$ is unitary, and $\alpha_{\mathrm{\Re}}\in \R\cap\mathbb{S}$;
\begin{equation}
\mathcal{M}_{{\lambda}}\left[\left(\mathrm{F}_1\ast
\mathrm{F}_2\right);\alpha_{\mathrm{\Re}}\right]= \mathcal{M}_{{\lambda}}\left[
\mathrm{F}_1;\alpha_{\mathrm{\Re}}\right]\mathcal{M}_{{\lambda}}\left[ \mathrm{F}_2;\alpha_{\mathrm{\Re}}\right]
\end{equation}
and
\begin{equation}
\mathcal{M}_{{\lambda}}\left[\left(\mathrm{F}_1\star
\mathrm{F}_2\right);\alpha_{\mathrm{\Re}}\right]= \mathcal{M}_{{\lambda}}\left[
\mathrm{F}_1;\alpha_{\mathrm{\Re}}\right]\mathcal{M}_{{\lambda}}\left[ \mathrm{F}_2;1-\alpha_{\mathrm{\Re}}\right]\;.
\end{equation}
\end{lemma}
\emph{Proof}:
If $G_{\lambda}^\C$ is abelian,
\begin{equation}
\rho((gh)^\alpha)=\rho(e^{\alpha(\log g+\log h)})
=\rho(e^{\alpha\log h}e^{\alpha\log g})
=\rho(h^\alpha)\rho( g^\alpha)\;.
\end{equation}
Since $\rho$ is unitary, $\rho'(\mathfrak{g})=\rho'(\log_G g)$ is anti-Hermitian and therefore $\rho(g^{-\alpha_{\mathrm{\Re}}})^\ast=e^{-\alpha^\ast_{\mathrm{\Re}}\rho'(\mathfrak{g})^\ast}
=e^{\alpha^\ast_{\mathrm{\Re}}\rho'(\mathfrak{g})}=\rho(g^{\alpha_{\mathrm{\Re}}})$. Hence,
\begin{eqnarray}
\int_{G^\C\times G^\C} \mathrm{F}_1(\tilde{g})(\mathrm{F}_2^\ast)^\ast(g(\tilde{g} g)^{\alpha})
\mathcal{D}_{{\lambda}}{\tilde{g}}\mathcal{D}_{{\lambda}}
g\notag\\
&&\hspace{-1.5in}=\int_{\overline{G}_{{\lambda}}^\C\times \overline{G}_{{\lambda}}^\C}
f_1(\tilde{g}_{{\lambda}})\rho(\tilde{g}_{{\lambda}}^{\alpha})
\rho(g_{{\lambda}}^{-\alpha})^\ast (f_2^\ast)({g}^{-1}_{{\lambda}})^\ast\Delta({g}^{-1}_{{\lambda}})
\;d\nu_{{{\lambda}}}(\tilde{g}_{{\lambda}})d\nu_{{\lambda}} (g_{{\lambda}}) \notag\\
&&\hspace{-1.5in}=\int_{G^\C\times G^\C}
\mathrm{F}_1(\widetilde{g}\widetilde{g}^{\alpha})\,(\mathrm{F}_2^\ast)^\ast({g}g^{\alpha})
\,\mathcal{D}_{{\lambda}}\widetilde{g} \mathcal{D}_{{\lambda}}{{g}}\notag\\
&&\hspace{-1.5in}=\int_{G^\C\times G^\C}
\mathrm{F}_1(\widetilde{g}\widetilde{g}^{\alpha})\,\mathrm{F}_2({g}g^{\alpha})\,\mathcal{D}_{{\lambda}}
\widetilde{g} \mathcal{D}_{{\lambda}}{{g}}\;.
\end{eqnarray}
The  first equality follows from $\rho(\tilde{g}^{-\alpha_{\mathrm{\Re}}})^\ast=\rho(\tilde{g}^{\alpha_{\mathrm{\Re}}})$ and the definition of involution.  If instead $\rho$ is assumed real, $\alpha_{\mathrm{\Re}}$ gets replaced by $\alpha_{\Im}\in i\R\cap\mathbb{S}$. If $\rho(g)\in Z(\mathfrak{C}^\ast)$ (e.g. $\rho(g)=\det g$) then there is no restriction on $\alpha\in\mathbb{S}$. The proof for the $\star$-convolution follows similarly.
$\QED$

Finally, for the most general case of non-commutative $\mathfrak{C}^\ast$ and non-abelian $G^\C$, we must restrict
to $\alpha=1$ for unitary $\rho$ (or $\pm i$ for real $\rho$) to get an algebra representation:
\begin{lemma}
If $\mathfrak{C}^\ast$ is non-commutative and $G^\C$ is non-abelian, but $\rho$ is unitary, then
\begin{equation}
\mathcal{M}_{{\lambda}}\left[\left(\mathrm{F}_1\ast
\mathrm{F}_2\right);1\right]= \mathcal{M}_{{\lambda}}\left[
\mathrm{F}_1;1\right]\mathcal{M}_{{\lambda}}\left[ \mathrm{F}_2;1\right]
\end{equation}
and
\begin{equation}
\mathcal{M}_{{\lambda}}\left[\left(\mathrm{F}_1\star
\mathrm{F}_2\right);1\right]= \mathcal{M}_{{\lambda}}\left[
\mathrm{F}_1;1\right]\mathcal{M}_{{\lambda}}\left[ \mathrm{F}_2;0\right]\;.
\end{equation}
\end{lemma}
\emph{Proof}: Use $\rho((gh)^{-1})^\ast=\rho(gh)=\rho(g)\rho(h)
=\rho(g)\rho(h^{-1})^\ast$ in the previous argument.  $\QED$

\begin{proposition}\label{prop 4.1}
With $\alpha\in\mathbb{S}$ suitably restricted according to the previous lemmas, $\mathcal{M}_\lambda$ is a $\ast$-homomorphism.
\end{proposition}
\emph{Proof}: Given the preceding lemmas, we only need to show
\begin{eqnarray}
\mathcal{M}^\ast_\lambda\left[\mathrm{F};\alpha\right]:=(\mathcal{M}_\lambda\left[\mathrm{F};\alpha\right])^\ast
&=&\left(\int_{G^\C}(\mathrm{F}(gg^{\alpha}))\mathcal{D}_\lambda g\right)^\ast\notag\\
&=&\int_{G_{\lambda}^\C}\left(f(g_\lambda g_\lambda^{\alpha})\right)
^\ast\;d\nu(g_\lambda)\notag\\
&=&\int_{G_{\lambda}^\C}\rho(g_\lambda^{\alpha})^\ast
\left(f(g_\lambda)\right)^\ast
\;d\nu(g_\lambda)\notag\\
&=&\int_{G_{\lambda}^\C}\rho(g_\lambda^{-\alpha})^\ast
f(g^{-1}_\lambda)^\ast
\Delta(g^{-1}_\lambda)\;d\nu(g_\lambda)\notag\\
&=&\int_{G^\C}\mathrm{F}^\ast(gg^{\alpha}) \mathcal{D}_\lambda g\notag\\
&=&\mathcal{M}_\lambda\left[{\mathrm{F}}^\ast;\alpha\right]
\end{eqnarray}
where we used
$f^\ast(g_\lambda g^\alpha_\lambda)=\rho(g^{-\alpha}_\lambda)^\ast f(g^{-1}_\lambda)^\ast
\Delta(g^{-1}_\lambda)=\rho(g^{-\alpha}_\lambda)^\ast f^\ast(g_\lambda)$. In particular $\mathrm{Id}^\ast=\mathrm{Id}$ if $\mathfrak{B}$ is a $\ast$-algebra.
 $\QED$

For later use we state;
\begin{corollary}\label{Plancherel}
If $\mathfrak{C}^\ast$ is commutative and $\alpha=1/2\pm i\sigma$ with $\sigma\in\R_+$,
\begin{equation}
\mathcal{M}_{{\lambda}}\left[\left(\mathrm{F}\ast
\mathrm{F}^\ast\right);1/2\pm i\sigma\right]
=\left|\mathcal{M}_{{\lambda}}\left[\mathrm{F};1/2\pm i\sigma\right]\right|^2
=\mathcal{M}_{{\lambda}}\left[\left(\mathrm{F}\star
\mathrm{F}^\ast\right);1/2\mp i\sigma\right]\;.
\end{equation}

If $\mathfrak{C}^\ast$ is non-commutative but $G^\C$ is abelian with $\rho$ unitary and $\alpha=1/2$,
\begin{equation}
\mathcal{M}_{{\lambda}}\left[\left(\mathrm{F}\ast
\mathrm{F}^\ast\right);1/2\right]
=\left|\mathcal{M}_{{\lambda}}\left[\mathrm{F};1/2\right]\right|^2
=\mathcal{M}_{{\lambda}}\left[\left(\mathrm{F}\star
\mathrm{F}^\ast\right);1/2\right]\;.
\end{equation}
\end{corollary}

It is important to ensure that $\mathcal{M}_\lambda$ is a $\ast$-homomorphism if we want to use functional Mellin in quantum physics, because then Mellin functional integrals will supply a representation of $\mathbf{F}_{\mathbb{S}}(G^\C)$.
\begin{theorem}\label{prop 4.2}
Let $\mathbb{S}_{\mathcal{R}}$ denote the fundamental region with $\alpha\in\mathbb{S}$ sufficiently restricted to render $\mathcal{M}_\lambda$ a $\ast$-homomorphism. Then $\mathbf{F}_{\mathbb{S}_{\mathcal{R}}}(G^\C)$ is a Banach $C^\ast$-algebra --- when
 endowed with an involution defined by $\mathrm{F}^\ast(g^{1+\alpha}):=\mathrm{F}(g^{-1-\alpha})^\ast\Delta(g^{-1})$
 and a suitable topology.
\end{theorem}
\emph{Proof}: Linearity and $(\mathrm{F}^\ast)^\ast=\mathrm{F}$ are obvious.  Next,
\begin{eqnarray}
\left(\mathrm{F}_1^\ast\ast \mathrm{F}_2^\ast\right)_{{\lambda}}(g^{1+\alpha})
&:=&\int_{G_\lambda}f^\ast_1(\tilde{g}_\lambda)
f^\ast_2(\tilde{g}_\lambda^{-1}g^{1+\alpha}_\lambda)
\;d\nu(\tilde{g}_\lambda)\notag\\
&=&\int_{G_\lambda}\left(f_2(g_\lambda^{-1-\alpha}\tilde{g}_\lambda)
\Delta(g_\lambda^{-1}\tilde{g}_\lambda)f_1(\tilde{g}^{-1}_\lambda)
\Delta(\tilde{g}_\lambda^{-1})\right)^\ast
\;d\nu(\tilde{g}_\lambda)\notag\\
&=&\left(\int_{G_\lambda}f_2(g_\lambda^{-1-\alpha}\tilde{g}_\lambda)
 f_1(\tilde{g}^{-1}_\lambda)\;d\nu(\tilde{g}_\lambda)
\right)^\ast\Delta(g_\lambda^{-1})\notag\\
&=&\left((\mathrm{F}_2\ast \mathrm{F}_1)_{{\lambda}}(g^{-1-\alpha})\right)^\ast\Delta(g^{-1})\notag\\
&=&{\left(\mathrm{F}_2\ast \mathrm{F}_1\right)_{{\lambda}}}^\ast(g^{1+\alpha})
\end{eqnarray}
using left-invariance of the Haar measure to put $\tilde{g}_\lambda\rightarrow g^{1+\alpha}_\lambda\tilde{g}_\lambda$ in the fourth line. This gives $\mathcal{M}_\lambda\left[\left(\mathrm{F}_1^\ast\ast \mathrm{F}_2^\ast\right);\alpha\right]
=\mathcal{M}_\lambda\left[{\left(\mathrm{F}_2\ast \mathrm{F}_1\right)}^\ast;\alpha\right]$. Lastly, since $\mathfrak{C}^\ast$ is a $C^\ast$-algebra in any case, it follows that $\|\mathrm{F}\|_{\mathbb{S}}=\|\mathrm{F}^\ast\|_{\mathbb{S}}$ and the lemmas imply $\|\mathrm{F}\ast\mathrm{F}^\ast\|_{\mathbb{S}}=\|\mathrm{F}\|^2_{\mathbb{S}}$.
$\QED$

It is convenient to denote by $\mathit{\Pi}^{(\alpha)}_\lambda:=\pi(\mathcal{M}_\lambda)$ the $\pi$-representation of functional Mellin and $\mathit{\Pi}^{(\alpha)}(\mathbf{F}_{\mathbb{S}_{\mathcal{R}}}(G^\C))$ its corresponding image in $L_B(\mathcal{H})$ under the various conditions that render $\mathcal{M}_\lambda$ a $\ast$-homomorphism.
\begin{corollary}
$\mathit{\Pi}^{(\alpha)}_\lambda$ is a $\ast$-representation on Hilbert $\mathcal{H}$.
\end{corollary}
For example, if $\mathfrak{C}^\ast$ is non-commutative and unital, then $\mathit{\Pi}^{(\alpha_{\Re})}_\lambda$ denotes
 a functional Mellin $\ast$-representation for an abelian group and unitary $\rho$. Similarly, $\mathit{\Pi}^{(1)}_\lambda$ is a $\ast$-representation for non-commutative
   $\mathfrak{C}^\ast$ and non-abelian $G^\C$. In fact, as already mentioned, $\mathit{\Pi}^{(1)}(\mathbf{F}_{\mathbb{S}_{\mathcal{R}}}(G^\C))$ is closely related to a crossed product\cite{W}.

\section{Mellin functional tools}\label{tools}
Before extracting useful tools from the functional Mellin transform, it is useful to gain some experience and insight by analyzing its reduction to finite-dimensional integrals under various conditions. Appendix \ref{exercises} contains several examples. They suggest how to define Mellin functional counterparts of resolvents, traces, logarithms, and determinants. For the most part, these are familiar objects and many have been constructed and extensively analyzed using a variety of approaches in the literature --- in particular they include resolvents, complex powers of operators, and zeta functions. Our purpose here is to establish them in the functional integral context and to show consistency.

From now on, to further clean up notation no distinction will
be made between $g$, $g_\lambda$, and $\rho(g_\lambda)$ when it will not cause
confusion. Also, instead of always detailing integrable conditions, we will often assume continuous Mellin integrable functions from the beginning.

\subsection{Functional resolvent}
\begin{definition}\label{resolvents}
Let
$\mathrm{A},\mathrm{Z}\in \mathrm{Mor}_C(G^\C,\mathfrak{C}^\ast)$  where $\mathrm{Z}\in Z(\mathbf{F}_{\mathbb{S}}(G^\C))$ is a central element and $\mathrm{E}^{-(\mathrm{A}-\mathrm{Z})}\in \mathbf{F}_{\mathbb{S}}(G^\C)$. Assume $(\mathrm{A}-\mathrm{Z})(g)$ with fixed $\mathrm{A}$ but variable $\mathrm{Z}$ is invertible. Define the functional resolvent of $\mathrm{A}$ by
\begin{equation}
\mathrm{R}_{\lambda}(\mathrm{A},\mathrm{Z};\alpha)
\equiv((\mathrm{A}-\mathrm{Z}))^{-\alpha}_\lambda
:=\mathcal{M}_\lambda\left[\mathrm{E}^{-(\mathrm{A}-\mathrm{Z})};\alpha\right]\;.
\end{equation}
\end{definition}

The following example develops useful realizations of the functional resolvent.
\begin{example}\label{resolvent example}
Consider $(\textsf{a}-\textsf{z})\in G^\C$ (which is invertible) and the associated real one-parameter subgroup ${\phi}_{\mathfrak{a}-\mathfrak{z}}(\R)\subseteq G^\C$ where $\mathrm{exp}_{G^\C}(\mathfrak{a}-\mathfrak{z})=(\textsf{a}-\textsf{z})$. For $g\in{\phi}_{\mathfrak{a}-\mathfrak{z}}(\R)$, define the functional $\mathrm{E}^{-(\mathrm{A}-\mathrm{Z})}(g):=e^{-(\mathrm{A}-\mathrm{Z})(g)}
:=e^{-\rho(\textsf{a}-\textsf{z})\rho(g)}$ such that $\rho(\textsf{a}-\textsf{z})=(A-zId)\in\mathfrak{C}^\ast$ with $z\in\C$.
The functional resolvent for $z\notin\sigma(A)$ is
\begin{equation}\label{resolvent}
\mathrm{R}_{\lambda}(\mathrm{A},\mathrm{Z};\alpha) = \int_{{\phi}_{\mathfrak{a}-\mathfrak{z}}(\R)} e^{-(A-zId)\rho(g)}\,\rho(g^\alpha)\;\mathcal{D}_\lambda g=(A-zId)^{-\alpha}
\;,\;\;\;\;\;\alpha\in\langle0,\infty\rangle\;.
\end{equation}
Note that, since $z\notin\sigma(A)$ is a regular value, $(A-zId)$ is invertible and commutes with all $\rho(g)$, and so it can be extracted from the integral using the invariance of the Haar measure.  After extraction, the remaining integral is a normalization we absorb into the measure; explicitly, we require $\int_{{\phi}_{\mathfrak{a}-\mathfrak{z}}(\R)} e^{-\rho(g)}\rho(g^\alpha)\;\mathcal{D}_\lambda g:=Id\in\mathfrak{C}^\ast$ for all $\lambda\in\Lambda$.  On the other hand, if $z\in\sigma(A)$, then $(A-zId)^{-\alpha}$ can no longer be extracted and the associated integral formally corresponds to a fractional derivative of an imaginary delta functional according to \emph{\cite{LA1}}.\footnote{When $(\mathrm{A}-\mathrm{Z})$ is linear on $G^\C$ it is useful to generalize for any
$\mathfrak{C}^\ast$ the delta functional defined in \cite{LA1} and
to \emph{formally} write (under appropriate conditions on $\mathrm{A}$) $\mathrm{R}_{\lambda}(\mathrm{A},\mathrm{Z};\alpha)=\mathrm{Pv}({A}-{Z})_\lambda^{-\alpha}+i
\delta^{(\alpha-1)'}({A}-{Z})_\lambda$.
Consequently, at $\alpha=1$, $\mathrm{Pv}\,({A}-{Z})_\Gamma^{-1}$ and
$\delta({A}-{Z})_\Gamma$ correspond to the  resolvent set and
 spectrum respectively.  The appearance of
distributions here motivates extending the theory of Mellin transforms of
distributions to the functional context.}

Emphasize \emph{(\ref{resolvent example})} is to be understood as a functional operator identity. That is, $\mathrm{R}_{\lambda}(\mathrm{A},\mathrm{Z};\alpha)$ represents a family of operators in $\mathfrak{C}^\ast$ that can be queried for measurable quantities. Let us exhibit two specific instantiations. To be concrete, let $\mathfrak{C}^\ast\equiv L_B(\mathcal{H})$ with $\mathcal{H}$ separable.

Assume $(A-zId)^{-\alpha}\in L_B(\mathcal{H})$ is diagonalizable and choose the topological localization $\lambda_{\digamma}:{\phi}_{\mathfrak{a}-\mathfrak{z}}(\R)\rightarrow \digamma$ where $\digamma\equiv(\C^\times)^d$ with $\C^\times=\C\backslash\{0\}$ when $d=\mathrm{dim}(\mathcal{H})<\infty$. When $\mathcal{H}$ is infinite-dimensional, $\digamma\equiv(\C^\times)^\C$  where $(\C^\times)^\C$ denotes the abelian group underlying the space of continuous, measurable functions $g_{\lambda_\digamma}:\C\rightarrow\C^\times\cong\R_+\times\mathrm{S}^1$ which must be localized further, say by expected values $\langle i|\rho(g)|j\rangle$ with $|i\rangle,|j\rangle\in\mathcal{H}$. Since $\rho(g)$ commutes with $(A-zId)$, it is also diagonalizable. In a diagonal basis then, \emph{(\ref{resolvent example})} localizes to
\begin{equation}\label{first resolvent}
\mathrm{R}_{\digamma,N}(\mathrm{A},\mathrm{Z};\alpha) = \int_{\digamma} e^{-(A-zId)\rho(g)}\,\rho(g^{\alpha})\;d\nu(g_N)=(A-zId)^{-\alpha}
\;,\;\;\;\;\;\;\;\;\alpha\in\langle0,\infty\rangle
\end{equation}
where $\nu(g_N)=(\log r \times\theta)/N$ with $g=re^{i\theta}$ and $\theta\in[-\pi/2,\pi/2]$ or $\theta\in[\pi/2,3\pi/2]$ depending on whether $\sigma(A-zId)\in\C_+$ or $\sigma(A-zId)\in\C_-$ respectively. The normalizations are $N=(\pi\Gamma(\alpha))^d$  and $N=\int_{\C^\times}\langle i|e^{-g}g^{\alpha}|j\rangle\;d\log r\,d\theta=\pi\Gamma(\alpha)$ in the finite-dimensional and infinite-dimensional case respectively.

Now suppose $A$ is self-adjoint. Since ${\phi}_{\mathfrak{a}-\mathfrak{z}}(\R)$ is a real, one-dimensional abelian topological group, the previous paragraph suggests a finer localization $\lambda_{\R_\pm}:\phi_{\mathfrak{a}-\mathfrak{z}}(\R)\rightarrow (\R_+\cup\R_-)^d$  where $\R_+\cup\R_-:=\R_+\times\{1,-1\}$.\footnote{ Observe that $\R^\times/\{-1,1\}\cong\R_{+}$ so $\R^\times$ is a double-cover of $\R_{+}$. We write $\lambda_{\R_\pm}:\phi_{\mathfrak{a}-\mathfrak{z}}(\R)\rightarrow \R_{+}\cup\R_{-}$ instead of $\lambda_{\R^\times}:\phi_{\mathfrak{a}-\mathfrak{z}}(\R)\rightarrow \R^\times$ because, in a physics context when $e^{-Ag}$ is unitary, we want to interpret the one-parameter subgroup as the forward \emph{and} reverse time-evolution operator, i.e. as the union of two evolution semi-groups. The minus sign for the $\R_{-}$ integral comes from the relative phase of $\pi$ radians between the arguments of elements in $\phi_{\mathfrak{a}-\mathfrak{z}}(\R_{+})$ and $\phi_{\mathfrak{a}-\mathfrak{z}}(\R_{-})$. Essentially, this is just the previous $\digamma$ case with measurable functions $g:\C\rightarrow\R_+\times \mathrm{S}^1$  restricted to $g:\C\rightarrow\R_+\times \{1,-1\}$.} To maintain unit normalization, choose the Haar measure $\nu(g_\Gamma):=\nu(g)/\Gamma(\alpha)=\log g/\Gamma(\alpha)$. In the diagonal basis, the matrix elements of the functional resolvent for self-adjoint $A$ reduce to
\begin{equation}\label{second resolvent}
\langle i|\mathrm{R}_{\R_\pm,\Gamma}(\mathrm{A},\mathrm{Z};\alpha)| j\rangle
=\frac{1}{\Gamma(\alpha)}\left(\int_0^\infty-\int^0_{-\infty}\right)\langle i|j\rangle e^{-(A-z)_jg_j}g_j^{\alpha}\;d\log g_j
=(A-zId)^{-\alpha}_{ij}\delta_{ij}
\end{equation}
where we choose a suitable $\log$ branch and only one of the integrals will contribute depending on whether $\sigma (A-zId)\subset\C_+$ or $\sigma (A-zId)\subset\C_-$. This verifies the operator identity $\mathrm{R}_{\R_\pm,\Gamma}(\mathrm{A},\mathrm{Z};\alpha)=(A-zId)^{-\alpha}$. Evidently, for both $\lambda_\digamma$ and $\lambda_{\R_\pm}$ we could just as well have localized \textbf{after} extracting $(A-zId)^{-\alpha}$ from $\mathrm{R}_{\lambda}(\mathrm{A},\mathrm{Z};\alpha)$.
\end{example}
Definition {\ref{resolvents}} is a ``good'' definition in the sense that it agrees with the standard resolvent when $(A-zId)\in L_B(\mathcal{H})$ and $\alpha=1$ according to (\ref{second resolvent}) and the spectral theorem. Moreover,
\begin{equation}
\frac{1}{\Gamma(n)}\frac{d^{n-1}}{dz^{n-1}}(A-zId)_{ii}^{-1}=\frac{1}{\Gamma(n)}\int_0^{\pm\infty} e^{-(A_i-z_i)g}g^{n}\;d\log g=(A-zId)_{ii}^{-n}
\end{equation}
so the objects agree at all allowed integer values of $\alpha$.

Specializing the functional resolvent defines a functional inverse
power:
\begin{definition}
The functional inverse power of $\mathrm{A}\in \mathrm{Mor}_C(G^\C,\mathfrak{C}^\ast)$ such that $\mathrm{E}^{-\mathrm{A}}\in \mathbf{F}_{\mathbb{S}}(G^\C)$ is
defined by
\begin{equation}
A_\lambda^{-\alpha}\equiv({\mathrm{A}})^{-\alpha}_\lambda
:= \mathrm{R}_{\lambda}(\mathrm{A},0;\alpha)
=\mathcal{M}_\lambda\left[\mathrm{E}^{-\mathrm{A}};\alpha\right]\;.
\end{equation}
\end{definition}
For example, if $\mathfrak{C}^\ast= L_B(L^2(\R^d))$ then $\mathit{\Delta}_\lambda^{-1}=\mathcal{M}_{\lambda}\left[\mathrm{E}^{-\Delta};1\right]$ is the inverse of the Laplacian on $\R^d$. Example \ref{elementary kernel} is the archetype inverse power of the Laplacian on $\R^d$ localized by position-to-position boundary conditions, i.e. $\langle \mathrm{x}_{a'}|\mathrm{R}_{\mathrm{sH}}(\Delta,0;1)|\mathrm{x}_{a}\rangle
=\langle \mathrm{x}_{a'}|\mathit{\Delta}_\mathrm{sH}^{-1}|\mathrm{x}_{a}\rangle
={K}(\mathrm{x}_{a'},\mathrm{x}_{a})$ is the elementary kernel/propagator. (Subscript $\mathrm{sH}$ denotes the standard Haar measure.)

\begin{example}\label{inverse power product}
Consider the same set-up from \emph{Example \ref{resolvent example}} with localization $\lambda_{\digamma}$; except  now with $\mathfrak{z}=0$ and some invertible $B\in\mathfrak{C}^\ast$. Then, for example if $\sigma(A)\subset\C_+$ and $\sigma(B)\subset\C_+$,
\begin{eqnarray}
\mathcal{M}_{\lambda}
\left[\mathrm{E}^{-\mathrm{B}}\ast\mathrm{E}^{-\mathrm{A}};\alpha\right]
&=&\int_{\phi_{\mathfrak{a}}(\R)}\int_{\phi_{\mathfrak{a}}(\R)}\,e^{-B \tilde{g}}\,e^{- A\tilde{g}^{-1}g}\,g^{\alpha}
\;\mathcal{D}_\lambda \tilde{g}\,\mathcal{D}_\lambda g\notag\\
&\stackrel{\lambda_{\digamma}}{\longrightarrow}&
\frac{1}{N^2}\int_{\digamma}\,e^{-B\tilde{g}}(\tilde{g}^{-\alpha})^\ast
A^{-\alpha}\;d\log \tilde{g}\notag\\
&=:&(BA)_{N^2}^{-\alpha}\;,\;\;\;\;\;\;\alpha\in \mathbb{S}_{N,N}
\end{eqnarray}
where we used Fubini, involution $\mathrm{F}(g(\tilde{g}g)^\alpha)=\rho((\tilde{g}g)^{-\alpha})^\ast\mathrm{F}^\ast(g^{-1})^\ast\Delta(g^{-1})$,  and the fact that $A$ commutes with both $g$ and $\tilde{g}$ in the first line. Since $\rho(g)$ commutes with $A^{-\alpha}$ in this case, by \emph{Lemma \ref{noncommutative}} we get $(BA)_{N^2}^{-\alpha}
=\mathcal{M}_{\digamma,N}\left[\mathrm{E}^{-\mathrm{B}};\alpha\right]
\mathcal{M}_{\digamma,N}\left[\mathrm{E}^{-\mathrm{A}};\alpha\right]
=B^{-\alpha}_{N}A^{-\alpha}_{N}=B^{-\alpha}A^{-\alpha}$.\footnote{The fundamental strip of the product is the intersection of the fundamental strips of the two factors. Also, we used $O^{-\alpha}=e^{-\alpha\rho'(\mathfrak{o})}$ for $O\in\mathfrak{C}^\ast$ and $\mathfrak{o}\in\mathfrak{G}^\C$.}
\end{example}

Complex inverse powers are generically only valid for $\Re(\alpha)>0$. Moreover, given some $A\in\mathfrak{C}^\ast$, the functional form $(\mathrm{A})^{-\alpha}_\lambda=:A^{-\alpha}_\lambda$ will not resemble $A^{-\alpha}$ except in some limited cases. This means we can't profitably define a positive power simply by $\mathcal{M}_\lambda[\mathrm{E}^{-{(\mathrm{A}^{-1})}};\alpha]$ when $A$ is invertible, because $((\mathrm{A}^{-1}))^{-\alpha}_\lambda=:(A^{-1})^{-\alpha}_\lambda\neq A^\alpha$ in general for any Haar measure normalization. We will come back to positive powers later, but for now we continue to build tools with inverse powers.

Definition \ref{resolvents} and the characterization of delta functionals in \cite{LA1}, suggest that the functional inverse complex power of $\mathrm{A}$ implicitly includes derivatives of Dirac delta
functionals if $0\in\sigma(A)$. On the other hand, if $A$ has $0\notin\sigma(A)$, then we can define a functional zeta.
\begin{definition}\label{zeta}
If $A_\lambda^{-\alpha}$ is trace class with $0\notin\sigma(A)$, then the functional zeta is defined by
\begin{equation}
\zeta_{{A}_{\lambda}}(\alpha)
:=\mathrm{tr}\,{A}_\lambda^{-\alpha}
=\mathrm{tr}\,\mathcal{M}_\lambda\left[\mathrm{E}^{-\mathrm{A}};\alpha\right]\;.
\end{equation}
\end{definition}
Of course this trace is not always a well-defined object for all $\alpha$ in the fundamental region of
${A}_\lambda^{-\alpha}$. Rather, the fundamental region containing $\alpha$ is dictated by $\lambda$ and is
often restricted. An effective strategy to quantify the restriction is to move the trace inside the integral.
Presumably the trace of the integrand will be well-defined for certain choices of $\lambda$ and the properties of the
functional Mellin transform will allow the domain of $\alpha$ to be determined. This strategy leads us to the next
subsection.

\subsection{Functional trace, logarithm, and determinant}
In this subsection we define and explore the functional Mellin analogs of trace, logarithm, and determinant.

\begin{definition}
Let $\mathrm{A}\in \mathrm{Mor}_C(G^\C,\mathfrak{C}^\ast)$ with
$\mathrm{E}^{-\mathrm{A}}\in \mathbf{F}_{\mathbb{S}}(G^\C)$ for some $\lambda$-dependent fundamental region
$\alpha\in\mathbb{S}_\lambda$, and let $A\in\mathfrak{C}^\ast$ such that $0\notin\sigma(A)$. The functional trace $\mathrm{Tr}\,\mathrm{A}\in\mathrm{Mor}_C(G^\C,\C)$ is defined by
\begin{equation}
\mathrm{Tr}\,{A}_\lambda^{-\alpha}\equiv(\mathrm{Tr}\,{\mathrm{A}})^{-\alpha}_\lambda
:=\mathcal{M}_\lambda\left[\mathrm{Tr}\,\mathrm{E}^{-\mathrm{A}};\alpha\right]
:=\int_{G^\C}\mathrm{tr}\left( e^{-\mathrm{A}(g)}g^\alpha\right)\,\mathcal{D}_\lambda g\;.
\end{equation}
\end{definition}

As a consequence of
Proposition \ref{banach integration}, the interchange of the ordinary trace and
functional integral is valid \emph{only for
$\mathrm{E}^{-\mathrm{A}}\in\mathbf{F}_{\mathbb{S}}(G^\C)$ and appropriate
$\lambda$}. Then according to the definition,
\begin{eqnarray}\label{trace}
\mathrm{tr}\,{A}_\lambda^{-\alpha}
&=&\mathrm{tr}\,\mathcal{M}_\lambda\left[\mathrm{E}^{-\mathrm{A}};\alpha\right]\notag\\
&=&\mathrm{tr}\,\left(\int_{G^\C}e^{-\mathrm{A}(g)}g^\alpha\;\mathcal{D}_\lambda g\right)\;,
\;\;\;\;\;\alpha\in\mathbb{S}_\lambda\notag\\
&=&\int_{G^\C}\mathrm{tr}\left( e^{-\mathrm{A}(g)}g^\alpha\right)\,\mathcal{D}_\lambda g\;,
\;\;\;\;\;\alpha\in\widetilde{\mathbb{S}}_\lambda\subseteq\mathbb{S}_\lambda\notag\\
&=&(\mathrm{Tr}\,{\mathrm{A}})^{-\alpha}_\lambda\notag\\
&\equiv &\mathrm{Tr}\,{A}_\lambda^{-\alpha}\;.
\end{eqnarray}
Evidently, the ordinary trace $\mathrm{tr}$ and functional trace $\mathrm{Tr}$ possess the same functional form. But the fundamental region of the functional trace depends on
the chosen normalization, and taking the ordinary trace inside the integral \emph{often}
requires a restriction on the fundamental region of
$\mathcal{M}_\lambda\left[\mathrm{Tr}\,\mathrm{E}^{-\mathrm{A}};\alpha\right]$ as in the third line of (\ref{trace}). The
point is we can turn the calculation around and (with appropriate
normalization/regularization) give meaning to the object
$\mathcal{M}_\lambda\left[\mathrm{Tr}\,\mathrm{E}^{-\mathrm{A}};\alpha\right]$
through the ordinary trace and an adjustment to $\mathbb{S}_\lambda$.

In particular, the zeta functional can be represented as
\begin{equation}\label{multiple zeta}
\zeta_{A_{\lambda}}(\alpha)
\equiv\mathcal{M}_\lambda\left[\mathrm{Tr}\,\mathrm{E}^{-\mathrm{A}};\alpha\right]
\end{equation}
but only for appropriate $\lambda$ and $\alpha$.\footnote{Recall that $\alpha$ is a multi-index with respect to a basis in the Lie algebra of $G^\C_\lambda$, so (\ref{multiple zeta}) can represent a multiple zeta functional if $\mathrm{dim}_\R(G^\C_\lambda)>1$.} For example\cite{NAK},
\begin{example}\label{zeta example}
Consider a counting functional $\mathrm{N}$ such that $\mathrm{N}(g)=Ng$. Let $\mathcal{H}$ be a separable Hilbert space. Suppose $N\in \mathfrak{C}^\ast=L_B(\mathcal{H})$ is self-adjoint with spectrum $\sigma(N)=\mathbb{Z}_+$, and let
$\{|i\rangle,\varepsilon_i\}$ with $i\in\{1,\ldots,\infty\}$ denote
the set of orthonormal eigenvectors and associated eigenvalues of
$N$. Choose $\lambda_{\R_\pm}:G^\C\rightarrow \R_+\cup\R_-$  and $\rho(g)=gId$ where (abusing notation) $g\equiv g_\lambda\in\R_+\cup\R_-$. The Riemann zeta function associated with
$N$ can be represented
 by
\begin{equation}
\zeta_{{N}_{\Gamma}}(\alpha)
=\mathrm{tr}\int_{\R_+\cup\R_-}e^{-Ng}\rho(g^\alpha)
\;d\nu(g_{\Gamma})\;.
\end{equation}
In the orthonormal basis this is
\begin{eqnarray}
\zeta_{{N}_{\Gamma}}(\alpha)
&=&\sum_{i=1}^\infty\,\int_{\R_+}\,e^{-\varepsilon_i g+\alpha(\log g)}\langle i | i \rangle
\;d\nu(g_{\Gamma})\;,\;\;\;\;\;\alpha\in\langle0,\infty\rangle\notag\\
&=&\int_{\R_+}\sum_{i=1}^\infty\,e^{-\varepsilon_i g+\alpha(\log g)}
\;d\nu(g_{\Gamma})\;,\;\;\;\;\;\alpha\in\langle1,\infty\rangle_{\Gamma}\notag\\
&=&\int_0^\infty\frac{1}{e^{\,g}-1}\,g^{\alpha} \;d\nu(g_{\Gamma})
\;,\;\;\;\;\;\alpha\in\langle1,\infty\rangle_{\Gamma}\notag\\
&=&\mathcal{M}_{\R_\pm,\Gamma}
\left[\mathrm{Tr}\,\mathrm{E}^{-\mathrm{N}};\alpha\right]=\zeta(\alpha)
\;,\;\;\;\;\;\alpha\in\langle1,\infty\rangle_{\Gamma}
\end{eqnarray}
where $\nu(g_{\Gamma}):=\nu(g)/\Gamma(\alpha)=\log g/\Gamma(\alpha)$. Note the integral in the first line is valid for
$\alpha\in\langle0,\infty\rangle$ and the integral over $\R_-$ does not converge for any $\alpha$. So exchanging summation and integration in the second line comes with the price of
restricting the fundamental strip.

If instead the localization is $\lambda_{\mathcal{C}}:G^\C\rightarrow \mathcal{C}$ a smooth contour
in $\C^\times\cong \C\setminus\{0\}$ with the branch cut for $\log$ along the positive real axis, this has the well-known representation
\begin{eqnarray}
\zeta_{{N}_{\Gamma_{\mathcal{C}}}}(\alpha)
&=&\int_\mathcal{C}\frac{1}{e^{\,g}-1}\,g^{\alpha}
\;d\nu(g_{\Gamma_{\mathcal{C}}}),
\;\;\;\;\;\alpha\in\langle0,\infty\rangle_{\Gamma_{\mathcal{C}}}\backslash\{1\}\notag\\
&=&\mathcal{M}_{\mathcal{C},\Gamma_{\mathcal{C}}}
\left[\mathrm{Tr}\,\mathrm{E}^{-\mathrm{N}};\alpha\right]=\zeta(\alpha) ,
\;\;\;\;\;\alpha\in\langle0,\infty\rangle_{\Gamma_{\mathcal{C}}}\backslash\{1\}
\end{eqnarray}
where $d\nu(g_{\Gamma_{\mathcal{C}}})
:=\frac{\pi\csc(\pi\alpha)}{2\pi\imath\,\Gamma(\alpha)}\frac{dg}{g}$ and
$\mathcal{C}$ starts at $+\infty$ just below the real axis, passes
around the origin clockwise, and then continues back to
$+\infty$ above the real axis. This is an explicit illustration that $\mathbb{S}$ depends on $\lambda$ through both the normalization and the group. One can also define the trace of a winding/degree functional $\mathrm{W}$ given by $\mathrm{W}(g)=Wg$ where $\sigma(W)=\mathbb{Z}\backslash\{0\}$:
\begin{eqnarray}
\zeta_{{W}_{\Gamma}}(\alpha)
&:=&\int_{\R_+\cup\R_-}\mathrm{tr}\left(e^{-Wg}\,\rho(g^\alpha)\right)
\;d\nu(g_{\Gamma})\;,\;\;\;\;\;\alpha\in\langle1,\infty\rangle_\Gamma\notag\\
 &=&\mathcal{M}_{\R_\pm,\Gamma}
\left[\mathrm{Tr}\,\mathrm{E}^{-\mathrm{W}};\alpha\right]
=2\zeta(\alpha)\;,\;\;\;\;\;\alpha\in\langle1,\infty\rangle_\Gamma\;.
\end{eqnarray}
Here each integral contributes depending on the sign of $\sigma(W)$.
\end{example}

\begin{example}\label{eta example}
Again for $N\in L_B(\mathcal{H})$ but now with $\mathrm{N}(g)=N(g\pm i\pi)$, the Dirichlet eta
function associated with $N$ can be represented by the functional
\begin{eqnarray}
\eta_{{N}_{\Gamma}}(\alpha)
&=&\int_{\R_+}\sum_{i=1}^\infty\,e^{-\varepsilon_i (g\pm i\pi)+\alpha(\log g)}
\;d\nu(g_{\Gamma})\notag\\
&=&\int^{\infty}_0\frac{-1}{e^{g}+1}\,g^{\alpha} \;d\nu(g_{\Gamma})
 \;,\;\;\;\;\;\alpha\in\langle0,\infty\rangle_{\Gamma}\notag\\
 &=&\mathcal{M}_{\R_\pm,\Gamma}
 \left[\mathrm{Tr}\,\mathrm{E}^{-\mathrm{N}};\alpha\right]=\eta(\alpha)
 ,\;\;\;\;\;\alpha\in\langle0,\infty\rangle_{\Gamma}
\end{eqnarray}
where  $\nu(g_{\Gamma}):=\log g/\Gamma(\alpha)$. The second equality uses the fact that $\sum\int e^{-\varepsilon_i g}g^\alpha\;d\nu(g_{\Gamma})$ converges for $\alpha\in\langle0,\infty\rangle$.
\end{example}

Continuing the strategy of defining functional analogs, define the functional logarithm:
\begin{definition}\label{functional log}
Let $\mathrm{A}\in \mathrm{Mor}_C(G^\C,\mathfrak{C}^\ast)$ be invertible and suppose that $\mathrm{E}^{-\mathrm{A}}\in
\mathbf{F}_{\mathbb{S}}(G^\C)$ for some
fundamental region $\alpha\in\mathbb{S}_\lambda$. The functional logarithm $\mathrm{Log}\,{\mathrm{A}}$ is
defined by
\begin{eqnarray}
\mathrm{Log}\,{A}^{-(\alpha+1)}_\lambda\equiv(\mathrm{Log}\,{\mathrm{A}})^{-(\alpha+1)}_\lambda
:=\frac{d}{d\alpha}\mathcal{M}_\lambda
\left[\mathrm{E}^{-\mathrm{A}};\alpha\right]
&=&\int_{G^\C}
e^{-\mathrm{A}(g)}\,g^{\alpha}\,\mathrm{log}_\lambda\,g \,\mathcal{D}_\lambda g\notag\\
&=&\int_{G^\C}
e^{-\mathrm{A}(g)}\,g^{\alpha} \,\widehat{\mathcal{D}}_\lambda g\notag\\
&=:&\widehat{\mathcal{M}}_\lambda
\left[\mathrm{E}^{-\mathrm{A}};\alpha\right]
\end{eqnarray}
where $g^\alpha\,\widehat{\mathcal{D}}_\lambda g=\frac{d}{d\alpha}(g^\alpha \,d\nu(g_\lambda))$ verifies $\int_{G^\C}
e^{-g}\,g^{\alpha}\,\widehat{\mathcal{D}}_\lambda g=0$ for the chosen normalization. Note that we allow $0\in\sigma (A^{-1}_\lambda)$ here if the $\alpha\rightarrow 0^+$ limit exists.
\end{definition}
For this to be well-defined requires a suitable definition of the ordinary logarithm $\log A^{-1}$ of operators $A^{-1}\in\mathfrak{C}^\ast$. Choosing the standard series or the spectral representation of (\ref{log def}) is convenient and typical, but their convergence is limited to $\|A^{-1}-Id\|<1$ relative to the norm on $\mathfrak{C}^\ast$. Perhaps a better choice for invertible $(A-Id)\in\C^{m\times m}$  is to use the Cayley transform $C(A):=(A+Id)(A-Id)^{-1}$. In this case,
\begin{equation}\label{log definition}
\log A^{-1}=2\sum_{m=3}^\infty\,\frac{(-1)^{m+1}}{m}C(A^{-1})^m
\end{equation}
converges if $\sigma(A^{-1})\subseteq\C_+$.\cite{HIG}

Notably, since functional $\mathrm{Log}$ is a derivative, it commutes with functional $\mathrm{Tr}$ if $\alpha\in\widetilde{\mathbb{S}}_\lambda$. To see this note that
\begin{equation}
\left(\mathrm{Log}\,\mathrm{Tr}\,\mathrm{A}\right)^{-(\alpha+1)}_\lambda
:=\frac{d}{d\alpha}\mathcal{M}_\lambda
\left[\mathrm{Tr}\,\mathrm{E}^{-\mathrm{A}};\alpha\right]
=\zeta'_{A_\lambda}(\alpha)\;,
\end{equation}
and
\begin{equation}\label{trace/log}
\left(\mathrm{Tr}\,\mathrm{Log}\,\mathrm{A}\right)^{-(\alpha+1)}_\lambda
:=\mathrm{tr}\,\widehat{\mathcal{M}}_\lambda
\left[\mathrm{E}^{-\mathrm{A}};\alpha\right]
=\widehat{\mathcal{M}}_\lambda
\left[\mathrm{Tr}\,\mathrm{E}^{-\mathrm{A}};{\alpha}\right]
=\zeta'_{A_\lambda}({\alpha})\;.
\end{equation}
But, $\left(\mathrm{Log}\,\mathrm{Tr}\,\mathrm{A}\right)^{-(\alpha+1)}_\lambda
\neq\log\,\left(\mathrm{Tr}\,\mathrm{A}\right)^{-(\alpha+1)}_\lambda$; indicating that functional $\mathrm{Log}$ and ordinary $\log$ are applied very differently. However, they can be equivalent:

\begin{example}\label{log example}
Return to \emph{Example \ref{resolvent}} but now with $\mathfrak{z}=0$ and  $\sigma(A)\subseteq\C_+$. Then calculate
\begin{eqnarray}
\mathrm{Log}\,{A}_{\lambda}^{-1}
&=&\lim_{\alpha\rightarrow0^+}\int_{\phi_{\mathfrak{a}(\R)}} \,e^{- A
g}g^{\alpha}\,\log_\lambda g\;\mathcal{D}_\lambda g\notag\\
&=&\lim_{\alpha\rightarrow0^+}\int_{\phi_{\mathfrak{a}(\R)}} \,e^{-
g}(A^{-1}g)^{\alpha}\,\log_\lambda (A^{-1}g)\;\mathcal{D}_\lambda g\notag\\
&\stackrel{\lambda_{\digamma}}{\rightarrow}&
\lim_{\alpha\rightarrow0^+}(A^{-1})^{\alpha}\int_{\digamma} \,e^{-
g}\,g^{\alpha}\;d\widehat{\nu}(g_{N})\notag\\
&=&
\lim_{\alpha\rightarrow0^+}(A^{-1})^{\alpha}\int_{\digamma} \,e^{-
g}\,g^{\alpha}(\log g+\log A^{-1}-\psi(\alpha))\;d\nu(g_{N})\notag\\
&=&\log A^{-1}
\end{eqnarray}
where the third line follows because $A^{-1}$ and $\rho(g)$ commute and the $\R_-$ integral does not converge for any $\alpha$. (See \emph{Appendix B} and \emph{(\ref{log measure})} for the definition and motivation for $d\widehat{\nu}(g_{N})$.) Note that $\mathrm{Log}\,{A}_{\Gamma}^{-(\alpha+1)}=(A^{-1})^{\alpha}\log A^{-1}$. Evidently, functional $\mathrm{Log}$ behaves like an ordinary logarithm only for $\lim\alpha\rightarrow0^+$; e.g. formally $({\mathrm{Log}\,({\mathrm{E}^{-\mathrm{A}}})})^{-1}_\lambda
=({\mathrm{E}^{-\mathrm{Log}\,\mathrm{A}}})^{-1}_\lambda$.
\end{example}

The final component of our triad is the functional determinant of an inverse power.
\begin{definition}\label{determinant}
Let $\mathrm{A}\in \mathrm{Mor}_C(G^\C,\mathfrak{C}^\ast)$  with $\mathrm{A}(g)\in\mathfrak{C}^\ast$ trace class and $0\notin\sigma(\mathrm{A}(g))$. Suppose $\mathrm{E}^{-\mathrm{Tr}\,\mathrm{A}}
\in\mathbf{F}_{\mathbb{S}}(G^\C)$ for $\alpha\in\mathbb{S}_\lambda$. The functional determinant $\mathrm{Det}\,{\mathrm{A}}\in \mathrm{Mor}_C(G^\C,\C)$ is defined by
\begin{eqnarray}
\mathrm{Det}\,{A}_\lambda^{-\alpha}\equiv(\mathrm{Det}\,{\mathrm{A}})^{-\alpha}_\lambda :=\mathcal{M}_{\lambda}
\left[\mathrm{Det}\,\mathrm{E}^{-\mathrm{A}};\alpha\right]
&:=&\int_{G^\C}\det\left( e^{-\mathrm{A}(g)
}g^\alpha\right)\;\mathcal{D}_\lambda g\notag\\
&=&\int_{G^\C}e^{-\mathrm{tr}\left(\mathrm{A}(g)\right)}
\,\det g^\alpha\;\mathcal{D}_\lambda g\notag\\
\end{eqnarray}
where $\det \rho(g^\alpha)=\det e^{\alpha\rho'(\mathfrak{g})}
=e^{\mathrm{tr}(\alpha\rho'(\mathfrak{g}))}
=(e^{\mathrm{tr}\,\rho'(\mathfrak{g})})^\alpha
=(\det \rho(g))^\alpha$ if $\arg(e^{\mathrm{tr}\,\rho'(\mathfrak{g})})=0$.
\end{definition}

In contrast to functional $\mathrm{Log}$ and $\mathrm{Tr}$, functional $\mathrm{Log}$ and $\mathrm{Det}$ do not generally commute;
\begin{equation}
\left(\mathrm{Log}\,\mathrm{Det}\,\mathrm{A}\right)^{-(\alpha+1)}_\lambda
:=\frac{d}{d\alpha}
\mathcal{M}_\lambda[\mathrm{Det}\,\mathrm{E}^{-\mathrm{A}};\alpha]
\neq \widehat{\mathcal{M}}_\lambda
[\mathrm{Det}\,\mathrm{E}^{-\mathrm{A}};\alpha]
=:\left(\mathrm{Det}\,\mathrm{Log}\,\mathrm{A}\right)^{-(\alpha+1)}_\lambda
\end{equation}
since $\frac{d}{d\alpha}\det \rho(g^{\alpha}_\lambda)
\neq\det (\rho(g^{\alpha})\log_\lambda \rho(g))$ unless $\rho$ is one-dimensional.

It is important to realize that, \emph{at the functional level},
\begin{proposition}
If $\mathrm{A}_1,\mathrm{A}_2\in\mathbf{F}_{\mathbb{S}_{\mathcal{R}}}(G^\C)$ such that $\mathrm{A}_1(g)$ and $\mathrm{A}_2(g)$ are trace class, then
\begin{equation}
(\mathrm{Det}( \mathrm{A}_1\ast  \mathrm{A}_2))^{-\alpha}_{\lambda}
=(\mathrm{Det} \mathrm{A}_1\ast \mathrm{Det}\mathrm{A}_2)^{-\alpha}_{\lambda}
=\mathrm{Det}\,{ A_1}_\lambda^{-\alpha}\,\mathrm{Det}\,{ A_2}_\lambda^{-\alpha}\;.
\end{equation}
\end{proposition}
\emph{Proof}\begin{eqnarray}
\mathcal{M}_{\lambda}
\left[\mathrm{Det}(\mathrm{E}^{-\mathrm{A}_1}\ast\mathrm{E}^{-\mathrm{A}_2});\alpha\right]
&=&\int\int_{G^\C}\det\left( e^{-\mathrm{A}_1(\tilde{g})}e^{-\mathrm{A}_2(\tilde{g}^{-1}g)}g^\alpha\right)\;\mathcal{D}_\lambda \tilde{g}\,\mathcal{D}_\lambda g\notag\\
&=&\int\int_{G^\C} e^{-\mathrm{tr}(\mathrm{A}_1(\tilde{g}))}
e^{-\mathrm{tr}(\mathrm{A}_2(\tilde{g}^{-1}g))}\,\det g^\alpha\;\mathcal{D}_\lambda \tilde{g}\,\mathcal{D}_\lambda g\notag\\
&=&\int\int_{G^\C}\det\left( e^{-\mathrm{A}_1(\tilde{g})}e^{-\mathrm{A}_2(g)}\right)
\,\det\left[(\tilde{g}g)^\alpha\right]\;\mathcal{D}_\lambda \tilde{g}\,\mathcal{D}_\lambda g\notag\\
&=&\mathrm{Det}\,{ A_1}_\lambda^{-\alpha}\,\mathrm{Det}\,{ A_2}_\lambda^{-\alpha}\;.
\end{eqnarray}
In line three we used the fact that BCH implies $e^{\alpha\,\mathrm{tr}(\log(\rho(gh)))}
=e^{\alpha\,\mathrm{tr}(\rho'(\mathfrak{g})+\rho'(\mathfrak{h}))}$.$\QED$

Consequently, the functional determinant possesses the multiplicative property if the pertinent functional determinants have overlapping critical strips. However, \emph{at the function level} precipitated by a specific $\lambda$, the determinant \emph{might not} have the multiplicative property with respect to
$\alpha$. For starters, if $A$ is not in the multiplier algebra $M_s(\mathfrak{C}^\ast)$, the functional determinant $\mathrm{Det}\,A_\lambda^{-\alpha}$ doesn't have the same form as $\mathrm{det}\,A^{-\alpha}$. But even if $\mathrm{A_1}(g)$ and $\mathrm{A_2}(g)$ are linear in $g$, there may not exist a consistent choice of $\lambda$ that renders both ${ A_1}_{\lambda}$ and
${ A_2}_{\lambda}$ simultaneously analytic at a common value of $\alpha$; even if their convolution is analytic there. Moreover,
if such a $\lambda$ does exist, the localization implicit in $\lambda$ that achieves the reduction
$\mathrm{Det}\rightarrow\mathrm{det}$ may introduce a ``multiplicative anomaly"\footnote{For example, if one chooses zeta function
regularization to effect the reduction $\mathrm{Det}\rightarrow\mathrm{det}$, it is well-known that a non-vanishing
Wodzicki residue at $\alpha=0$ leads to a multiplicative anomaly. }:
\begin{proposition}\label{funtional determinant}
If $\mathrm{A}(g)=\rho(ag)=: A \rho(g)$ where $ a\in G^\C$ and $ A \rho(g)$ is trace class, then
\begin{equation}
\mathrm{Det}\,{A}_{\lambda}^{-\alpha}
=\mathcal{N}_\lambda(\alpha)\,|\det
A^{-1}|^{\alpha}\,e^{i\varphi_ A(\alpha)}
\,,\;\;\;\;\;\alpha\in\mathbb{S}_\lambda
\end{equation}
where $\mathcal{N}_\lambda(\alpha)$ is a $\lambda$-dependent
normalization and $\varphi_{ A}(\alpha)=2\alpha(\arg(e^{\mathrm{tr}(\log A^{-1})})+n\pi)$.
\end{proposition}
\emph{Proof}:
\begin{eqnarray}
\mathrm{Det}\,{A}_{\lambda}^{-\alpha} =\mathcal{M}_{\lambda}
\left[\mathrm{Det}\,\mathrm{E}^{-\mathrm{A}};\alpha\right]
&=&\int_{G_{\lambda}^\C}\, e^{-\mathrm{tr}(g)}\,
\det[( A^{-1}g)^\alpha]\;d\nu(g_{\lambda})\notag\\
&=&\int_{G_{\lambda}^\C}\, e^{-\mathrm{tr}( g)}\,
(\det A^{-1})^{\alpha}e^{i\alpha\arg(\det( A^{-1}))}\det g^{\alpha}\;d\nu(g_{\lambda})\notag\\
&=&|\det A^{-1}|^{\alpha}\,e^{i\varphi_{ A}(\alpha)}
\int_{G_{\lambda}^\C}\, e^{-\mathrm{tr}(g)}\,\det
g^\alpha \;d\nu(g_{\lambda})\notag\\
&=:&|\det A^{-1}|^{\alpha}\,e^{i\varphi_{ A}(\alpha)}\,
\mathcal{N}_\lambda(\alpha)\,,\;\;\;\;\alpha\in\mathbb{S}_\lambda\,,\;\;\;\;\alpha\in\mathbb{S}_\lambda
\end{eqnarray}
where $\varphi_{ A}(\alpha)=2\alpha(\arg(e^{\mathrm{tr}(\log A^{-1})})+n\pi)$. $\QED$

The $\lambda$-dependent normalization $\mathcal{N}_\lambda(\alpha):=\mathrm{det}\,{{Id}}_\lambda^{-\alpha}$ requires some scrutiny. The definition of functional determinant assumes that $\mathrm{A}(g)$ is trace class. However, $\mathrm{A}(g)=\rho(ag)$ will not be trace class for generic $G^\C$. If it is not, then we can try to regulate $\mathcal{N}_\lambda(\alpha)$ with a positive-definite invertible fixed element $R\in G^\C_{\lambda}$ such that $Rg$ is trace class and $e^{-\mathrm{tr}(Rg)}\in \mathbf{F}_{\mathbb{S}}(G^\C)$.

Let $\mathcal{H}$ furnish a representation of $\pi(\mathfrak{C}^\ast)$. Pick a basis in $\mathcal{H}$ for which $R$ is diagonal. Then
\begin{eqnarray}
  && \mathcal{N}_\lambda(R;\alpha):=\int_{G^\C_{\lambda}} e^{-\mathrm{tr}(Rg)}\,
\det g^\alpha \;d\nu(g_{\lambda})=\prod_{i=1}^d r_i^{-\alpha}\;,\;\;\;\;\;\alpha\in\mathbb{S}_\lambda\;.
\end{eqnarray}
where $\mathrm{dim}(\mathcal{H})=d$. Even if $d=\infty$, this product \emph{potentially} can be rendered finite and well-defined \emph{if} a suitable regulator $R$ exists.

Alternatively, we can simply choose the normalization/regularization associated with the choice of $\lambda$ to set $\mathcal{N}_\lambda(R;\alpha)=1$ --- which amounts to formally dividing out this factor from the functional determinant (similarly to what is done with $\Gamma(\alpha)$). The corresponding regularized functional determinant of an operator $O$ on $\mathcal{H}$ can then be defined by
\begin{eqnarray}\label{regularized determinant}
 {\mathrm{Det}_{R}}\,{O}_{\lambda}^{-\alpha}
 &:=&\frac{{\mathrm{Det}}\,Ad(R)O_{\lambda}^{-\alpha}}{\mathcal{N}_\lambda(R;\alpha)}\notag\\
&:=&\frac{1}{\mathcal{N}_\lambda(R;\alpha)}\int_{G^\C_{\lambda}} e^{-\mathrm{tr}(ROR^{-1}g)}
\det g^\alpha\,d\nu(g_\lambda)\notag\\
&=&\frac{1}{\mathcal{N}_\lambda(R;\alpha)}\left(\mathrm{det}( R^{\alpha})\,\mathrm{det} (O^{-\alpha})\right)
\int_{G^\C_{\lambda}}  e^{-\mathrm{tr}(Rg)}
\det g^\alpha\,d\nu(g_\lambda)\notag\\
&=&e^{i\varphi_{O,R}(\alpha)}\left|\frac{\mathrm{det} (O^{-1})}{\mathrm{det} (R^{-1})}\right|^{\alpha}\notag\\
&=:&(\mathrm{det}_R\,O^{-1})^\alpha\;.
\end{eqnarray}
This is common practice: One knows how $R$ acts on $\mathcal{H}$ and then defines the regularized determinant of $A$ by ${\mathrm{det}_{R}}\,A:=\det A/\det R$. A familiar example is the harmonic oscillator where $R=d^2/dx^2$, $A=d^2/dx^2-\omega^2$ and $G^\C_{\lambda}$ is the space of eigenfunctions.

So in particular, if $\arg \mathrm{tr}(\log (RA^{-1}))=0\mod2\pi$ and $\alpha=1\in\mathbb{S}_\lambda$, then
\begin{equation}
{\mathrm{Det}_{R}}\, {A}_{\lambda}^{-1}\,{\mathrm{Det}_{R}}\,{A}^{-1}_{\lambda}
={\mathrm{Det}_{R}} \,{A}_{\lambda}^{-2}=(\det A^{-1}/\det R^{-1})^{2}
\end{equation}
and the regularized determinant enjoys the usual multiplicative property. However, if instead $\arg( e^{\mathrm{tr}(\log( RA^{-1}))})\neq0\mod2\pi$, a multiplicative anomaly obtains.

Yet another perspective on the determinant comes from Lemma \ref{pi}  along with the fact that $\mathrm{det}:G^\C_\lambda\rightarrow\C^\times$ is a (possibly projective) representation. Suppose $A$ is self-adjoint and bounded. Then choose the localization $\lambda_{\R_\pm}:G^\C_\lambda\rightarrow\R_+\cup\R_-$ to get
\begin{equation}
\mathcal{M}_{\R_\pm,\Gamma}[\mathrm{Det}\,\mathrm{E}^{-\mathrm{A}};\alpha]
=\det\left(\mathcal{M}_{\R_\pm,\Gamma}[\mathrm{E}^{-\mathrm{A}};{\alpha}]\right)
=\mathrm{det}\,A^{-{\alpha}}_{\R_\pm,\Gamma}
=\det A^{-{\alpha}}\;,\;\;\;\;\;\alpha\in\widetilde{\mathbb{S}}_\lambda
\end{equation}
This sidesteps the potentially thorny issue of $\mathcal{N}_\lambda(\alpha)$ --- at least for self-adjoint $A$.

\section{More Mellin functional tools}\label{positive powers}
Conspicuously absent from the development so far have been functional operators $A_\lambda^\alpha$ with $\alpha\in\C_+$  --- which we will refer to as positive powers. This section remedies the deficiency.
\subsection{Positive powers}
Recall the notion that $(A-z)^{-\alpha}\sim d^{\alpha-1}(A-z)^{-1}/dz^{\alpha-1}$. We want to view an inverse power in the same way. That is, formally $A^{-\alpha}\sim d^{\alpha-1}(A-z)^{-1}/dz^{\alpha-1}|_{z=0}=(A^{-1})^{(\alpha-1)'}$. From this perspective, there should be associated functionals in $\mathbf{F}_{\mathbb{S}}(G^\C)$ that roughly correspond to complex derivatives of the resolvent at the point $z=0$. Positive powers are based on these objects.
\begin{definition}\label{positive def}
Suppose $\mathrm{A}\in\mathrm{Mor}_C(G^\C,\mathfrak{C}^\ast)$ is invertible and $\mathrm{E}^{-(\mathrm{A}-\mathrm{Z})}\in \mathbf{F}_{\mathbb{S}}(G^\C)$.
Let $\beta$ be a point in the fundamental strip of the functional resolvent of $\mathrm{A}$. Define the functional $(\mathrm{Id}-\mathrm{A}^{-1})_{(\beta)}\in\mathrm{Mor}_C(G^\C,\mathfrak{C}^\ast)$ by $g\mapsto\left({\mathrm{A}^{-1}}(g)\right)^{-\beta}\left(\mathrm{A}(g)-Id\right)^{-\beta}
=\left(Id-{\mathrm{A}^{-1}}(g)\right)^{-\beta}$. The functional positive power of $\mathrm{A}$ can be defined by \emph{(cf. \cite{LAM,BAL,KOM})}
\begin{eqnarray}
A_{\lambda,\beta}^{\alpha}\equiv({\mathrm{A}})^{\alpha}_{\lambda,\beta}
&:=&\mathcal{M}_\lambda
\left[(\mathrm{Id}-\mathrm{A}^{-1})_{(\beta)};\alpha\right]\notag\\
&:=&\int_{G^\C}\,\left(Id-{\mathrm{A}^{-1}}(g)\right)^{-\beta}g^\alpha\;\mathcal{D}_\lambda g\;,\;\;\;\;\;\alpha\in\mathbb{S}_{\lambda,\beta}\;.
\end{eqnarray}
\end{definition}
Notice the $\beta$ dependence of $A^\alpha_{\lambda,\beta}$. It exerts its influence by restricting the fundamental strip of $\alpha$.
\begin{example}\label{positive power}

Return to the context of \emph{Example \ref{resolvent example}}.  As before, ${\mathrm{A}^{-1}}(g)={A}^{-1}\rho(g)$, and to simplify matters assume $A^{-1}\in L_B(\mathcal{H})$ is positive-definite. Here we choose the standard Haar normalization and the localization $\lambda_{\R^\times}:\phi_{\mathfrak{a}}(\R)\rightarrow\R^\times$ since the integrand derives from the resolvent which characterizes the spectrum of $\mathfrak{a}$ and we assume invertible $A$. We calculate
\begin{eqnarray}\label{positive example}
A_{\lambda,\beta}^{\alpha}
&=&\int_{\phi_{\mathfrak{a}}(\R)}
\,(Id-A^{-1}\rho(g))^{-\beta}\,\rho(g)^{\alpha}\;\mathcal{D}_\lambda g
\;,\;\;\;\;\;\alpha\in\mathbb{S}_\beta\notag\\
&=&A^{\alpha}\int_{\phi_{\mathfrak{a}}(\R)}
\,(Id-\rho(g))^{-\beta}\,\rho(g)^{\alpha}\;\mathcal{D}_\lambda g
\;,\;\;\;\;\;\alpha\in\mathbb{S}_\beta\notag\\
&\langle i|\stackrel{\lambda_{\R^\times}}{\rightarrow}|j\rangle&
A_{ij}^{\alpha}\delta_{ij}\int_{\R^\times}\, (1-g)^{-\beta}\,g^{\alpha}\;d\nu(g_\mathrm{sH})
\;,\;\;\;\;\;\alpha\in\mathbb{S}_{\mathrm{sH},\beta}\notag\\
&=&A_{ij}^{\alpha}\delta_{ij}\left(\int_{-\infty}^0+\int_{0}^1+\int_{1}^\infty\right)\, (1-g)^{-\beta}\,g^{\alpha}\;d \log g
\;,\;\;\;\;\;\alpha\in\mathbb{S}_{\mathrm{sH},\beta}
\end{eqnarray}
where $A^\alpha=e^{\alpha\log A}$ and the second line follows by the measure-invariant left translation $\rho(g)\rightarrow \rho(\textsf{{a}}g)$ and from $A\rho(g)=\rho(g)A$. The third and fourth lines follow after localization and taking matrix elements in a diagonal basis.

Consider first the case $\Re(\beta)\geq1$ and save the analysis of $0<\Re(\beta)<1$ for the next subsection. Since we assume $A^{-1}$ positive-definite, only the first integral contributes because the next two do not converge for any $\alpha$ when $\Re(\beta)\geq1$. (Conversely, if $A$ is negative definite only the integral over $\R_+$ contributes.)  Change integration variable $g\rightarrow-g$ and use the integral relation\emph{(\cite[3.194(3.)]{GR})}
\begin{equation}\label{table integral}
\int_0^\infty(1+\xi x)^{-\nu} x^{\mu-1}\;dx=\xi^{-\mu} \mathrm{B}(\mu,\nu-\mu)\;\;\;\;\;\;\;\;|\arg\xi|<\pi\,;\;\;\;\;0<\Re(\mu)<\Re(\nu)
\end{equation}
to find
\begin{equation}
A^\alpha_{\R^\times,\mathrm{sH},\beta}=(-1)^{\alpha-1}\,\frac{\Gamma(\alpha)\Gamma(\beta-\alpha)}{\Gamma(\beta)}\cdot A^\alpha
=(-1)^{\alpha-1}\,\mathrm{B}(\alpha,\beta-\alpha)\cdot A^\alpha\;\;\;\;\;\;\;\;\alpha\in\langle0,\Re(\beta)\rangle
\end{equation}
where $\mathrm{B}(\cdot,\cdot)$ is the beta function and we put $\alpha=\mu-1$ and $\beta=\nu$. One can instead elect to use Haar measure $\nu(g_{\mathrm{B}}):=\log g/[(-1)^{\alpha-1}\,\mathrm{B}(\alpha,\beta-\alpha)]$ to get the presentation
\begin{equation}
A_{\R^\times,\mathrm{B},\beta}^{\alpha}=A^\alpha\;.
\end{equation}

Alternatively, if $A^{-1}$ is in the unit ball of $\mathbf{1}\in\mathfrak{C}^\ast$, one can localize by $\lambda_{\mathcal{C}}:\phi_{\mathfrak{a}}(\R)\rightarrow \mathcal{C}$ where $\mathcal{C}$ is a  smooth contour enclosing $\sigma(A^{-1})$ without crossing any branch cut. For $g\rightarrow g^{-1}$ and choosing Haar measure $\nu(g_{\mathcal{C}_{\mathrm{B}}}):=\frac{\pi\csc(\pi\alpha)}{2\pi i\,\mathrm{B}(\alpha,\beta-\alpha)}\frac{dg}{g}$, the holomorphic functional calculus yields (strictly for $\beta\in\mathbb{Z}_+$ but formally for $\beta\in\mathbb{S}$)
\begin{equation}\label{contour integral}
A_{\mathcal{C},\mathcal{C}_\mathrm{B},\beta}^{\alpha}
=\int_{\mathcal{C}}\, (\rho(g)-A^{-1})^{-\beta}\,\rho(g)^{\beta-\alpha}\;d\nu(g_{\mathcal{C}_{\mathrm{B}}})
=A^\alpha\;,\;\;\;\;\;\alpha\in\langle0,\Re(\beta)\rangle\;.
\end{equation}
 \end{example}

Given the definition of positive powers, we can define the positive-power analogs of functional trace, log, and determinant for $\alpha\in\langle0,\Re(\beta)\rangle$:

\begin{definition}\label{positive TLD}
Let $\mathrm{A}\in \mathrm{Mor}_C(G^\C,\mathfrak{C}^\ast)$ be invertible such that
$(\mathrm{Id}-\mathrm{A}^{-1})_{(\beta)}\in \mathbf{F}_{\mathbb{S}}(G^\C)$ is trace class. The positive-power functional trace, log, and determinant are defined by
\begin{eqnarray}
\mathrm{Tr}\,{A}_{\lambda,\beta}^{\alpha}\equiv(\mathrm{Tr}\,{\mathrm{A}})^{\alpha}_{\lambda,\beta}
&:=&\mathcal{M}_\lambda
\left[\mathrm{Tr}\,(\mathrm{Id}-\mathrm{A}^{-1})_{(\beta)};\alpha\right]\notag\\
&=&\int_{G^\C}\mathrm{tr}\left[(Id-{\mathrm{A}^{-1}}(g))^{-\beta}\,g^\alpha\right]\,\mathcal{D}_\lambda g\;,
\end{eqnarray}

\begin{eqnarray}
\mathrm{Log}\,{A}_{\lambda,\beta}^{\alpha+1}
\equiv(\mathrm{Log}\,\mathrm{A})^{\alpha+1}_{\lambda,\beta}
&:=&\frac{d}{d\alpha}\mathcal{M}_\lambda
\left[(\mathrm{Id}-\mathrm{A}^{-1})_{(\beta)};\alpha\right]\notag\\
&=:&\widehat{\mathcal{M}}_\lambda
\left[{(\mathrm{Id}-\mathrm{A}^{-1})}_{(\beta)};\alpha\right]\;,
\end{eqnarray}

\begin{eqnarray}
\mathrm{Det}\,{A}_{\lambda,{\beta}}^{\alpha}
\equiv(\mathrm{Det}\,{\mathrm{A}})^{{\alpha}}_{\lambda,{\beta}} &:=&\mathcal{M}_{\lambda}
\left[\mathrm{Det}\,(\mathrm{Id}-\mathrm{A}^{-1})_{(\beta)};{\alpha}\right]\notag\\
&=&\int_{G^\C}\det\left[ (Id-{\mathrm{A}^{-1}}(g))^{-{\beta}}\,g^{{\alpha}}\right]\,\mathcal{D}_\lambda g\notag\\
&=&\int_{G^\C}\det\left[(Id-{\mathrm{A}^{-1}}(g))^{-{\beta}}\right]\,\det g^{{\alpha}}\,\mathcal{D}_\lambda g\;.
\end{eqnarray}
\end{definition}

As in the negative power case, convergence of the integrals and the $\alpha$-limit are particularly sensitive to the chosen normalization. Recall $\Gamma(\alpha)$ played a crucial role in normalizing negative powers, and $\mathrm{B}(\alpha,\beta-\alpha)$ will do the same for positive powers. It is a useful exercise to compare against previous examples. For instance, one finds $\mathrm{Tr}\,{M^{-1}}^\alpha_{\mathrm{B},\beta}=\zeta(-\alpha)$ where $\alpha_{\Re}\in(0,\Re(\beta))$ (use $M^{-1}$ since the trace of $M$ obviously diverges) and $\mathrm{Log}\,A_{\mathrm{B},\beta}=\log A$.

\subsection{Some loose ends}\label{loose ends}

Return to Example \ref{positive power} for the case $0<\Re(\beta)<1$ and restrict to real $\alpha,\beta$. Perform  transformations $g\rightarrow \frac{g-1}{g}$ and $g\rightarrow \frac{-1}{g-1}$ (which are $SL(2,\mathbb{Z})$) in the second and third integrals respectively. Again using (\ref{table integral}), all three integrals contribute to give
\begin{equation}\label{string amplitude}
A_{\R^\times,\mathrm{sH},\beta}^{\alpha}
=[A_1^{\alpha}\mathrm{B}(\alpha,\beta-\alpha)+A_2^{\alpha}\mathrm{B}(\alpha,1-\beta)
+A_3^{\alpha}\mathrm{B}(1-\beta,\beta-\alpha)]
\end{equation}
where $0<\alpha,\beta<1$ and we have defined $A_k^{\alpha}:=A^{\alpha}e^{i\pi\phi_k}$ with $\phi_1=\alpha-1$, $\phi_2=0$, and $\phi_3=-\beta$.
Introduce real $a,b>0$ and restrict to $\alpha\equiv a$ and $\beta\equiv 1-b$ such that $\beta-\alpha=c$ with $c$ a constant. Notice that $a+b+c=1$. Then
\begin{equation}
A_{\R^\times,\mathrm{sH},1-b}^{a}
=A_1^{a}\mathrm{B}(a,c)+A_2^{a}\mathrm{B}(a,b)
+A_3^{a}\mathrm{B}(b,c)\;.
\end{equation}
Of course there is nothing preventing the $\alpha,\beta$ assignments with $a\leftrightarrow b$. Including this contribution gives
\begin{eqnarray}\label{tachyon scattering}
A(a,b):=A_{\R^\times,\mathrm{sH},1-b}^{a}+A_{\R^\times,\mathrm{sH},1-a}^{b}&=&\left[e^{i\pi\phi_1}\mathrm{B}(a,c)
+e^{i\pi\phi_2}\mathrm{B}(a,b)+e^{i\pi\phi_3}\mathrm{B}(b,c)\right]\,(A^{a}+A^{b})\;.\notag\\
\end{eqnarray}

\begin{example}
Let $\Psi_{i,j}$ with $i\neq j$ be asymptotic scattering sates in some Hilbert space $\mathcal{H}$. Restrict to $i,j\in\{1,2,3,4\}$ and consider the trivial functional ${\mathrm{Id}^{-1}}(g)\equiv\rho(g)$ so that $A=Id$. We claim that $\langle\Psi_{4,3}|Id(a,b)|\Psi_{2,1}\rangle$ mimics  the \textbf{kinematics} of  tree-level open string scattering in the sense that it reproduces the four-point tachyon string scattering amplitude on the unit disk up to normalization and momentum conservation --- with vertex interchange automatically included.

To see this, represent the disk by the upper-half $\C$-plane and fix three of the four tachyon vertex operators $\mathcal{V}_{k_1}(g)$ at $-\infty$, $\mathcal{V}_{k_2}(g)$ at $0$, and $\mathcal{V}_{k_4}(g)$ at $1$. The vertex ordering $\{1234\}$ and its $(24)$ permutation $\{1432\}$ associated with $a\leftrightarrow b$ corresponds to the second term in \emph{(\ref{tachyon scattering})}. Then the $(23)$ permutation of this ordering given by $\{1324\}$ along with its $(34)$ permutation associated with $a\leftrightarrow c$ corresponds to the first term, and the $(34)$ permutation of $\{1234\}$ along with its $(23)$ permutation associated with $b\leftrightarrow c$ corresponds to the third term in \emph{(\ref{tachyon scattering})}. The relative phases account for the non-cyclic permutations of the vertex orderings. Explicitly, $\langle\Psi_{4,3}|e^{i\pi\phi_1}(Id^a+Id^b)|\Psi_{2,1}\rangle
=\langle\Psi_{4,2}|\Psi_{3,1}\rangle$ and $\langle\Psi_{4,3}|e^{i\pi\phi_3}(Id^a+Id^b)|\Psi_{2,1}\rangle
=\langle\Psi_{3,4}|\Psi_{2,1}\rangle$ which can be seen by following the minus signs accrued when transforming $g$ to the corresponding integral domains. We get
\begin{equation}
\langle\Psi_{4,3}|Id(a,b)|\Psi_{2,1}\rangle
=\langle\Psi_{4,3}|\mathrm{B}(a,b)|\Psi_{2,1}\rangle+\langle\Psi_{4,2}|\mathrm{B}(a,c)|\Psi_{3,1}\rangle
+\langle\Psi_{3,4}|\mathrm{B}(b,c)|\Psi_{2,1}\rangle\;.
\end{equation}
\end{example}

This example is not surprising since the integral we are solving is essentially the string scattering amplitude integral without the absolute values. Still, it is curious that  positive powers, with $\alpha,\beta\in (0,1)$ and $\beta-\alpha$ fixed, appear to model scattering channels with crossing symmetry. It gets more curious.

\begin{example}
In the same vein as \emph{Example \ref{inverse power product}}, calculate the product of positive powers $\mathcal{M}_{\R^\times,\mathrm{sH}^2}
\left[\left((\mathrm{Id}-\mathrm{B}^{-1})_{(\beta)}
\ast(\mathrm{Id}-\mathrm{A}^{-1})_{(\gamma)}\right);\alpha\right]
=B_{\R^\times,\mathrm{sH},\beta}^{\alpha}
A_{\R^\times,\mathrm{sH},\gamma}^{\alpha}$ for $0<\Re(\beta),\Re(\gamma)<1$ where
\begin{eqnarray}\label{closed string}
B_{\lambda,\beta}^{\alpha}
A_{\lambda,\gamma}^{\alpha}
&=&\int_{\phi_{\mathfrak{a}}(\R)}\left[\int_{\phi_{\mathfrak{a}}(\R)}\,(Id-B^{-1}\tilde{g})^{-\beta}
(Id-A^{-1}\tilde{g}^{-1}g)^{-\gamma}\,g^{\alpha}
\;\mathcal{D}_\lambda \tilde{g}\right]\,\mathcal{D}_\lambda g\notag\\
\end{eqnarray}
and the fundamental strip depends on $\lambda$  and the choice of $\beta,\gamma$. To simplify matters, assume $[B,A]=0$. Then, by now familiar manipulations bring the right-hand side into the form
\begin{equation}
\stackrel{\lambda_{\R^\times}}{\rightarrow}B^{\alpha}
A^{\alpha}\int_{\R^\times}
\left[\int_{\R^\times}\,(1-\tilde{g})^{-\beta}\,\tilde{g}^{\alpha}\,
(1-g)^{-\gamma}
\,g^{\alpha}\;d\log \tilde{g}\right]d\log g
\;,\;\;\;\;\;\alpha\in\mathbb{S}_{\beta,\gamma}\;.
\end{equation}
 Unlike the previous example, the double integral here is very different from the corresponding string integral. It is most easily evaluated by expressing the inverse operators as Mellin transforms
\begin{eqnarray}
\int_{(\R_+\cup\R_-)\times\R^\times}
\left[\int_{(\R_+\cup\R_-)\times\R^\times}\frac{e^{-(Id-\tilde{g})t}}{\Gamma(\beta)}t^{\beta} \tilde{g}^{\alpha}
\frac{e^{-(Id-g)s}}{\Gamma(\gamma)}s^{\gamma} g^{\alpha}\;d\log t\,d\log \tilde{g}\right]\,d\log s\,d\log g\;.\notag\\
\end{eqnarray}
Assuming $\mathbb{S}_{\beta,\gamma}$ is not empty, the integrals converge for some $\alpha\in\mathbb{S}_{\beta,\gamma}$ and the integrands are continuous functions on their various domains, so we can interchange $t,\tilde{g}$ and $s,g$ integration order.  Integrating over $\tilde{g},g$ (in that order)\footnote{Reversing the order gives the same integrals but with the $<>$ conditions on $t$ instead of $s$.} gives four integrals:
\begin{eqnarray*}
I_1&=&\frac{\Gamma(\alpha)^2}{\Gamma(\beta)\Gamma(\gamma)}
\int_{(\R_+\cup\R_-)^2}e^{-Id(s+t)}t^{\beta-1}(-t)^{-\alpha}
s^{\gamma-1}(-s)^{-\alpha}\;dt\,ds\;,\;\;\;\;\;0<\alpha\,,\,s<0\notag\\
I_2&=&(-1)^{\alpha-1}\frac{\Gamma(\alpha)^2}{\Gamma(\beta)\Gamma(\gamma)}
\int_{(\R_+\cup\R_-)^2}e^{-Id(s+t)}t^{\beta-1-\alpha}
s^{\gamma-1}(-s)^{-\alpha}\;dt\,ds\;,\;\;\;\;\;0<\alpha\,,\,s<0\notag\\
I_3&=&(-1)^{\alpha-1}\frac{\Gamma(\alpha)^2}{\Gamma(\beta)\Gamma(\gamma)}
\int_{(\R_+\cup\R_-)^2}e^{-Id(s+t)}t^{\beta-1}(-t)^{-\alpha}
s^{\gamma-1-\alpha}\;dt\,ds\;,\;\;\;\;\;0<\alpha\,,\,s>0\notag\\
I_4&=&(-1)^{2\alpha}\frac{\Gamma(\alpha)^2}{\Gamma(\beta)\Gamma(\gamma)}
\int_{(\R_+\cup\R_-)^2}e^{-Id(s+t)}t^{\beta-1-\alpha}
s^{\gamma-1-\alpha}\;dt\,ds\;,\;\;\;\;\;0<\alpha\,,\,s>0\;.
\end{eqnarray*}
$I_1\,,I_2$ don't converge for either $t\in\R_+$ or $t\in\R_-$, and $I_3\,,I_4$ don't converge for $t\in\R_-$. Finally, $I_3+I_4$ integrated over $\R_+^2$ yields
\begin{equation}
((-1)^{2\alpha}-1)\frac{\Gamma(\alpha)\Gamma(\beta-\alpha)}{\Gamma(\beta)}
\frac{\Gamma(\alpha)\Gamma(\gamma-\alpha)}{\Gamma(\gamma)}\;,\;\;\;\;\;0<\alpha<\Re(\beta)<1
\end{equation}
and we get
\begin{eqnarray}\label{positive power product}
B_{\R^\times,\mathrm{sH},\beta}^{\alpha}
A_{\R^\times,\mathrm{sH},\gamma}^{\alpha}&=&\frac{2\pi i e^{i\pi\alpha}}{\pi\csc(\pi\alpha)}\,
\mathrm{B}(\alpha,\beta-\alpha)\mathrm{B}(\alpha,\gamma-\alpha)B^{\alpha}
A^{\alpha}\;.
\end{eqnarray}
Observe the right-hand side is symmetric under $\beta\leftrightarrow\gamma$. This was to be expected since we assumed $[B,A]=0$ so the integration order of $\tilde{g},g$ could be reversed due to Fubini. We can then conclude that $0<\alpha<\Re(\beta),\Re(\gamma)<1$.

Now suppose $A=Id$ and $B=(e^{-i\pi}Id)=:\widetilde{Id}$. Returning to $a+b+c=1$ and putting $\alpha=b$, $\beta=1-a=b+c$, and $\gamma=1-c=a+b$ in \emph{(\ref{positive power product})} yields
\begin{equation}\label{closed}
\widetilde{Id}_{\R^\times,\mathrm{sH},1-c}^{\,b}Id_{\R^\times,\mathrm{sH},1-a}^{\,b}=2i\sin(\pi b)\,\mathrm{B}(b,c)\mathrm{B}(b,a)\,Id
=2i\mathrm{B}(a,b,c)\,Id
\end{equation}
whose trace is recognized as the four-point tachyon scattering of closed strings on the unit sphere up to normalization and momentum conservation. Note that we need $\arg(\det A)$ and $\arg(\det B)$ to be $\pi$ out of phase to cancel the $e^{i\pi\alpha}$ term in \emph{(\ref{positive power product})} to get this result.

We learn that the algebraic product of positive-power operators with $0<\alpha,\beta<1$  \textbf{formally} parallels ``$\mathrm{gravity}=(\mathrm{gauge})^2$''. Explicitly, since functional Mellin is a representation in this case,
\begin{eqnarray}
\mathit{\Pi}_{\lambda}^{(b)}\left((\mathrm{Id}-\mathrm{O}^{-1})_{(a+b)}
\ast(\mathrm{Id}-\mathrm{O}^{-1})_{(c+b)})\right)&&\notag\\
&&\hspace{-1.5in}=\mathit{\Pi}_{\lambda}^{(b)}\left((\mathrm{Id}-\mathrm{O}^{-1})_{(a+b)})\right)
\cdot\mathit{\Pi}_{\lambda}^{(b)}\left((\mathrm{Id}-\mathrm{O}^{-1})_{(c+b)})\right)
\end{eqnarray}
where $a+b+c=1$.
\end{example}

Similarly, for the product of positive-power and negative-power operators, using
\begin{equation}
\stackrel{\lambda_{\R^\times},\lambda_{\R_\pm}}{\rightarrow}\int_{\R^\times}\left[\int_{\R_+\cup\R_-}\,(Id-B^{-1}\tilde{g})^{-\beta}
\,e^{-A\tilde{g}^{-1}g}\,g^\alpha\;d\nu(\tilde{g}_H)\right]\,d\nu(g_\Gamma)
\end{equation}
and again expressing the inverse operator as a Mellin transform yields
\begin{eqnarray}
\mathcal{M}_{\R^\times,\R_\pm,\mathrm{sH}^2}
\left[{(\mathrm{Id}-\mathrm{B}^{-1})_{(\beta)}}\ast\mathrm{E}^{-\mathrm{A}};\alpha\right]
=B_{\R^\times,\mathrm{sH},\beta}^{\alpha}A_{\R_\pm,,\mathrm{sH}}^{-\alpha}
=(-1)^{\alpha-1}\mathrm{B}(\alpha,\beta-\alpha)\Gamma(\alpha)\,B^{\alpha}A^{-\alpha}\;.\notag\\
\end{eqnarray}
So $Id_{\R^\times,\mathrm{sH},a+b}^{\,b}Id_{\R_\pm,\mathrm{sH}}^{-b}
=(-1)^{b-1}\mathrm{B}(a,b)\Gamma(b)\,Id$, and recall Example \ref{inverse power product} gives the product of two negative-power identity operators $Id_{\R_\pm,\mathrm{sH}}^{-b}Id_{\R_\pm,\mathrm{sH}}^{-b}=\Gamma(b)^2Id$. Along with (\ref{closed}), these three functional products highlight the fundamentally different algebraic underpinnings of operators in $\mathfrak{C}^\ast$ with ``positive" versus ``negative" powers.

\begin{example}\label{multiple resolvent}
Evidently, a positive power operator $A^\alpha_\lambda$ probes an associated resolvent through functional Mellin. However, so far we have restricted to a one-parameter real subgroup $\phi_{\mathfrak{a}}(\R)$ and localized to a very simple abelian group $\R^\times$ identified as the domain of the spectrum of operator $A$ relative to states located at the end-points of evolution-time intervals in $\R$. It is easy to imagine less drastic simplifications that presumably probe more properties of the resolvent by accessing more of the group structure.

As an obvious example, relax the restriction to $\R$ for the domain of the one-parameter subgroup. Then $\phi_{\mathfrak{a}}(\C)$ generates a complex one-parameter subgroup, and the spectrum of $A$ relative to states located at marked points on the boundaries of $1$-dimensional compact manifolds in $\C$ motivates localization to $\C^\times$. The countable family $G^\C_\Lambda=\{G^\C_\lambda, \lambda\in\Lambda\}$ that partially characterizes the functional integral will then represent a sum of integrals over all Riemann surfaces with boundary relevant to the operator $A$.
\end{example}

The resemblance to string scattering amplitudes for fractional positive powers is intriguing: But of course there is no string \emph{physics} coming from functional Mellin. Nevertheless, the mathematical nature of $\mathbf{F}_{\mathbb{S}}(G^\C)$ (being a $C^\ast$-algebra) appears to capture at least some algebraic structure of perturbative string theory; which is not too surprising since Mellin-type integrals feature in scattering amplitude and $n$-point correlation calculations in a variety of theories---especially CFTs. This suggests the resemblance of fractional positive powers to scattering amplitudes is likely more than just coincidence. We return to this idea in the next section in the context of QFT.

\section{Functional Mellin and QFT}\label{compare}
We aim to show that functional Mellin can represent some relevant objects in QFT normally constructed from functional Fourier. This can be seen already at a scopic level: Functional Fourier in QFT is defined on vector spaces of complex-valued fields and their topological duals, so their underlying groups are abelian under point-wise addition. Essentially, this means that functional Fourier in QFT derives from the Lie algebra associated with functional Mellin in the special case of an abelian group and $\alpha=1$.

\subsection{Generating functional}\label{generating functional}
Here we construct the QFT generating functional first by functional Fourier methods and then by functional Mellin methods for comparison.

Let us  review the context of the functional Fourier transform.\cite{LA1} Begin with the topological vector space $\mathcal{P}_a\C^{m}$ of piece-wise continuous, pointed maps $ z:(\mathbb{T},\ti_a)\rightarrow (\C^m,\mathrm{z}_a)$. The involution and complex structure on $\C^m$ induce an involution and complex structure on $\mathcal{P}_a\C^{m}$. Use this to  complexify $(\mathcal{P}_a\C^{m})^\C\cong\mathcal{P}_a\C^{m}\oplus i\mathcal{P}_a\C^{m}$. Let $Z_a\cong X_a\oplus iY_a$ denote the underlying complex abelian group (under point-wise addition) of $(\mathcal{P}_a\C^{m})^\C$, and denote its dual by $Z_a'$. Given is a nondegenerate linear operator $G:Z_a'\rightarrow Z_a$ and its inverse $D:Z_a\rightarrow Z_a'$ on an appropriate domain excluding the set of zero modes $\{\zz:D\zz=0\}$. These define a quadratic form $\mathrm{Q}_{\mathrm{B}}:Z_a\times Z_a\rightarrow \C$ by $(z_1,z_2)\mapsto-\frac{1}{2}\langle(D+D^\dag)z_1,z_2\rangle=:-\frac{1}{2}\langle Qz_1,z_2\rangle$ rendered self-adjoint by a suitable boundary term $\mathrm{B}:Z_a\times Z_a\rightarrow\C$. We will assume $\mathrm{Q}_{\mathrm{B}}$ is positive-definite. The complex structure $\mathrm{J}$ on $Z_a$ allows the $\mathbb{Z}_2$-graded decomposition $Z_a=Z_a^+\oplus Z_a^-$ relative to the inner product $(z_1|z_2)_{Z_a}:=\mathrm{Id}_{\mathrm{B}}(z_1,z_2)$, and it determines associated maps $J:Z_a\rightarrow Z_a'$ and $J':Z'_a\rightarrow Z_a$ by $-\frac{1}{2}\langle Jz_1,z_2\rangle:=(\mathrm{J}z_1|z_2)_{Z_a}$ and $J'J=-\mathrm{Id}_{Z_a}$. A subspace $X_a\subset Z_a$ is real relative to the complex structure if $X_a\cap\mathrm{J} X_a=\{0\}$ and $Z_a=X_a\oplus\mathrm{J} Y_a$. Extend $\mathrm{Q}_{\mathrm{B}}$ to $Z_a\times Z_a'$ by duality.\footnote{Explicitly, $\mathrm{Q}_{\mathrm{B}}$ maps $(z_1\times z_1',z_2\times z_2')\mapsto-\frac{1}{2}\langle Q z_1\times z_1',z_2\times Q'z_2'\rangle$ where $\langle z_1',Q'z_2'\rangle:=\langle Qz_1,z_2\rangle$.} $Z_a\times Z_a'$ is also $\mathbb{Z}_2$-graded and decomposes as
 \begin{eqnarray}\label{decomposition}
  Z_a\times Z'_a={\bigoplus}_\pm\left[\left(Z^\pm_a\times Z'^\pm_a\right)\oplus \left(Z^\pm_a\times Z'^\mp_a\right)\right]
  =:W^{\mathbf{e}}_a\oplus W^{\mathbf{o}}_a=:W_a\;.
\end{eqnarray}
Note that $W^{\mathbf{e}}_a$ and $W^{\mathbf{o}}_a$ are even and odd respectively under topological duality.

Functional Fourier of the Gaussian functional induced by $\mathrm{Q}_{\mathrm{B}}$ on $W^{\mathbf{e}}_a$ is defined to be\cite{LA1}
\begin{equation}
\int_{W^{\mathbf{e}}_a}e^{2\pi i \langle  {w^{\mathbf{e}}}',w^{\mathbf{e}}\rangle-(\pi/\s) \mathrm{Q}_\mathrm{B}( w^{\mathbf{e}}-\bar{w}^{\mathbf{e}})}\;\mathcal{D}_\lambda w^{\mathbf{e}}
:=e^{2\pi i \langle  {w^{\mathbf{e}}}',\bar{w}^{\mathbf{e}}\rangle}\mathrm{Det}_\lambda (\s{\mathrm{W}_\mathrm{B}})^{1/2}e^{-\pi\s \mathrm{W}_\mathrm{B}( {w^{\mathbf{e}}}')}
\end{equation}
where $\mathrm{W}_\mathrm{B}({w_1^{\mathbf{e}}}',{w_2^{\mathbf{e}}}'):=-2\langle W {w_1^{\mathbf{e}}}',{w^{\mathbf{e}}_2}'\rangle$ and $W$ is inverse to $Q$. At ${w^{\mathbf{e}}}'=0$ this decomposes
\begin{equation}\label{integral def}
\int_{ W^{\mathbf{e}}_a}e^{-(\pi/\s) \mathrm{Q}_\mathrm{B}( w^{\mathbf{e}}-\bar{w}^{\mathbf{e}})}\;\mathcal{D}_\lambda w^{\mathbf{e}}
=\int\!\!\!\!\!\!\!\!\sum_{\{\bar{w}^{\mathbf{e}}\}}\int_{ W^{\mathbf{e}}_{0}}e^{-(\pi/\s) \mathrm{Q}_\mathrm{B}(w^{\mathbf{e}})}\;\mathcal{D}_\lambda w^{\mathbf{e}}
=\int\!\!\!\!\!\!\!\!\sum_{\{\bar{w}^{\mathbf{e}}\}} \mathrm{Det}_\lambda
 (\s \mathrm{W}_\mathrm{B})^{1/2}
\end{equation}
where $ W^{\mathbf{e}}_{0}$ is a {Banach} space of pointed maps $ w^{\mathbf{e}}:(\mathbb{T},\ti_a)\rightarrow (\C^{2m},0)$.\footnote{Since $ W^{\mathbf{e}}_{0}$ is Banach, its integrator $\mathcal{D}_\lambda w^{\mathbf{e}}$ is translation invariant. Together with $W^{\mathbf{e}}_a\ni w^{\mathbf{e}}=\bar{w}^{\mathbf{e}}+G{w^{\mathbf{e}}}'$ for all ${w^{\mathbf{e}}}'$ in the dual space of $W^{\mathbf{e}}_a\backslash\mathrm{Ker}(D)$, this allows the decomposition expressed by the first equality.}

Being an abelian group, $W^{\mathbf{e}}_a$ doesn't support scalar multiplication of $w^{\mathbf{e}}\in W^{\mathbf{e}}_a$, so direct field renormalization as practised in QFT is not available. However, the factor $\s\in\C_+$ scales the quadratic form $\mathrm{Q}_\mathrm{B}$ and hence, indirectly, the argument $w^{\mathbf{e}}-\bar{w}^{\mathbf{e}}$. Further, adjusting any parameters in $\mathrm{Q}_\mathrm{B}$ is tantamount to defining a new quadratic form $\widetilde{\mathrm{Q}}_\mathrm{B}$. Otherwise said; the scale $\s$ and defining quadratic form $\mathrm{Q}_\mathrm{B}$ are part of the specification of $\mathcal{D}_\lambda w^{\mathbf{e}}$. The same can be said for a general action functional $\mathrm{S}_{\mathrm{B}}=\mathrm{Q}_{\mathrm{B}}+\mathrm{V}$. Therefore, off hand it appears $G_\Lambda$ can accommodate the renormalization program of QFT, but we will not pursue the details here.

Transcribing to QFT we have: I) $W^{\mathbf{e}}_a$ corresponds to the underlying additive abelian group $\Bold{\phi}$ of the Banach space of bosonic fields (see \cite[application A.4]{LA1} for details). Because of the $\mathbb{Z}_2$ structure on $W^{\mathbf{e}}_a$, elements $w^{\mathbf{e}}\in W^{\mathbf{e}}_a$ correspond to 2-tuples of \emph{independent} degrees of freedom $\Bold{\phi}\ni(\phi^\ast,\phi):\R^{3,1}\rightarrow \C^{2m}$ where $\phi\sim z^+\times z'^+$ and $\phi^\ast\sim z^-\times z'^-$. II) Elements ${w^{\mathbf{e}}}'\in{W^{\mathbf{e}}_a}'$ correspond to 2-tuples of external sources\footnote{Not to be confused with the complex-induced map $J$.} with $(J_{\phi^\ast},J_\phi)\sim 2\pi {w^{\mathbf{e}}}'$. III) $D$ is a linear differential operator. IV) $\lambda_{vac}:\Bold{\phi}\rightarrow\C^{2m}$ with $\bar{w}^{\mathbf{e}}=0$ for the specific instance of vacuum-to-vacuum transitions. V) Lastly, $\s\rightarrow\pi i\hbar$. This yields (dropping the subscript on $\mathrm{Q}_{\mathrm{B}}$ to reflect vacuum-to-vacuum boundary conditions)
\begin{equation}
\langle0|0\rangle_{\mathrm{bos}}:=\int_{\Bold{\phi}}e^{\frac{i}{\hbar} \mathrm{Q}(\phi^\ast,\phi)}\;\mathcal{D}_{\lambda_{vac}}(\phi^\ast,\phi)
=\mathrm{Det}_{\lambda_{vac}}
 (i \hbar\mathrm{W})^{1/2}
 :=\mathrm{det}(i \hbar Q)^{-1/2}
\end{equation}
where the form $\mathrm{Q}(\phi^\ast,\phi)=-\frac{1}{2}\langle Q \phi^\ast ,\phi\rangle$, and $W$ is the inverse of $Q$. For ${w^{\mathbf{e}}}'\neq0$, we get
\begin{equation}
Z(J):=\langle0|0\rangle_{J_\Phi}
=\int_{\Bold{\phi}}e^{ i \langle  J_{\phi^\ast},\phi^\ast\rangle+i \langle  J_\phi,\phi\rangle+(i/\hbar) \mathrm{Q}(\phi^\ast,\phi)}\;\mathcal{D}_{\lambda_{vac}}(\phi^\ast,\phi)
=\mathrm{det}(i\hbar{Q})^{-1/2}e^{(i\hbar/2) \mathrm{W}(J_{\phi^\ast},J_\phi)}
\end{equation}
which must be supplemented with interaction terms in the form $\mathrm{S}_{\mathrm{B}}=\mathrm{Q}_{\mathrm{B}}+\mathrm{V}$ and then perturbative methods must be applied to yield the generating functional employed by QFT for bosonic fields. Remark that $\det Q=\det Q_\Re^2$ with $Q_\Re:=Q|_{X_a}$ where $Z_a=X_a\oplus\mathrm{J} Y_a$.

For fermionic fields $\Bold{\psi}\ni(\overline{\psi},\psi):\R^{3,1}\rightarrow \C^{2m}$, construct functional Fourier for a \emph{skew-Gaussian} functional induced by a symplectic form $\Omega_{\mathrm{B}}$ on $Z_a\backslash\mathrm{Ker}(D)$ (see \cite[\S~3.2]{LA1}). Transcription to QFT parallels the bosonic case where now $(\overline{\psi},\psi)\sim w^{\mathbf{o}}=(z^-\times z'^+,z^+\times z'^-)$ and $(J_{\overline{\psi}},J_\psi)\sim 2\pi {w^{\mathbf{o}}}'$. This gives
\begin{equation}
\langle0|0\rangle_{\mathrm{ferm}}:=\int_{\Bold{\psi}}e^{\frac{1}{\hbar} \Omega(\overline{\psi},\psi)}\;\mathcal{D}_{\lambda_{vac}}(\overline{\psi},\psi)
=\mathrm{Pf}_{\lambda_{vac}}
 (\hbar\mathrm{M})^{-1}
 :=\mathrm{pf}(\hbar \mathit{\Omega})
\end{equation}
where the form $\Omega(\overline{\psi},\psi):=-\frac{1}{2}\langle\mathit{\Omega}\overline{\psi},\psi\rangle$ and $\mathrm{M}=\Omega^{-1}$, and the generating functional
\begin{equation}
Z(J):=\langle0|0\rangle_{{J}_\Psi}
=\int_{\Bold{\psi}}e^{ i \langle  {J}_{\overline{\psi}},\overline{\psi}\rangle
+i \langle  {J}_\psi,\psi\rangle+(1/\hbar) \Omega(\overline{\psi},\psi)}\;\mathcal{D}_{\lambda_{vac}}(\overline{\psi},\psi)
=\mathrm{pf}(\hbar\mathit{\Omega})e^{(1\hbar/2) {\mathrm{M}}(J_{\overline{\psi}},J_\psi)}
\end{equation}
where ${\mathrm{M}}(J_{\overline{\psi}},J_\psi)=-\frac{1}{2}\langle J_\psi,\mathit{\Omega}^{-1}J_\psi\rangle$. Note that $\overline{\psi}$ and $\psi$ are contained in orthogonal Lagrangian subspaces determined by $\Omega$, and in this sense they are (dynamical) conjugate degrees of freedom. The interpretation, therefore, is that $\mathcal{D}_{\lambda_{vac}}(\overline{\psi},\psi)$ quantifies quantum dynamics while $\mathcal{D}_{\lambda_{vac}}(\phi^\ast,\phi)$  quantifies quantum correlations.

Now we construct the generating functional using functional Mellin. In the context of QFT, restrict to abelian $G^\C$ and take $\rho(g)\in L_B(\mathcal{H})$ and $\mathrm{Q}(g)=-\frac{i}{\hbar}\rho(g)^\dag Q \rho(g)\in L_B(\mathcal{H})$ where $Q$ is positive-definite and $\rho$ is unitary.

As we did for Fourier, extend $\mathrm{Q}$ to $G^\C\times\tilde{G}^\C$ where $\tilde{G}^\C$ is the Pontryagin dual and decompose into even and odd parts according to the complex grading
\begin{equation}
  G^\C\times \tilde{G}^\C={\bigoplus}_\pm\left[\left(G^\pm\times \tilde{G}^\pm\right)\oplus \left(G^\pm\times \tilde{G}^\mp\right)\right]
  =:G^{\mathbf{e}}\oplus G^{\mathbf{o}}\;.
\end{equation}
Identify  $\mathbf{F}_{\mathbb{S}}(G^{\mathbf{e}})$ with bosonic observables. This gives
\begin{eqnarray}
Q^{-\alpha_{\Re}}_\lambda=\int_{G^{\mathbf{e}}} e^{-\mathrm{Q}(g)}g^{\alpha_{\Re}}\;\mathcal{D}_\lambda g
\;,\;\;\;\;\;{\alpha_{\Re}\in\mathbb{S}}\;.
\end{eqnarray}
By analogy with the Fourier case, sources could be introduced via the Pontryagin dual group of $G^{\mathbf{e}}$ but this is not particularly useful or necessary as will be evident shortly.

To include interactions, consider  $\mathrm{S}(g)=\mathrm{Q}(g)+\mathrm{V}(g)$ such that $[\mathrm{Q}(g),\mathrm{V}(g)]=0$. Expand $e^{-\mathrm{V}(g)}$ as a formal power series and write $e^{-\mathrm{S}(g)}=\sum_{m=0}^\infty a_m\, \mathrm{V}(g) ^m e^{-\mathrm{Q}(g)}$ where $a_m$ are real constants multiplying $\mathrm{V}(g)^m$. Hence, the integral for a non-quadratic action functional is
\begin{equation}
S_{\lambda}^{-{\alpha_{\Re}}}
\equiv\int_{G^{\mathbf{e}}} e^{-{\mathrm{S}}(g)}g^{\alpha_{\Re}}\;\mathcal{D}_\lambda g
=\int_{G^{\mathbf{e}}} \sum_{m=0}^\infty a_m \mathrm{V}(g)^m e^{-\mathrm{Q}(g)}g^{\alpha_{\Re}}\;\mathcal{D}_\lambda g
\end{equation}
which gives rise to a perturbative loop expansion up to order $M$ given by
\begin{equation}\label{loop expansion}
{S^{(M)}}_{\lambda}^{-{\alpha_{\Re}}}:=\sum_{m=0}^M a_m\int_{G^{\mathbf{e}}} \mathrm{V}(g)^me^{-\mathrm{Q}(g)}g^{\alpha_{\Re}}\;\mathcal{D}_\lambda g\;.
\end{equation}

Using the same QFT identifications as in the preceding Fourier construction yields an ${\alpha_{\Re}}$-dependent, bosonic generating functional defined by
\begin{equation}\label{generating functional}
Z_{\lambda_{vac}}(\alpha_{\Re})
:=\langle0|Q^{-\alpha_{\Re}}_\lambda|0\rangle_{\mathrm{bos}}
=\int_{G^{\mathbf{e}}} \langle0|e^{-\mathrm{Q}(g)}g^{\alpha_{\Re}}|0\rangle\;\mathcal{D}_\lambda g
=:\int_{\Bold{\Phi}} e^{-\mathrm{Q}(\Phi)}\Phi^{\alpha_{\Re}}\;\mathcal{D}_{\lambda_{vac}} \Phi
\end{equation}
with a perturbative generating functional up to order $M$ for a non-quadratic action functional $\mathrm{S}(g)=\mathrm{Q}(g)+\mathrm{V}(g)$ given by
\begin{equation}\label{perturbation series}
{Z_{\lambda_{vac}}^{(M)}}(\alpha_{\Re})
:=\sum_{m=0}^M {a}_m\int_{\Bold{\Phi}}
\,\mathrm{V}(\Phi)^me^{-\mathrm{Q}(\Phi)}\Phi^{\alpha_{\Re}}\;\mathcal{D}_{\lambda_{vac}} \Phi\;.
\end{equation}
In equation (\ref{generating functional}) we identified $G^{\mathbf{e}}$ with the underlying \emph{additive} abelian group of a Banach space of complex scalar bosonic fields $\Phi\in\Bold{\Phi}$ relative to the vacuum $|0\rangle\in\mathcal{H}$ such that $\langle0|e^{-\mathrm{S}(g)}g^{\alpha_{\Re}}|0\rangle
\equiv e^{-\mathrm{S}(\Phi)}\Phi^{\alpha_{\Re}}$. According to the Fourier construction, the vacuum expectation corresponds to the determinant representation $\det:L_B(\mathcal{H})\rightarrow\C$ so we have normalized $\mathcal{D}_{\lambda_{vac}}\Phi$ by $Z_{\lambda_{vac}}(0)=\det(i\hbar Q)^{-1/2}$. Note that here $\mathcal{D}_{\lambda_{vac}}\Phi$ is translation invariant as befits an integrator over an additive abelian group.

To make contact with vacuum $n$-point functions in QFT: First, restrict to ${\alpha_{\Re}}\in\mathbb{N}_+$. Observe that ${Z^{(M)}}_{\lambda}(\alpha_{\Re})=0$ for $V(\Phi)=0$ and  $\alpha_{\Re}$ an odd number since $Q$ is quadratic and $\mathcal{D}_\lambda \Phi$ is translation invariant. Also, $\Phi^{\alpha_{\Re}}:=e^{{\alpha_{\Re}}(\log \Phi^\ast+\log \Phi)/2}=\frac{1}{2}(\Phi^\ast \Phi+\Phi\Phi^\ast)^{{\alpha_{\Re}}/2}$ follows from  BCH  since the \emph{group} commutator $[\Phi^\ast,\Phi]_{\Bold{\Phi}}$ is a $c$-number and $\log \Phi$ is Hermitian.\footnote{Since $[\Phi^\ast,\Phi]_{\Bold{\Phi}}$ is a $c$-number and $\log \Phi$ is Hermitian, then $[\log\Phi^\ast,\log \Phi]=0$. The multiplication here is defined by $\Phi^\ast\Phi=Q(\Phi)=Q(\Phi)^\ast=Q(\Phi^\ast)$.} Next, localize by $\lambda_{vac}:\Bold{\Phi}\rightarrow \C$ according to
\begin{equation}
[\Phi^\ast,\Phi]_{\Bold{\Phi}}\stackrel{\lambda_{vac}}{\longrightarrow} [\Phi^\ast,\Phi]_{\Bold{\Phi}}(p,p')
=[\Phi^\ast(p),\Phi(p')]\;\;\;\;\;\;p,p'\in\R^{3,1}\;.
\end{equation}
This is tantamount to $\Phi^2\mapsto e^{\left(\log\Phi^\ast(p)+\log\Phi(p')\right)}
=\frac{1}{2}(\Phi^\ast(p)\Phi(p')+\Phi(p)\Phi^\ast(p'))=:\Phi(p,p')$ and by iteration $\Phi^{2n}\mapsto\Phi(p_1,p'_1)\Phi(p_2,p'_2)\cdots\Phi(p_n,p'_n)$ up to different pairings of $p_i,p'_j$. Momentum $n$-point correlation functions for general QFT operators $\mathcal{O}(\Phi)$ obtain from
$\mathcal{M}_\lambda[\mathrm{O}^{\mathrm{S}};\alpha]$ with $\mathrm{O}^{\mathrm{S}}(g)\equiv O(\Phi)e^{-{S}(\Phi)}$.

For the fermionic generating functional, follow the Fourier construction. Replace Hermitian $\mathrm{Q}$ with skew-Hermitian $\Omega$ and $G^{\mathrm{e}}$ with $G^{\mathrm{o}}$. Identify $G^{\mathrm{o}}$ with the underlying additive abelian group of a Banach space of complex scalar fermionic fields $\Psi\in\Bold{\Psi}$. This leads to the fermionic generating functional
\begin{equation}
Z_{\lambda}(\alpha_{\Re})
:=\langle0|i\mathit{\Omega}_\lambda^{-\alpha_{\Re}}|0\rangle_{\mathrm{ferm}}
=\int_{\Bold{\Psi}} e^{-i\mathit{\Omega}(\Psi)}\,\Psi^{{\alpha_{\Re}}}\;\mathcal{D}_{\lambda_{vac}} \Psi
\end{equation}
where $\mathcal{D}_{\lambda_{vac}} \Psi$ is a translation-invariant skew-Gaussian integrator\cite{LA1} that transforms obversely to $\mathcal{D}_{\lambda_{vac}} \Phi$ and is normalized so that $Z_{\lambda_{vac}}(0)=\mathrm{pf}(\hbar\mathit{\Omega})$. Interactions are included in parallel with the bosonic case.

Crucially, the group commutator on $G^{\mathrm{e}}\oplus G^{\mathrm{o}}$ is graded because the product on $G^{\mathrm{o}}$ is defined by $\Omega(g_1,g_2)=-\Omega(g_2,g_1)$. This implies $\overline{\Psi}\Psi
=-\frac{1}{2}\langle\mathit{\Omega}\overline{\Psi},\mathrm{J}\Psi\rangle
=\frac{1}{2}\langle\mathit{\Omega}\Psi,\mathrm{J}\overline{\Psi}\rangle=-\Psi\overline{\Psi}$ since $\overline{\Psi}\sim \mathrm{J}\Psi$ and $[\mathit{\Omega},\mathrm{J}]=0$. Hence by BCH, $\Psi^2=e^{(\log \overline{\Psi}+\log\Psi)} =\frac{1}{2}(\overline{\Psi} \Psi-\Psi\overline{\Psi})$ because the group anti-commutator $\{\overline{\Psi},\Psi\}_{\Bold{\Psi}}$ is a $c$-number. Our chosen topological localization then gives $\Psi^2(p,p')=\frac{1}{2}(\overline{\Psi}(p) \Psi(p')-\Psi(p)\overline{\Psi}(p'))$ and
\begin{equation}
\{\overline{\Psi},\Psi\}_{\Bold{\Psi}}\stackrel{\lambda_{vac}}{\longrightarrow}\{\overline{\Psi},\Psi\}_{\Bold{\Psi}}(p,p')
=\{\overline{\Psi}(p),\Psi(p')\}
\end{equation}
which gives the two-point correlation function $\langle0|\{\overline{\Psi}(p),\Psi(p')\}|0\rangle$.

The functional Mellin objects for non-quadratic action functionals presented in this subsection should be considered preliminary since we did not adequately deal with renormalization and gauge symmetry.

\subsection{Mellin scattering}\label{scattering}
Mellin-type integrals are ubiquitous in QFT and string theory scattering amplitudes: In QFT, Schwinger's trick leads to the parametric representation of Feynman diagrams \cite{IZ}, which motivates the world-line formalism \cite{SCH}, which in turn can be seen as the infinite-tension limit of string scattering \cite{BK}. In this subsection we want to explore how these representations fit into the functional Mellin formalism.

\subsubsection{point-to-point Green's functions}\label{point-to-point}
Here we calculate Green's functions for the Klein-Gordon operator of massive states in $L^2(\R^{1,3},\mathcal{W})\equiv\mathcal{H}$ where $\mathcal{W}$ furnishes the spin $0,1/2,1$ representations of $\R^{1,3}\rtimes SL(2,\C)$. Of course the Green's functions can be alternatively realized as correlation functions via functional integrals over second quantized fields according to QFT in the usual way, but it has long been recognized that simple point-particle path integrals can also do the job---even with background gauge fields included.\cite{STR} We include the calculation here to illustrate our methods.

The first step is to calculate the free scalar elementary kernel for evolution from the state $|\psi_{x_a}\rangle$ at evolution-time $t_a$ to the state $|\psi_{x_b}\rangle$ at evolution-time $t_b$.\footnote{We follow \cite{PDM,LA2} which provide a QM path integral realization of elementary kernels/Green's functions of linear, second order partial differential equations.} The equation to solve is
$\left(\square+V(x)\right){K}(x_a,x_b)={\delta}(x_a-x_b)$
where $V(x)=m^2$ in our case, $x_b,\;x_a\in\R^{1,3}$, and the position-to-position \emph{Dirichlet} elementary kernel (a.k.a. Green's function) is
\begin{eqnarray}\label{covariance}
{K}(x_b,x_a)
&:=&\langle \psi_{x_b}|{K}_\Gamma|\psi_{x_a}\rangle\notag\\
&=&\langle \psi_{x_b}|\mathcal{M}_\Gamma[\mathrm{E}^{-(\square+m^2)};1]|\psi_{x_a}\rangle \notag\\ &=&\mathcal{M}_\Gamma[\langle \psi_{x_b}|\mathrm{E}^{-(\square+m^2)}| \psi_{x_a}\rangle;1]\;.
\end{eqnarray}
Localizing according to $\mathrm{E}^{-(\square+m^2)}(g)
\stackrel{\lambda_{i\R_+}}{\longrightarrow}e^{-(\square+m^2)\Delta t}$ with\footnote{Since $(\square+m^2)$ is self-adjoint, $\rho(g)$ must be imaginary. To ensure a unitary representation $\rho$, we need skew-Hermitian $\log(i|\Delta t|)\in i\R$ which implies $\Delta t^\dag=-i|\Delta t|^{-1}=(\Delta t)^{-1}$ so that $\rho(g)^\dag=\rho(g)^{-1}$. } $\rho(g)\equiv\Delta t:= t_b-t_a\in i\R_+$ gives
\begin{eqnarray}\label{Dirichlet kernel}
\langle \psi_{x_b}|e^{-(\square+m^2)\Delta t}|\psi_{x_a}\rangle\notag\\
&&\hspace{-1.65in}=\int_{\C^4}\int_{\C^4}
\psi^\ast_{x_b}(z_b)(z_b;\Bold{\mu}'|
e^{-(\square+m^2)\Delta t}|z^\ast_a;\Bold{\mu})\psi_{x_a}(z^\ast_a)
\;dz_b\,dz_a\notag\\
&&\hspace{-1.65in}\stackrel{real\,submanifold}{\longrightarrow}\int_{\R^{1,3}}\int_{\R^{1,3}}
\delta_\epsilon({x_b},\Re(z_b)){K}_{U_{\Delta t}}(z_b,z_a^\ast)
\delta_\epsilon({x_a},\Re(z_a))\;dz_b\,dz_a\notag\\
\end{eqnarray}
with $\Bold{\mu}$ a basis in $\mathcal{W}$, $U_{\Delta t}:=e^{-(\square+m^2)\Delta t}$ the evolution operator, $z_b:=z(t_b)$ and $z^\ast_a:=z^\ast(t_a)$ in $\C^4$, and fixed-point states $\psi_{x_b}(z_b)=\delta_\epsilon({x_b},\Re(z_b))$ and $\psi_{x_a}(z^\ast_a)=\delta_\epsilon({x_a},\Re(z_a))$. Here, $\delta_\epsilon$ denotes a \emph{regularized} delta distribution. The chosen regularization must respect covariance with respect to $\R^{1,3}\rtimes SL(2,\C)$, and the limit $\epsilon\rightarrow0$ is only taken after integration.

The reproducing kernel ${K}_{U_{\Delta t}}(z_b,z_a^\ast)$ derives from a parametrized Gaussian functional integral\cite{LA1} associated with pointed maps $Z_a\ni\zeta:(\mathbb{T},t_a)\rightarrow(\C^4,0)$ where $\mathbb{T}\equiv[t_a,t_b]\subset i\R$.
\begin{equation}\label{distribution}
{K}_{U_{\Delta t}}(z_b,z^\ast_a)
:=\int_{Z_a}{\delta}(z(t_b,\zeta)-z_b)e^{- S(z(t_b,\zeta))}
\;\mathcal{D}\zeta
\end{equation}
with $z(t,\zeta)=z^\ast_a\cdot\mathit{\Sigma}(t-t_a,\zeta)$ where $\mathit{\Sigma}(t-t_a,\zeta)$ is a global transformation on $\R^{1,3}$ such that $z(t_a,\zeta)=z^\ast_a\cdot\mathit{\Sigma}(0,\zeta)=z^\ast_a\,$, and the Poincar\'{e} invariant and (gauge fixed) time-reparametrization invariant action is
\begin{equation}
 S(z(t_b,\zeta)
 =Q(z(t_b,\zeta)+\int_{t_a}^{t_b}m^2\;dt
 =\int_{t_a}^{t_b}
z^\ast_\mu(t,\zeta)\left(\eta^{\mu\nu}\frac{d^2}{dt^2}\right)z_\nu(t,\zeta)\;dt
+\int_{t_a}^{t_b}m^2\;dt\;.
\end{equation}
Explicitly, the parametrization $z(t,\zeta)=z_{cr}(t)+\zeta(t)$ is a variation about the critical path $z_{cr}(t)$ with fixed end-points  $z^\ast_a,z_b$ given by $z_{cr}(t)=z^\ast_a+\left(\frac{z_b-z^\ast_a}{t_b-t_a}\right)(t-t_a)$. Substituting into the functional integral yields the well-known heat kernel
\begin{eqnarray}
 \langle z_b|e^{-(\square+m^2)\Delta t}|z_a\rangle
 &=&e^{-\left(\frac{(z_b-z_a^\ast)^2}{4\Delta t}-m^2\Delta t\right)}
 \int_{Z_a}\delta(\zeta(t_b))e^{-\int_{t_a}^{t_b}
(\dot{\zeta}(\tau))^2\;d\tau}\;\mathcal{D}\zeta\notag\\
&=&e^{-\left(\frac{\pi(z_b-z_a^\ast)^2}{4\Delta t}-m^2\Delta t\right)}(2\Delta t)^{-2}
\end{eqnarray}
where we used
\begin{equation}
\delta(\zeta(t_b))
=\int_{\R^{1,3}}e^{- i\langle u',\zeta(t_b)\rangle}\;du'
=\int_{\R^{1,3}}e^{- i\langle u'\delta_{t_b},\zeta\rangle}\;du'
\end{equation}
and skipped a few trivial steps that explicate $\delta_\epsilon$ and calculate the function integral.\footnote{The skipped steps include the localization $\lambda:Z_a\rightarrow\R^{1,3}$ by $\zeta\mapsto \zeta(t_b)=:{u}$ for ${u}\in\R^{1,3}$. By duality then, $\zeta'\mapsto {u}'\delta_{t_b}$ which renders the variance $W({u}'\delta_{t_b})=|{u}'|^2\Delta t$.  Remark that, following localization, ${K}(z_b,z^\ast_a)$ should be regarded as a distribution; in which case the action functional must localize to an honest function and this will typically involve regularization/gauge fixing and renormalization in more complicated examples. Also, for the regularized delta distributions in (\ref{Dirichlet kernel}) we used $\delta_\epsilon(x)=(\epsilon)^{-1/2}e^{-\pi x^2/\epsilon}$.} Finally, functional Mellin of $\langle \psi_{x_b}|e^{-(\square+m^2)\Delta t}|\psi_{x_a}\rangle$ renders the familiar modified Bessel function representation of the scalar Klein-Gordon propagator.

Similarly, by Fourier transform
\begin{equation}
\langle {p_b}|e^{-(\square+m^2)\tau}|{p_a}\rangle=\delta(p_b-p_a^\ast)e^{-\tau(p_b^2- m^2)}
\end{equation}
where $p_b,p_a^\ast\in \C^4$ (recall $z\in\C^4$) and $\tau:=\Delta t$ is a Lorentz scalar and interpreted as an evolution-time interval.
This gives the (complex) momentum-to-momentum realization (recall Example \ref{resolvent example})
\begin{eqnarray}\label{momentum propagator}
 \langle{p_b}|(\square+m^2)^{-1}_\Gamma|{p_a}\rangle
 &=&\int_{i\R_+\cup i\R_-}\langle{p_b}| e^{-(\square+m^2)(\tau)}|{p_a}\rangle \tau\;d\log (\tau)\notag\\
&&\hspace{-1.25in}=\left(\int_{0}^{i\infty}+\int_{0}^{-i\infty}\right)e^{-S(p_a^\ast,t_a;p_b,t_b)}\tau
\;d\log (\tau)\notag\\
&&\hspace{-1.25in}=\left(\int_{0}^{i\infty}+\int_{0}^{-i\infty}\right)
e^{-\tau(p_b^2- m^2)}\tau\;d\log (\tau)\notag\\
\end{eqnarray}
with $p_a^\ast=p_b$. A functional inverse power such as $(\square+m^2)^{-1}_\Gamma$ in a complex momentum representation comprises a principle value for $|p_b|^2> m^2$ along with a delta function for $|p_b|^2=m^2$.(see \cite[def.~5.5]{LA3} and footnote 6 in \S~\ref{resolvents}) Consequently
\begin{equation}
{K}(p_b,p_a^\ast)=\langle{p_b}|(\square+m^2)^{-1}_\Gamma|{p_a}\rangle
=\frac{1}{|p_b|^2-m^2}\pm i\delta(|p_b|^2-m^2)\;\;\;\;\mathrm{for}\;i\R_\pm\;.
\end{equation}
Choosing the opposite sign of the delta function exchanges $i\R_+\leftrightarrow i\R_-$.

Notice the integral-convergence conditions $\Im(p_b^2-m^2)\gtrless0$ for $i\R_\pm$ naturally incorporate the Feynman prescription for the Fourier transformed propagator. Moreover, unitarity of the representation $\rho$ in the Mellin functional integral implies  the position-to-position as well as the momentum-to-momentum $i\R_\pm$ kernels are invariant under $\tau\rightarrow\tau^{-1}$ as one can confirm by direct calculation. This is a reflection of unitarity and the multiplicative-group nature of the evolution-time \emph{interval}. Because $\tau$ is interpreted as an evolution-time interval, invariance under $\tau\rightarrow \tau^{-1}$ appears to indicate any UV/IR dichotomy is not fundamental. Meanwhile, evolution-time reversal (which means $\tau=i|\Delta t|\rightarrow -i(|-\Delta t|)=\tau^\ast$) exchanges $i\R_+\leftrightarrow i\R_-$. This suggests one should study system evolution parametrized by \emph{complex} one-parameter subgroups with localization $\lambda_{\C^\times}:\phi_{\mathfrak{a}}(\C)\rightarrow \C^\times$ --- a kind of complex world-line formalism.

\begin{remark}
As a consistency check we calculate the inverse off-shell momentum-to-momentum propagator when $p_a^\ast=p_b$ and $\Re(p_b^2)> m^2$
\begin{eqnarray}
{K}^{-\alpha}(p_b,p_a^\ast)
&=&\langle{p_b}|\mathcal{M}_\Gamma[\mathrm{E}^{\frac{-1}{(p_b^2-m^2)}};\alpha]|
{p_a}\rangle\notag\\
&=&\frac{1}{\Gamma(\alpha)}\int_{i\R_+\cup i\R_-}e^{\frac{-1}{p_b^2-m^2}\tau}\tau^\alpha\;d\log (\tau)\notag\\
&=&\left(\frac{1}{p_b^2-m^2}\right)^{-\alpha}\,
\;,\;\;\;\Re(\alpha)>0\;.
\end{eqnarray}
In particular, for $\alpha=1$, this yields ${K}^{-1}(p_b,p_a^\ast)=(p_b^2-m^2)$ which verifies the identity $\int{K}^{-1}(p',p''){K}(p'',p)\;dp''=\delta(p',p){Id}$ for $p_b^2\neq m^2$.
\end{remark}

To go beyond the spin-0 case, consider the operator
\begin{equation}
\mathcal{M}_\Gamma[\mathrm{E}^{-(\square+m^2)};\alpha]
=\int_{\phi_{\mathfrak{a}}(\R)} e^{-(\square+m^2)(g)}\,\rho(g^\alpha)\;d\nu(g_{\Gamma})\;.
\end{equation}
If we want to talk about \emph{particle} transitions, we need $\rho$ to be a unitary \emph{irreducible} representation, and we need for it to act trivially on state vectors corresponding to elementary particles since the notion of `elementary' implies invariance under evolution.

Accordingly, we now localize by $\lambda_{A_{\mathfrak{a}}}:\phi_{\mathfrak{a}}(\R)\rightarrow A_{\mathfrak{a}}$ where $A_{\mathfrak{a}}$ is abelian. This induces a unitary irreducible representation $g_\lambda\equiv a\rightarrow\rho(a)$ with $a\in A_{\mathfrak{a}}$. In light of this, for each relevant representation $r$ let us take for the Hamiltonian operator
\begin{equation}
\mathrm{H}^{(r)}(a)=(p^2-m^2)\rho^{(r)}(a)=(p^2-m^2)\,\tau P_{\|}^{(r)}/|P_{\|}^{(r)}{P_{\|}^{(r)}}^\dag|
\end{equation}
where $\tau P_{\|}^{(r)}\in L(\mathcal{H})$, $\tau\in i\R_+\cup i\R_-$, and $P_{\|}^{(r)}$ is a projection onto an irreducible subspace of $\mathcal{H}$ furnished by $\mathcal{W}$.\footnote{We got this projection idea from \cite{FOF}.} To verify the unitarity of representations $\rho^{(r)}$, notice that $\log (i|\tau|)\in i\R$ implies $\tau^\dag=\tau^{-1}=-i|\tau|^{-1}$ so that
\begin{equation}
{\rho^{(r)}(a)}^\dag={(\tau P_{\|}^{(r)}/|P_{\|}^{(r)}{P_{\|}^{(r)}}^\dag|)}^\dag
=-i\frac{|\tau|^{-1}}{|P_{\|}^{(r)}{P_{\|}^{(r)}}^\dag|}{P_{\|}^{(r)}}^\dag
\end{equation}
 and therefore $\rho^{(r)}(a){\rho^{(r)}(a)}^\dag=Id$. (Recall that $\tau^\dag\neq \tau^\ast$.)

The relevant objects to analyze are the total Hamiltonian
\begin{equation}
H(a):=\bigoplus_r \mathrm{H}^{(r)}(a)
=\bigoplus_r(p^2-m^2)\frac{\tau P_{\|}^{(r)}}{|P^{(r)}_{\|}{P^{(r)}_{\|}}^\dag|}
\end{equation}
and its elementary kernel
\begin{equation}
\left({K}(p_b,p_a^\ast)\right)_{\Bold{\mu}'\Bold{\mu}}
=\bigoplus_r\left({K}^{(r)}(p_b,p_a^\ast)\right)_{\mu'\mu}\;.
\end{equation}

For higher spin state vectors labelled by $(j',j)$, the work to find the projection operators has already been done many times over and we just state the $(1/2,0)\oplus(0,1/2)$ and $(1/2,1/2)$ results.  For Dirac spinors we take
\begin{equation}
\frac{\tau P_{\|}^{(1/2)}}{|P_{\|}^{(1/2)}(P_{\|}^{(1/2)})^\dag|}=\frac{\tau}{p^2-m^2}(\slashed{p}\pm m)\;.
\end{equation}
which gives
\begin{eqnarray}
\left({K}^{(1/2)}(p_b,p_a^\ast)\right)_{\alpha\dot{\alpha}}
&=&\left(\slashed{p}_b\pm m\right)_{\alpha\dot{\alpha}}\,\langle {p_b}|\mathrm{E}^{-\mathrm{H}^{(0)}}{p_a}\rangle\notag\\
&=&\frac{\left(\slashed{p}_b\pm m\right)_{\alpha\dot{\alpha}}}{p_b^2-m^2}\,\delta(p_b,p_a^\ast)\;.
\end{eqnarray}
For vector bosons, the operator $\square+m^2$ corresponds to the unitary gauge so we choose $P_{\|}^{(1)}=(\eta-(p\cdot p)/m^2)$ which yields
\begin{eqnarray}
\left({K}^{(1)}(p_b,p_a^\ast)\right)_{\mu\nu}
&=&\left(\eta-(p\cdot p)/m^2\right)_{\mu\nu}\,\langle {p_b}|\mathrm{E}^{-\mathrm{H}^{(0)}}{p_a}\rangle\notag\\
&=&\frac{\left(\eta-(p\cdot p)/m^2\right)_{\mu\nu}}{p_b^2-m^2}\,\delta(p_b,p_a^\ast)\;.
\end{eqnarray}
Both calculations used the Haar measure to re-scale $\frac{\tau}{|P_{\|}^{(r)}(P_{\|}^{(r)})^\dag|}\rightarrow \tau$ and the operator identity
\begin{equation}
e^{-(p^2-m^2)\tau P_{\|}^{(r)}}\,\tau P_{\|}^{(r)}=e^{-(p^2-m^2)\tau}\,\tau P_{\|}^{(r)}
\end{equation}
which can be seen by expanding the exponential and using ${P_{\|}^{(r)}}^2=P_{\|}^{(r)}$.

\subsubsection{Green's functions on graphs}\label{graphs}
We have shown the point-to-point elementary kernel $\langle b|{K}^\alpha|a\rangle$ associated with a quadratic observable $\mathrm{S}\in\mathbf{F}_{\mathbb{S}_{\mathcal{R}}}(G^\C)$ for the spin $0,1/2,1$ representations of the Poincar\'{e} group are the Mellin transform $\mathcal{M}_\Gamma[\langle b|\mathrm{E}^{-\mathrm{S}}|a\rangle;\alpha]$ for the topological localization $\lambda_{A_{\mathfrak{a}}}:\phi_{\mathfrak{a}}(i\R)\rightarrow A_{\mathfrak{a}}$. The generalization to non-quadratic observables and $n$-point interaction in QFT leads to the machinery of perturbation theory, Feynman diagrams, and their associated graphs. From the functional Mellin approach, the generalization amounts to Mellin transforms of reproducing kernels associated with the graphs. Significantly, functional Mellin adds a new gadget to the transition amplitude machinery --- the complex power $\alpha$. In the next section we'll see expansion around $\alpha=1$ implements renormalization.

To illustrate the functional Mellin approach, momentarily restrict to the massless scalar case. Let $\Bold{\varepsilon}$ with components $[\Bold{\varepsilon}]_{iv}$ be the $I\times V$ incidence matrix of a connected graph $\mathrm{G}$ with  $V$ vertices, $L$ loops, and internal lines $I=L+(V-1)$. Note that $\Bold{\varepsilon}\Bold{\varepsilon}^{\mathrm{T}}$ is the graph Laplacian. One can show that $\mathrm{Rank}(\varepsilon)=V-1$ which implies conservation of external momenta.(see e.g. \cite{IZ}) For $n$-point interactions this can be written as $\sum_{i=1}^n p^i=0$ --- the absence of linear independence reflecting the degeneracy of $\Bold{\varepsilon}$ and its associated Laplacian.

But to proceed we require an invertible Laplacian. It can be deduced by considering (instead of the $n$-point momenta $p^i$) the {sum of momenta} incident at each vertex $v$ which we denote by $p^v$. Total momentum conservation during the evolution time interval, which is implied by $\mathrm{Rank}(\Bold{\varepsilon})=V-1$, demands $\sum_{r=1}^V p^{v_r}=0$. Hence, the function space of \emph{linearly independent momenta} localizes by $\lambda: P_a\rightarrow \C^4\times\R^{V-1}$ where $P_a$ is the space of piece-wise continuous, pointed maps $p^{v_r}:(\mathbb{T},\ti_a)\rightarrow(\C^4,p^{v_r}_a)$. Therefore the relevant generator of time evolution is the graph Laplacian of a truncation of the incidence matrix to $I\times (V-1)$. In the complex momentum representation, the associated quadratic form\footnote{We are using notation from \cite[\S3]{LA1}.} is\cite{IZ}
\begin{equation}
\mathrm{Q}_{\mathrm{G}}(\Bold{p},\Bold{\tau})
=\langle Q_\varepsilon(\Bold{\tau})
\Bold{p},\Bold{p}\rangle
:=\langle \left(\Bold{\varepsilon}\Delta(\Bold{\tau})\Bold{\varepsilon}^{\mathrm{T}}\right)^{-1}
\Bold{p},\Bold{p}\rangle
\end{equation}
where $\Bold{p}:=(p^{v_1},\ldots,p^{v_i},
\ldots,p^{v_{(V-1)}})$ with each $p^{v_r}\in\C^4$ a sum of external momenta incident at vertex $v_r$, the collection of evolution-time intervals is $\Bold{\tau}:=(\tau^1,\ldots,\tau^l,\ldots,\tau^I)\in i\R_+^I$, and $\Delta(\Bold{\tau})$ is the diagonal matrix
\begin{equation}
 \Delta(\Bold{\tau})=\left(
            \begin{array}{ccc}
              1/\tau_1 & \, &\, \\
              \, & \ddots & \, \\
              \, & \, & 1/\tau_I \\
            \end{array}
          \right)\;.
\end{equation}
Observe that $\langle Q_\varepsilon(\Bold{\tau})
\Bold{p},\Bold{p}\rangle$ is the many-point generalization of $\tau p_b^2$ in (\ref{momentum propagator}).

The heat kernel for free propagation on the graph $\mathrm{G}$ with $V$ incoming (conserved) momenta is the $n$-point analog of the reproducing kernel (\ref{distribution});
\begin{equation}
\langle\Bold{p}_b|e^{-(\mathrm{Q}_\mathrm{G}-m^2)\Delta\Bold{\tau}}|\Bold{p}^\ast_a\rangle
={K}_{U_{\Delta\Bold{\tau} }}(\Bold{p}_b,\Bold{p}_a^\ast)
:=\delta(\Bold{p}_b-\Bold{p}_a^\ast)\int_{P_a}{\delta}(\Bold{p}(\Bold{\tau}_b,\zeta)-\Bold{p}_b)e^{- \mathrm{Q}_{\mathrm{G}}(\Bold{p}(\Bold{\tau}_b,\zeta))}
\;\mathcal{D}\zeta
\end{equation}
with $\Bold{p}(\Bold{\tau},\zeta)=\Bold{\varepsilon}\Bold{p}(\Bold{\tau})+\zeta(\Bold{\tau})$. This evaluates to
\begin{equation}
{K}_{U_{\Delta\Bold{\tau} }}(\Bold{p}_b,\Bold{p}_a^\ast)
=\delta(\Bold{p}_b-\Bold{p}_a^\ast)
e^{-(\pi/\s)\langle Q_\varepsilon(\Bold{\tau})
\Bold{p}_b,\Bold{p}_b\rangle}\,\mathrm{det} (Q^{-1}_\varepsilon(\Bold{\tau}))^{-2}\;.
\end{equation}
The right-hand side can be written in terms of  the first and second Symanzik polynomials $\mathcal{U}(\Bold{\tau})$  and $\mathcal{F}(\Bold{p},\Bold{\tau})$ in the usual way.\cite{IZ}

Including now massive scalars, the transition amplitude for $\mathrm{G}$ is the functional Mellin transform $\mathcal{M}_\Gamma[\langle\Bold{p}_b|
\mathrm{E}^{-(\mathrm{Q}_\mathrm{G}-m^2)}|\Bold{p}^\ast_a\rangle;\Bold{\alpha}]$ where $\Bold{\alpha}=(\alpha_1,\ldots,\alpha_I)$ is a multi-index\footnote{Recall (\ref{rho}): Since $G^\C$ is abelian and $\mathfrak{C}^\ast$ is commutative in this case, $\rho(g^\alpha)=\otimes_l\rho(g_l^{\alpha_l})$.} with $\alpha_l\in\mathbb{S}$. Localizing by $\lambda:\phi_{\mathfrak{a}}(i\R)\rightarrow i\R_+^I$, this yields the $\Bold{\alpha}$-dependent parametric representation (with $\delta(\Bold{p}_b-\Bold{p}_a^\ast)$ implied); (c.f. \cite{WE})
\begin{eqnarray}\label{parametric Feynman}
K^{\Bold{\alpha}}(\Bold{p}_b,\Bold{p}_a^\ast)
&=&\int_{i\R_+^I}
e^{-\{\mathcal{F}(\Bold{p}_b,\Bold{\tau})/\mathcal{U}(\Bold{\tau})
-\mathrm{tr}(M^{2}\Delta^{-1}(\Bold{\tau}))\}}
\,\mathcal{U}(\Bold{\tau})^{-d/2}\Bold{\tau}^{\Bold{\alpha}-1}
\;d\Bold{\tau}_\Gamma\notag\\
&\stackrel{\lambda_{i\R_+^I}}{\leftarrow}&
\mathcal{M}_\Gamma[\langle\Bold{p}_b|
\mathrm{E}^{-(\mathrm{Q}_{\mathrm{G}}-\mathrm{M}^2)}|\Bold{p}_b\rangle;\Bold{\alpha}]\notag\\
&=&\langle\Bold{p}_b|
\mathcal{M}_\Gamma[\mathrm{E}^{-(\mathrm{Q}_{\mathrm{G}}-\mathrm{M}^2)};\Bold{\alpha}]|\Bold{p}_b\rangle
=\langle\Bold{p}_b|(\mathrm{Q}_\mathrm{G}-m^2)_\Gamma^{-\Bold{\alpha}}|\Bold{p}^\ast_a\rangle
\end{eqnarray}
where $\mathcal{F}(\Bold{p},\Bold{\tau})/\mathcal{U}(\Bold{\tau})=
\mathrm{Q}_{\mathrm{G}}(\Bold{p},\Bold{\tau})$, the variance $\mathcal{U}(\Bold{\tau})=\mathrm{det}(Q^{-1}_\varepsilon(\Bold{\tau}))\prod\tau^l$, $d=1+3$, mass-squared $M^2$ is an $I\times I$ diagonal matrix, and the volume form is $d\Bold{\tau}_\Gamma=\prod_{l=1}^I d\tau_l/\Gamma(\alpha_l)$.

Observe the integral has the schematic form $\int \mathrm{det}\,F(\Bold{\tau})\Bold{\tau}^{\Bold{\alpha}-d/2-1}\,d\tilde{\Bold{\tau}}_\Gamma$ where we define $d\tilde{\Bold{\tau}}:=\mathrm{det}(Q^{-1}_\varepsilon(\Bold{\tau}))^{-d/2}d{\Bold{\tau}}$ and $F(\Bold{\tau}):=e^{-P^2Q_\varepsilon(\Bold{\tau})+M^2\Delta^{-1}(\Bold{\tau})}$  with matrices $P^2:=|\Bold{p}\rangle\langle\Bold{p}|$, and $M^2:=|\Bold{m}\rangle\langle\Bold{m}|$. Evidently
\begin{equation}
\langle\Bold{p}_b|\mathcal{M}_\Gamma[\mathrm{Det}\,
\mathrm{F}
;\Bold{\alpha}]|\Bold{p}_b\rangle\stackrel{\lambda_{i\R_+^I}}{\rightarrow}
\delta(\Bold{p}_b-\Bold{p}_a^\ast) K^{\Bold{\alpha}}(\Bold{p}_b,\Bold{p}_a^\ast)\;,
\end{equation}
which says the $\Bold{\alpha}$-dependent Feynman amplitude in momentum space is the expectation of $\mathrm{Det}\,\mathrm{E}^{-(\mathrm{P}^2\mathrm{Q}-\mathrm{M}^2\Delta^{-1})}$ with conserved momenta and localized measure $\tilde{\Bold{\tau}}_\Gamma$.

This is the central object of interest in perturbative QFT. It can be viewed as a Mellin functional integral over the multiplicative group of positive-definite diagonal matrices whose integrand is the reproducing kernel of a graph $\mathrm{G}$ evaluated at a classical/critical point. The individual Mellin integrals yield Gamma function terms that depend on $\Bold{\alpha}$ with their concomitant pole structure. Hence, following standard practice, analytic regularization can be performed by suitable subtraction of pole terms in the Laurent expansion around $\Bold{\alpha}=1$. (see e.g. \cite{SP,LE} and \S~\ref{renormalization} below)

\subsubsection{positive-powers and scattering}
Previously we calculated the momentum-to-momentum realization of the \emph{negative-power} operator $(\square+m^2)_\Gamma^{-\alpha}$ and saw that it represents a scattering propagator at $\alpha=1$. Although not true in general, in an eigenbasis of the operator we get the same result for the \emph{positive power} of the inverse operator;
\begin{eqnarray}\label{positive-power scattering}
\langle{p_b}|\left([\square+m^2]^{-1}\right)_{\mathrm{sH},\beta}^\alpha|{p_a}\rangle
&=&
\int_{\R^\times}\left(1-(p_b^2-m^2)\tau\right)^{-\beta}\tau^\alpha\;d\log (\tau)\notag\\
&=&e^{i\pi(\alpha-1)}{\mathrm{B}(\alpha,\beta-\alpha)}(p_b^2-m^2)^{-\alpha}\,,\;\;\;\;\Re(\beta)\geq1
\end{eqnarray}
for $|p_b^2|> m^2$. Putting $\alpha=1$ and $\beta=2$ gives a massive-particle-Green's-function interpretation of a positive-power operator $A_{\mathrm{sH},\beta}^\alpha\:=\left([\square+m^2]^{-1}\right)_{\mathrm{sH},\beta}^\alpha
\in\mathit{\Pi}^{(\Bold{\alpha})}(\mathbf{F}_{\mathbb{S}_{\mathcal{R}}}(G^\C))$. The same can be done for $(\mathrm{Q}_\mathrm{G}+m^2)_\Gamma^{-\bold{\alpha}}$ and $n$-point scattering. 

This hints that suitable positive powers of inverse operators can codify effective actions in QFT. Interestingly, in \S\ref{positive powers} we saw that positive-power operators of the same form as (\ref{positive-power scattering}) but with $0<\Re(\alpha)<\Re(\beta)<1$ exhibit a structural similarity to $4$-point tree level tachyon string scattering. There is more overlap with string scattering.

 For example, the $4$-point amplitudes for three tachyons and one higher-spin massive string contain an integral of the schematic form\cite[pg.~4]{LLY}
\begin{eqnarray}
F_D^{(K)}(a:b_1,\ldots,b_K:c:x_1,\ldots,x_K)
\notag\\
&&\hspace{-2in}=\frac{\Gamma(c)}{\Gamma(a)\Gamma(c-a)}
\int_0^1(1-t)^{c-a-1}(1-x_1t)^{-b_1},\cdots(1-x_Kt)^{-b_K}t^{a-1}\;dt\;.
\end{eqnarray}
This has the same form as a positive-power operator when $(\mathrm{Id}-\mathrm{A}(g))^{-\beta}\stackrel{\lambda}{\longrightarrow}\prod_i(1-A_{ii}g)^{-\beta_i}$ and the normalized Haar measure is $\nu(g_\mathrm{B})=\log g/[(-1)^{\alpha-1}\mathrm{B}(a,c-a)]$.

Another source of overlap comes from the ``stringy canonical forms'' of \cite{AHL};
\begin{equation}\label{stringy form}
\mathcal{I}_{\{p\}}(\mathbf{X},\{c\})=(\alpha')^d
\int_{\R^d_+}\prod_Jp_J(\mathbf{x})^{-\alpha'c_I}\prod_{i=1}^d x_i^{\alpha'X_i}\;d\log x_i
\end{equation}
where $p_J(\mathbf{x})$ are Laurent polynomials. Hiding in (\ref{stringy form}) are $n$-point tree-level open string scattering amplitudes if $\prod_Jp_J(\mathbf{x})^{-\alpha'c_I}=\prod_{1\leq i
<j\leq n-2}p_{i,j}(\mathbf{x})^{\alpha's_{i,j}}$ where $n\geq4$ and three of the $x_i$ are fixed so that $d=n-3$.\cite[pg.~3]{AHL} Conjecturally, given a non-abelian group $G^\C$ of rank $n-3$, this integral could correspond to the localization of functional Mellin onto a Borel subgroup $Bor<G^\C$. Presumably, the $(n-3)(n-2)/2$ degree polynomials $\prod_{1\leq i
<j\leq n-2}p_{i,j}(\mathbf{x})^{\alpha's_{i,j}}$ would come from an operator $A\in Bor$ in the adjoint representation with $
\mathrm{Det}\,(\mathrm{Id}-\mathrm{A}(g))^{-\beta}\stackrel{\lambda}{\longrightarrow}\prod_{i\leq j}(1-Ad(A)g)^{-\beta_i}$ and $g\in\R_+^{n-3}$ parametrizing a maximal torus in $Bor$.

{If} complex-power operators, both negative and positive, underlie scattering processes this raises a question: Why does the parameter range $1\leq\alpha<\Re(\beta)$ correspond to particle scattering while $0<\Re(\alpha)<\Re(\beta)<1$ corresponds to string scattering? Of course there are more questions that need to be asked and answered to give credence to the notion that functional determinants of complex-power operators codify scattering processes. The next subsection takes some first steps by investigating the role of functional Mellin in renormalization.

\subsection{Renormalization}\label{renormalization}
Section \ref{scattering} taught two important lessons: 1) Feynman diagrams that are usually associated with QFT can be formulated and interpreted as Mellin scattering. 2) In the context of Mellin scattering, the $\tau$ parameter represents a (Lorentz scalar) time-interval as opposed to an instant of time $t$. The fact that $\tau$ is an element of an abelian group with a scale-invariant Haar measure suggests a connection with renormalization; to which we now turn.

There are two overarching aspects of renormalization; regularization and scaling. In the context of QFT, one approach to regularization goes  under the moniker analytic regularization which is closely related to dimensional regularization. Analytic regularization was applied to amplitudes of the form (\ref{parametric Feynman}) (and more generally) by \cite{SP} to define renormalized scattering amplitudes in QFT. The Mellin-type integrals that implement said analytic/dimensional regularization arise from a mathematical device to re-express divergences--- introduced into QFT by coincident interaction points in spacetime (or infinite momenta) --- as poles of a meromorphic function. By contrast, in the context of functional Mellin, amplitudes in the perturbative regime are described directly by (\ref{parametric Feynman}) which is the expectation  in a momentum-state eigenbasis of the operator $(\mathrm{Q_{\mathrm{G}}}+\mathrm{M}^2)^{-\Bold{\alpha}}_\Gamma$ localized to an abelian group of time-interval parameters along some graph $\mathrm{G}$. In both cases, constant terms in Laurent expansions constitute regularized Feynman amplitudes. That the two approaches lead to the same result follows from the happenstance that the transition amplitudes associated with scattering processes in the perturbative regime can be formulated either as (\ref{perturbation series}) or (\ref{covariance}).

The scaling aspect of renormalization in QFT is well understood. For functional Mellin, the scaling aspect stems from the time-interval parameters $\tau_i\in i\R_+$ along internal lines of a graph representing Mellin scattering. By the term Mellin scattering, we mean the operator $\mathcal{M}_\lambda[\mathrm{E}^{-\mathrm{H}_{\mathrm{I}}};\alpha]$ where $\mathrm{H}_{\mathrm{I}}=\mathrm{H}-\mathrm{H}_{0}$ is the interaction Hamiltonian responsible for non-trivial system evolution. If $\mathfrak{h}$ is an element in the Lie algebra of $G^\C$, then Mellin scattering is a probe of the one-parameter subgroup $\phi_{\mathfrak{h}}(i\R)$, and  $\mathcal{M}_\lambda[\mathrm{E}^{-\mathrm{H}_{\mathrm{I}}};\alpha]$, with $\alpha\in\R\cap\mathbb{S}$, characterizes the resolvent and spectrum of the observable $\mathrm{H}\in \phi_{\mathfrak{h}}(i\R)$ with respect to eigenstates of $\mathrm{H}_{0}$. To get quantitative information, $\lambda$ must be specified. As in the preceding negative power examples, choose $\lambda_{i\R_\pm}:\phi_{\mathfrak{h}}(\R)\rightarrow i\R_+\cup i\R_-$ and Haar measure $d\log{\tau}_\Gamma$.

If we want to consider expectations of $\mathcal{M}_{\Gamma}[\mathrm{E}^{-\mathrm{H}_{\mathrm{I}}};\alpha]$ relative to a basis of suitable eigenstates of $\mathrm{H}_{0}$, we are implicitly assuming the eigenstate basis is adiabatic with respect to a necessarily time-interval-dependent operator
$\mathrm{E}^{-\mathrm{H}_{\mathrm{I}}}(g)\stackrel{\lambda_{i\R_\pm}}{\longrightarrow}e^{-H_{I}(\tau)}
:=\mathcal{T}e^{-\langle \tau,H_I\rangle}$ where $\mathcal{T}$ denotes time ordering and $H_{I}(\tau)=\int_\tau H_I(t)dt=:\langle \tau,H_I\rangle$
is the pairing of the oriented $1$-chain $\tau$ and the $1$-form $H_I$ on $\R$ with values in $\mathfrak{C}^\ast$. That is, in $\hbar=1$ units, adiabaticity requires $|\tau|\Delta{H_{I}(\tau)}\slashed{\ll} 1$ where $\Delta{H_{I}(\tau)}$ is the root mean square deviation averaged over $\tau$ relative to the eigenstate basis: otherwise there will be no \emph{non-trivial} scattering. For QFT this means we assume the field characterization in terms of Lorentz varying free-particle degrees of freedom (with respect to $\mathrm{H}_{0}$) remains valid for any interaction time interval $|\tau|\in(0,\infty)$. In other words, in general the $\tau$-dependent amplitude $ \langle\Phi_f|U(\tau)|\Phi_i\rangle\equiv \langle\Phi_f|e^{-H_{I}(\tau)}|\Phi_i\rangle$, which depends on asymptotic free-particle states,
is not well defined for all $\tau$ if $H_{I}(\tau)$ acts diabatically on $|\Phi_i\rangle$; that is if $|\tau|\Delta{H_{I}(\tau)}\ll1$.

The point is, the $S$-matrix is determined by $\langle\Phi_f|\lim_{\tau\rightarrow\infty}U(\tau)|\Phi_i\rangle$ which is no longer meaningful for a diabatic process in the sense that a trivial transition amplitude can result from a non-trivial Hamiltonian. Evidently then, the Mellin integral that describes scattering is strictly valid for time intervals $\tau>\check{\tau}$ where $\check{\tau}$ is some minimum characteristic time interval, say $|\check{\tau}|\sim1/\Delta{H_{I}(\tau)}$. This means we are restricting to interactions such that $\Delta{H_{I}(\tau)}\gtrsim1/|{\tau}|$ for $\tau\in[\check{\tau},i\infty)$ which effectively acts as a UV cutoff. Now, transition rates go like $|\langle\Phi_f|\lim_{\tau\rightarrow\infty}U(\tau)|\Phi_i\rangle|^2$ and $\mathcal{M}_{\Gamma}[\mathrm{E}^{-\mathrm{H}_{\mathrm{I}}};\alpha]\rightarrow (\mathcal{M}_{\Gamma}[\mathrm{E}^{-\mathrm{H}_{\mathrm{I}}};\alpha])^\dag$ under $\tau\rightarrow \tau^\dag=\tau^{-1}$. Hence, as long as we are only concerned with transition rates, a UV cutoff also acts as an IR cutoff, and we should restrict the Mellin integral to say $\hat{\tau}\sim\check{\tau}^{-1}$ so long as $\Delta{H_{I}(\tau)}\gtrsim1/|\tau|$ for $\tau\in[\check{\tau},i]\cup[i,\hat{\tau}]$. Now, since the Haar measure associated with $\tau$ is scale invariant, we need this to hold for all intervals $[\check{\tau},\hat{\tau}]\subseteq (0,i\infty)$ and their associated $\Delta{H_{I}(\tau)}$.

According to the above discussion and following the ideas of \cite{BDH, BBS} as outlined and presented in \cite{CA/D-W3}, the integral $\int_0^{i\infty} U(\tau)\tau^\alpha\;d\log \tau$, which represents the \emph{forward} time-evolution operator for some Lorentz inertial observer\footnote{Note that the full time-evolution operator (forward and backward) $\int_{-i\infty}^{i\infty} U(\tau)\tau^\alpha\;d\log \tau$ is invariant under evolution-time rescaling \emph{and} reversal so it is valid for all Lorentz inertial observers.}, can be decomposed as
\begin{eqnarray}
\mathcal{M}_{\Gamma}[\mathrm{E}^{-\mathrm{H}_{\mathrm{I}}};\alpha]
=\int_0^{i\infty} U(\tau)\tau^\alpha\;d\log \tau
&=&\sum_{j=-\infty}^\infty\int_{\mathrm{s}^j\check{\tau}}^{\mathrm{s}^{j+1}\check{\tau}}U(\tau)\tau^\alpha\;d\log \tau
\notag\\
&=:&\sum_{j=-\infty}^\infty
\mathcal{M}_{\Gamma}[\mathrm{E}^{-\mathrm{H}_{\mathrm{I}}};\alpha,\mathrm{s}^j\check{\tau}]\;.
\end{eqnarray}
where $\mathrm{s}\in[1,\infty)$ and $\mathrm{E}^{-\mathrm{H}_{\mathrm{I}}}(g)
\stackrel{\lambda_{i\R_+}}{\longrightarrow} U(\tau)=\mathcal{T}e^{-\int_{\tau} H_I(t)dt}$. This can be viewed as an adiabatic decomposition of the scattering operator into scale-dependent components; akin to a decomposition into frequency components.
Crucially, under a re-scaling $\tau\rightarrow \sigma\tau$ with $\sigma\in\R_+$,
\begin{equation}\label{adiabatic decomposition}
\int_{\mathrm{s}^j\check{\tau}}^{\mathrm{s}^{j+1}\check{\tau}}U(\tau)\tau^\alpha\;d\log \tau
=\sigma^\alpha\int_{\frac{\mathrm{s}^j}{\sigma}\check{\tau}}^{\frac{\mathrm{s}^{j+1}}{\sigma}\check{\tau}}U(\sigma\tau)\tau^\alpha\;d\log \tau\;.
\end{equation}
If we define a scaling operator
\begin{eqnarray}
P_\sigma:\mathbf{F}_{\mathbb{S}_{\mathcal{R}}}(G^\C)
&\rightarrow&\mathbf{F}_{\mathbb{S}_{\mathcal{R}}}(G^\C)\notag\\
(P_\sigma\mathrm{F})(g)&\mapsto& \mathrm{F}(\sigma g)\,,
\end{eqnarray}
equation (\ref{adiabatic decomposition}) can be written as
\begin{equation}
\mathcal{M}_{\Gamma}[\mathrm{E}^{-\mathrm{H}_{\mathrm{I}}};\alpha,\mathrm{s}^j\check{\tau}]
=\sigma^\alpha\mathcal{M}_{\Gamma}[P_\sigma\mathrm{E}^{-\mathrm{H}_{\mathrm{I}}};\alpha,\mathrm{s}^j\check{\tau}/\sigma]\;.
\end{equation}
Accordingly, choosing $\sigma=\mathrm{s}^j$ for each $j$, the forward scattering operator $\mathcal{M}_{\Gamma}[\mathrm{E}^{-\mathrm{H}_{\mathrm{I}}};\alpha]$ has a simple decomposition
\begin{eqnarray}
\mathcal{M}_{\Gamma}[\mathrm{E}^{-\mathrm{H}_{\mathrm{I}}};\alpha]
&=&\sum_{j=-\infty}^\infty
(\mathrm{s}^j)^\alpha\mathcal{M}_{\Gamma}[P_{\mathrm{s}^j}\mathrm{E}^{-\mathrm{H}_{\mathrm{I}}};\alpha,\check{\tau}]
\end{eqnarray}
where we define a characteristic, scale-dependent scattering operator for each $j$
\begin{equation}
\mathcal{M}_{\Gamma}[P_{\mathrm{s}^j}\mathrm{E}^{-\mathrm{H}_{\mathrm{I}}};\alpha,\check{\tau}]
:=\int_{\check{\tau}}^{\mathrm{s}\check{\tau}}U(\mathrm{s}^j\tau)\tau^\alpha\;d\log\tau\;.
\end{equation}

Thanks to the adiabatic decomposition, one can make sense of $\mathcal{M}_{\Gamma}[\langle\Phi_f|\mathrm{E}^{-\mathrm{H}_{\mathrm{I}}}|\Phi_i\rangle;\alpha]$ by choosing stationary momentum eigenstates $|\Bold{p}_{\check{\tau}}\rangle$ at the characteristic time-interval so that
\begin{equation}
H_{I}(\tau)|\Bold{p}_{\check{\tau}}\rangle=
\left\{\begin{array}{cc}
p(\tau)|\Bold{p}_{\check{\tau}}\rangle & \tau\in[\check{\tau},\mathrm{s}\check{\tau}] \\
0 & \tau\in(0,\check{\tau})
\end{array}\right.\;.
\end{equation}
Presumably, the characteristic eigenstates $|\Bold{p}_{\check{\tau}}\rangle$ are a superposition
\begin{equation}
|\Bold{p}_{\check{\tau}}\rangle:=\left.\int_0^{\check{\tau}}U(\tau)|\Bold{p}_{0}\rangle\tau^\alpha\;d\log\tau\right|_{\alpha=1}
=\int_0^{\check{\tau}}U(\tau)|\Bold{p}_{0}\rangle\;d\tau
\end{equation}
where $|\Bold{p}_{0}\rangle$ represents (\emph{a priori} unknown) primary degrees of freedom including any relevant elementary particles and bound states associated with the spectrum of $\mathrm{H}$. The interpretation of $|\Bold{p}_{\check{\tau}}\rangle$ as ``effective'' degrees of freedom is evident.

The regularized $n$-point\footnote{Off hand, it appears one can exclude tadpole contributions by expanding $\mathrm{E}^{-\mathrm{H}_{\mathrm{I}}}$ in terms of Laguerre integrators as described in Appendix \ref{ast and star}. This follows by noting that the renormalization program can be carried out instead in the context of (\ref{perturbation series}) which is the standard approach. In this case, the Laguerre expansion of the quadratic $\mathrm{Q}(\Phi)$ would yield Hermite polynomials which encode normal ordering.} Mellin scattering amplitude for non-trivial scattering is the finite part of
\begin{equation}\label{renormalized amplitude}
\lim_{\Bold{\alpha}\rightarrow 1}\mathcal{M}_{\Gamma}[\langle\Bold{p}_{\check{\Bold{\tau}}}|
\mathrm{E}^{-\mathrm{H}_{\mathrm{I}}}|\Bold{p}_{\check{\Bold{\tau}}}\rangle;\Bold{\alpha}]
=\lim_{\Bold{\alpha}\rightarrow 1}\sum_{j=-\infty}^\infty
(\mathrm{s}^j)^\Bold{\alpha}\mathcal{M}_{\Gamma}
[\langle\Bold{p}_{\check{\Bold{\tau}}}| P_{\mathrm{s}^j}\mathrm{E}^{-\mathrm{H}_{\mathrm{I}}}
|\Bold{p}_{\check{\Bold{\tau}}}\rangle;\Bold{\alpha},\check{\Bold{\tau}}]
\end{equation}
which is approximated in the perturbative regime through analytic regularization of covariances $\mathcal{M}_{\Gamma}[\langle\Bold{p}_{\check{\Bold{\tau}}}
|\mathrm{E}^{-(\mathrm{Q_{\mathrm{G}}}+\mathrm{M}^2)}|\Bold{p}_{\check{\Bold{\tau}}}\rangle;\Bold{\alpha}]$ associated with relevant graphs $\mathrm{G}$ at each order of perturbation as presented in \S\ref{scattering}. The physically relevant finite part of such graphs can be obtained via the `analytic renormalization' scheme of \cite[ch. III]{SP}.

Borrowing from QFT, intuition regarding renormalized bare parameters suggests adding counter terms to $\langle\Bold{p}_{\check{\Bold{\tau}}}
|\mathrm{E}^{-(\mathrm{Q_{\mathrm{G}}}+\mathrm{M}^2)}|\Bold{p}_{\check{\Bold{\tau}}}\rangle$ that cancel pole terms in (\ref{renormalized amplitude}). But this interpretation is not required in Mellin scattering: We can just as well define all physical parameters that characterize $H_{I}(\Bold{\tau})$ to be determined by the {finite} part of (\ref{renormalized amplitude}) along with a suitable choice of Haar measure $\mathcal{M}_\Gamma\rightarrow\mathcal{M}_{\Gamma_{ren}}$ (corresponding to parameter measurements); with each order of perturbation giving a progressively better approximation to their values.

The above prescription is not quite complete because the physical parameters calculated by the finite part of the right-hand side of (\ref{renormalized amplitude}) depend on the choice of $\mathrm{s}$ directly through $P_{\mathrm{s}^j}\mathrm{E}^{-\mathrm{H}_{\mathrm{I}}}$ and indirectly through  stationary momentum eigenstates $|\Bold{p}_{\check{\Bold{\tau}}}\rangle$.\footnote{The physical parameters also depend on $\check{\tau}$, but this just reflects a choice of time-interval unit which can be chosen so that $|\check{\tau}|=1$ (in units of $\hbar$). This, of course, is the identity element of $\R_+$ and the fixed point of the function $|\tau|\mapsto|\tau^{-1}|$; hence a natural choice of time-interval unit --- but still a choice.} Since $\mathrm{s}$ is a matter of choice (or observer), a satisfactory scattering theory must describe how changing the scale $\mathrm{s}\rightarrow\mathrm{s}'$ will affect the physical parameters determined by the renormalized characteristic scattering operator $\mathcal{M}_{\Gamma_{ren}}[P_{\mathrm{s}^j}\mathrm{E}^{-\mathrm{H}_{\mathrm{I}}};\Bold{\alpha},\check{\Bold{\tau}}]$  assuming $|\Bold{p}_{\check{\Bold{\tau}}}\rangle$ remains adiabatic for $\Bold{\tau}\in[\check{\Bold{\tau}},\mathrm{s}'\check{\Bold{\tau}}]$.

To describe the effect of $\mathrm{s}\rightarrow\mathrm{s}'$, restrict attention to the massive scalar case with one coupling constant $\mathrm{g}$ for simplicity. The left-hand side of (\ref{renormalized amplitude}) does not depend on $\mathrm{s}$, but the right-hand side depends explicitly on $\mathrm{s}^j$ and implicitly on $\mathrm{s}^j$ through $P_{\mathrm{s}^j}\mathrm{E}^{-\mathrm{H}_{\mathrm{I}}}$ which induces $\mathrm{s}$-dependent parameters $\mathrm{g}(\mathrm{s})$ and $m(\mathrm{s})$. Assuming $|\Bold{p}_{\check{\Bold{\tau}}}\rangle$ remains adiabatic, scale independence of the scattering amplitude yields an exact (regularized) RGE
\begin{equation}
0=\lim_{\Bold{\alpha}\rightarrow 1}\mathrm{s}\frac{d}{d\mathrm{s}}\langle\Bold{p}_{\check{\Bold{\tau}}}|H_I^{-\Bold{\alpha}}|
\Bold{p}_{\check{\Bold{\Bold{\tau}}}}\rangle
=\lim_{\Bold{\alpha}\rightarrow 1}\sum_{j=-\infty}^\infty j\,\mathrm{s}^j\frac{d}{d\mathrm{s}}
\left\{\mathcal{M}_{\Gamma_{ren}}[\langle\Bold{p}_{\check{\Bold{\tau}}}|
\mathrm{E}^{-\mathrm{H}_{\mathrm{I}}}|\Bold{p}_{\check{\Bold{\Bold{\tau}}}}\rangle;
\Bold{\alpha},\mathrm{s}^j\check{\Bold{\tau}}]
\right\}\;.
\end{equation}

In the perturbative regime we can go further. For a given on-shell Feynman diagram with cutoff $\check{\Bold{\Bold{\tau}}}$, (\ref{parametric Feynman}) becomes
\begin{eqnarray}\label{Feynman diagram cutoff}\mathcal{M}_{\Gamma_{ren}}[\langle\Bold{p}_{\check{\Bold{\tau}}}|
\mathrm{E}^{-(\mathrm{Q_{\mathrm{G}}}+\mathrm{M}^2)}|
\Bold{p}_{\check{\Bold{\tau}}}\rangle;\Bold{\alpha}]
=\sum_{j=-\infty}^\infty
\mathcal{M}_{\Gamma_{ren}}\langle\Bold{p}_{\check{\Bold{\tau}}}|
\mathrm{E}^{-(\mathrm{Q_{\mathrm{G}}}+\mathrm{M}^2)}|
\Bold{p}_{\check{\Bold{\Bold{\tau}}}}\rangle;\Bold{\alpha},\mathrm{s}^j\check{\Bold{\Bold{\tau}}}]\;.
\end{eqnarray}
 Up to momentum conservation and normalization, this is a polynomial in $\mathrm{g}(s)$ and $m(s)$. Scale independence implies
\begin{equation}
0=\lim_{\Bold{\alpha}\rightarrow 1}\mathrm{s}\frac{d}{d\mathrm{s}}
\langle\Bold{p}_{\check{\Bold{\tau}}}|({Q_{\mathrm{G}}}+{M}^2)^{-\Bold{\alpha}}|
\Bold{p}_{\check{\Bold{\Bold{\tau}}}}\rangle
=\lim_{\Bold{\alpha}\rightarrow 1}\sum_{j=-\infty}^\infty j\,\mathrm{s}^j\frac{d}{d\mathrm{s}}
\left\{\mathcal{M}_{\Gamma_{ren}}
[\langle\Bold{p}_{\check{\Bold{\tau}}}| \mathrm{E}^{-(\mathrm{Q_{\mathrm{G}}}+\mathrm{M}^2)}
|\Bold{p}_{\check{\Bold{\tau}}}\rangle;\Bold{\alpha},\mathrm{s}^j\check{\Bold{\tau}}]\right\}\;.
\end{equation}
As the scaled covariance terms in the sum are self-similar by (\ref{adiabatic decomposition}), we get a perturbative RGE for the characteristic Feynman amplitude at any scale
\begin{equation}
\lim_{\Bold{\alpha}\rightarrow 1}\mathrm{s}\frac{d}{d\mathrm{s}}
\left\{(\mathrm{s})^\Bold{\alpha}\mathcal{M}_{\Gamma_{ren}}
[\langle\Bold{p}_{\check{\Bold{\tau}}}|
P_{\mathrm{s}}\mathrm{E}^{-(\mathrm{Q_{\mathrm{G}}}+\mathrm{M}^2)}|
\Bold{p}_{\check{\Bold{\Bold{\tau}}}}\rangle
;\Bold{\alpha},\check{\Bold{\tau}}]\right\}=0\;.
\end{equation}

\begin{remark}
Our treatment of the scattering observable $\mathrm{E}^{-\mathrm{H}_{\mathrm{I}}}$ and the subsequent renormalization of its associated scattering operator $H_I^{-\alpha}$ depends on our localization of the multiplicative abelian one-parameter subgroup $\phi_{\mathfrak{h}}(i\R)$ to the multiplicative abelian group $i\R_+\cup i\R_-$ and the localized functional $\mathrm{E}^{-\mathrm{H}_{\mathrm{I}}}(g)
\stackrel{\lambda_{i\R_\pm}}{\longrightarrow} U(\tau)=\mathcal{T}e^{-\int_{\tau} H_I(t)dt}$. We could instead follow the standard renormalization story by localizing the additive abelian group $\Bold{\Phi}$ of complex scalar fields to the additive abelian group $\C$ as in \emph{$\S$6.1} and put  $\mathrm{E}^{-\mathrm{H}_{\mathrm{I}}}(\Phi)
\stackrel{\lambda_{vac}}{\longrightarrow} e^{-\int_{\R^{3,1}}\mathrm{S}_{\mathrm{I}}(\Phi(x))\;dx}$. The two approaches are just different ways to probe the abstract observable $\mathrm{E}^{-\mathrm{H}_{\mathrm{I}}}$.\footnote{However, the two approaches have very different underlying physical interpretations regarding a cutoff: The Wilsonian approach requires a \emph{minimum} resolution of space-time while Functional Mellin requires a \emph{bounded} root-mean-square deviation of $\mathrm{H}_{\mathrm{I}}$. The latter implies bounded quantum fluctuations and does not require the auxiliary notion of space-time.} There are likely other useful probes: in particular we have in mind localizing the complex one-parameter subgroup $\phi_{\mathfrak{h}}(\C)$ to $\C^\times$ with $\mathrm{E}^{-\mathrm{H}_{\mathrm{I}}}(g)
\stackrel{\lambda_{\C^\times}}{\longrightarrow} U(\sigma)=\mathcal{P}e^{-\int_{\sigma}{H}_{{I}}(z)\;dz}=\mathcal{P}e^{-\langle \sigma, H_I\rangle}$ where $\sigma$ is an oriented $1(or\,2)$-chain in $\C^\times$. Both orientations are a priori allowed  and paired with a holomorphic, operator-valued $1(or\,2)$-form $H_I$. Remark that $\langle \sigma,H_I\rangle$ is linear in $\sigma$, and the set of $1(or\,2)$-chains is a vector space over $\R$ with an underlying abelian group structure.
\end{remark}

\subsection{Density operators, entropy, and the replica trick}\label{replica trick}
Since functional Mellin deals in complex groups, quantum system evolution encoded in $\mathrm{E}^{-\mathrm{H}}$ is perhaps more naturally represented and probed by complex one-parameter subgroups localized to $\C^\times$. We have already considered free-particle evolution by localizing a real one-parameter subgroup to $i\R_+\cup i\R_-$. To see what the orthogonal (real) direction of a complex evolution-time interval $\tau\in\C^\times$ characterizes, fix its imaginary component to $\Im(\tau)=0$. This restricts $\tau$ to two copies of the positive reals $\R_\pm\equiv\R_+\cup\R_-$ (of opposite orientation) which we interpret as parametrizing properties of \emph{fixed-energy} processes.

Consider the complex one-parameter subgroup $\phi_{\mathfrak{h}}(\C)$ with its related Hamiltonian $\mathrm{H}$ such that $\mathrm{H}(g)\in L_B(\mathcal{H})$. As we are restricting to the positive reals, we consider topological localization of $\mathrm{H}(g)$ to an operator-valued, de Rham homology/cohomology pairing $\langle r,H\rangle$ in $H_1(\R_\pm,\C^\times)\times H^1(\R_\pm,\C^\times)\otimes\mathfrak{C}^\ast$. The associated negative power operator is
\begin{equation}\
H^{-\alpha}_{{\R_\pm}}
=\mathcal{M}_{{\R_\pm}}[\mathrm{E}^{-\mathrm{H}};\alpha]
=\int_{\R_\pm}e^{-\langle r,H\rangle}r^{\alpha}d\log{r}
\;\;\;\;\;\;\alpha\in\R\cap\mathbb{S}\;.
\end{equation}
Also associated with $\mathrm{H}$ is an observable $\mathrm{Tr}\,\mathrm{E}^{-\mathrm{H}}$ and its zeta functional
\begin{equation}\label{H-zeta}
 \mathrm{Tr}\,H^{-\alpha}_{{\R_\pm}}
 =\mathcal{M}_{{\R_\pm}}[\mathrm{Tr}\,\mathrm{E}^{-\mathrm{H}};\alpha]
 =\int_{\R_\pm}\mathrm{tr}\left(e^{-\langle r,H\rangle}r^\alpha\right)d\log r
 =\zeta_{H}(\alpha)
\end{equation}
where $\mathrm{tr}(e^{-\langle r,H\rangle})$ is the partition function of the operator $H$.

It is convenient to put $r\rightarrow r^{-1}$ in the integrals and define
\begin{equation}
 \varrho^{-\alpha}_{{\R_\pm}}
 :=\frac{\int_{\R_\pm}e^{-\langle r^{-1},H\rangle}r^{-\alpha}d\log{r}}
 {\mathrm{Tr}\,H^{-1}_{{\R_\pm}}}\;.
\end{equation}
In an $H$-eigensystem $\{|i\rangle,\varepsilon_i\}$ with dimensionless $\varepsilon_i:=E_i/k_B\mathrm{T}_0$ for $E_i=\langle i|H|i\rangle$ and $\mathrm{T}_0$ some reference temperature, we have $\langle i|\varrho^{-\alpha}_{\lambda_{\R_\pm}}|i\rangle
=\varepsilon_i^{-\alpha}/\sum_i\varepsilon_i^{-1}$. Moreover, the integration variable can be associated with (dimensionless) energy/temperature and $\mathrm{Tr}\,\varrho^{-1}_{{\R_\pm}}=1$. Therefore we regard $\varrho^{-1}_{{\R_\pm}}$ as an inverse density operator with von-Neumann entropy
\begin{eqnarray}
\mathrm{tr}\,\mathrm{Log}\varrho^{-(1+1)}_{{\R_\pm}}
=\mathrm{tr}
\left.\frac{d}{d\alpha}\varrho^{-\alpha}_{{\R_\pm}}\right|_{\alpha=1}
&=&\frac{\widehat{\mathcal{M}}_{{\R_\pm}}[\mathrm{Tr}\,\mathrm{E}^{-\mathrm{H}};1]}
{\zeta_{H}(1)}\notag\\
&=&\frac{\zeta'_{H}(1)}{\zeta_{H}(1)}\notag\\
&=&\left.\frac{d}{d\alpha}\log\zeta_{H}(\alpha)\right|_{\alpha=1}\notag\\
&=&\mathrm{tr}\,(\varrho^{-1}_{{\R_\pm}}\log\varrho^{-1}_{{\R_\pm}})\;.
\end{eqnarray}

Alternatively, for invertible ${H}$, the associated density operator is the positive-power counterpart of $\varrho^{-1}_{{\R_\pm}}$ defined with the help of (\ref{contour integral})
\begin{equation}
\varrho_{\mathcal{C}_\mathrm{B}}^{\alpha}
=\frac{{\mathcal{M}}_{\mathcal{C}_\mathrm{B}}[(\mathrm{Id}-\mathrm{H})_{(\beta)};\alpha]}
{{\mathcal{M}}_{\mathcal{C}_\mathrm{B}}[\mathrm{Tr}\,(\mathrm{Id}-\mathrm{H})_{(\beta)};1]}
=\frac{\int_{\mathcal{C}}\, (\theta-H)^{-\beta}\,\theta^{\beta-\alpha}\;d\nu(\theta_{\mathcal{C}_{\mathrm{B}}})}
{\mathrm{Tr}\,(H^{-1})_{\mathcal{C}_\mathrm{B}}^{1}}\;.
\end{equation}
Here the integration variable is simply a parameter along a contour $\mathcal{C}\subset\C^\times$ enclosing the spectrum $\sigma(H)$.
From Definition \ref{positive TLD} with $\alpha=1$ and $\beta=2$, its von-Neumann entropy is
\begin{eqnarray}
\mathrm{tr}\,\mathrm{Log}\varrho_{\mathcal{C}_\mathrm{B}}^{1+1}
=\mathrm{tr}
\left.\frac{d}{d\alpha}\varrho_{\mathcal{C}_\mathrm{B}}^{\alpha}\right|_{\alpha=1}
&=&\frac{\mathrm{tr}\,\widehat{\mathcal{M}}_{\mathcal{C}_\mathrm{B}}[(\mathrm{Id}-\mathrm{H})_{(\beta)};1]}
{\mathrm{tr}\,{H}_{\mathcal{C}_\mathrm{B}}^{1}}\notag\\
&=&\frac{\widehat{\mathcal{M}}_{\mathcal{C}_\mathrm{B}}[\mathrm{Tr}\,(\mathrm{Id}-\mathrm{H})_{(\beta)};1]}
{{\mathcal{M}}_{\mathcal{C}_\mathrm{B}}[\mathrm{Tr}\,(\mathrm{Id}-\mathrm{H})_{(\beta)};1]}\notag\\
&=&\left.\frac{d}{d\alpha}
\log\,\mathrm{tr}\,(H^{-1})^{\alpha}
\right|_{\alpha=1}\notag\\
&=&\mathrm{tr}\,(\varrho_{\mathcal{C}_\mathrm{B}}\log\varrho_{\mathcal{C}_\mathrm{B}})\;.
\end{eqnarray}

Clearly $\varrho^{-1}$ (which is the Mellin transform of the partition function of $H$) and $\varrho$ (which is the Mellin transform of the resolvent of $H$) give the same von-Neumann entropy since they derive from normalized eigenvalues of $H$.\footnote{There is a close parallel between the von-Neumann entropy and $\log\zeta(\alpha)$ where $\zeta(\alpha)$ is the Riemann zeta function. The log of Riemann zeta acts like a spectral measure on $(\mathbb{Z}/\x\mathbb{Z})^\times$ when counting prime powers up to some cut-off integer $\x$, and it can be generalized to prime $k$-tuples.(see \cite[App. 4.8]{LA1}) It would be interesting to apply the (essentially) functional Mellin techniques developed in \cite{LA1} to count eigenvalues and $k$-tuple/$k$-correlated eigenvalues of $H$ using the ``spectral measure'' $\log\zeta_{H}(\alpha)$.}

To obtain the representation of the \emph{standard} density operator $\hat{\rho}$ in terms of the partition function, apply $\mathrm{H}\rightarrow\mathrm{E}^{\mathrm{H}}$ and $\alpha\rightarrow \mathrm{T}_0/\mathrm{T}$ to the positive power
\begin{equation}
(\mathrm{E}^{-\mathrm{H}})^{\alpha}_{\mathcal{C}_\mathrm{B}}
:={\mathcal{M}}_{\mathcal{C}_\mathrm{B}}
[(\mathrm{Id}-\mathrm{E}^{\mathrm{H}})_{(\beta)};\alpha]
=(e^{-H})^\alpha
\;\;\;\;\;\;\alpha<\beta\in\R\cap\mathbb{S}\;.
\end{equation}
Use this to define the $\alpha$-dependent partition functional\footnote{This is \emph{not} the usual Euclidean field theory partition \emph{function} $Z(\beta)$ where $\beta\in \R_\pm$ is inverse temperature. We used $Z(\beta)$ with $\beta\equiv r$ as the integrand in (\ref{H-zeta}) to calculate the zeta functional of $H$.}
\begin{equation}
\mathcal{Z}(\alpha):={\mathcal{M}}_{\mathcal{C}_\mathrm{B}}
[\mathrm{Tr}\,(\mathrm{Id}-\mathrm{E}^{\mathrm{H}})_{(\beta)};\alpha]=\mathrm{tr}\,(e^{-H})^\alpha\;,
\end{equation}
density operator
\begin{equation}
\hat{\rho}(\alpha):=(e^{-H})^\alpha/\mathcal{Z}(1)\;,
\end{equation}
and entropy functional
\begin{equation}
S(\alpha):=d\log\mathrm{tr}\,\hat{\rho}(\alpha)/d\alpha
=d\log\mathcal{Z}(\alpha)/d\alpha\;.
\end{equation}
This assigns a nice physical interpretation to the parameter $\alpha$, viz. $\alpha\equiv \mathrm{T}_0/\mathrm{T}$.
\begin{application}
The replica trick is a means to calculate $\log\mathcal{Z}$ and/or $\mathrm{tr}\,(\hat{\rho}\log\hat{\rho})$. It relies on replacing $\hat{\rho}^n$ for $n\in\mathbb{Z}_{\geq0}$ with $\hat{\rho}^r$ for $r\in\R_{\geq0}$ which is hard to rigorously justify. Fortunately, functional Mellin with its built-in $\alpha$-dependence provides automatic justification: From \emph{Definition \ref{positive TLD}}, $\mathrm{Log}\,\mathcal{Z}(0):=d\mathcal{Z}(\alpha)/d\alpha|_{\alpha=0^+}
\equiv\log \mathcal{Z}$ when the limit $\alpha\rightarrow 0^+$ exists and $S(1)=d\log\mathcal{Z}(\alpha)/d\alpha|_{\alpha=1}
=\mathrm{tr}\,(\hat{\rho}(1)\log\hat{\rho}(1))
\equiv\mathrm{tr}\,(\hat{\rho}\log\hat{\rho})$.
\end{application}

\section{Functional Mellin and Number Theory}\label{number theory}
Much like the physical toy model of primes and prime $k$-tuples presented in the
 companion paper \cite{LA1}, this section offers a physics perspective to formulate and understand certain $L$-functions and zeta functions in the context of scalar Mellin scattering studied in the previous section. There, we primarily dealt with negative-power operators and their relationship to scattering processes under the topological localization $\lambda_{i\R_\pm} :\phi_{\mathfrak{h}}(\R)\rightarrow i\R_+\cup i\R_-$. But we observed that it was perhaps more natural to consider complex one-parameter subgroups, and in \S \ref{replica trick} we calculated negative-power operators and some associated functionals for fixed-energy processes with the topological localization $\lambda_{\C^\times} :\phi_{\mathfrak{h}}(\C)\rightarrow \C^\times$ at $\Im(\tau) =0$. We will now relax this constraint and allow $\Im(\tau)\neq0$ but still constant.

For this localization, zeta functionals, which recall are functional traces, are closely related to Dirichlet $L$-functions; the latter of which can therefore be viewed as grand canonical partition functionals with an imaginary chemical potential.\footnote{Imaginary chemical potentials are a useful tool in condensed matter physics.\cite{AKW,DL}} When the Dirichlet series associated with a Dirichlet $L$-function is a sum over elements of a unique factorization domain (UFD) and the Dirichlet character is completely multiplicative, the Dirichlet $L$-function can be alternatively expressed as an Euler product over the prime elements in the UFD. Moreover, various other $L$-functions and even some zeta functions are defined in terms of Euler products (as opposed to traces) and hence are more aligned with determinants. From this vantage there seems to be no reason, other than convention, to maintain a distinction between zeta functions and $L$-functions since they are similarly characterized by traces and/or determinants. For our purposes then, we will use the term $L$-functional to mean both functional trace and functional determinant.\footnote{In a physics context, functional traces encode partition functions and hence statistical properties while functional determinants encode spectral properties of a quantum system. Apparently the two are related if the microstates can be put into correspondence with a UFD and the spectra correspond to prime elements.}

\subsection{Traces}
Consider a functional $\mathrm{F}$ associated with some Hermitian Hamiltonian observable $\mathrm{H}\in \phi_{\mathfrak{h}}(\C)$ and localize by $\lambda_{\C^\times} :\phi_{\mathfrak{h}}(\C)\rightarrow \R_\pm\times i\R_\pm\cong\C^\times$. Write $z=(r,\tau)\in\R_\pm\times i\R$ and choose a functional of the form $\mathrm{F}(\rho(g))\rightarrow e^{-\frac{1}{2}\langle z,H\rangle}f(\langle r,H_r\rangle)=:e^{-\frac{1}{2}\langle \tau,H_\tau\rangle}F(\langle r,H_r\rangle)$ where $\langle \cdot,\cdot\rangle$ denotes a Poincar\'{e} pairing $H_1(\phi_{\mathfrak{h}}(\C),\C)\times (H^1(\phi_{\mathfrak{h}}(\C),\C)\otimes\mathfrak{C}^\ast)$ of $z$ and $H=H_rdr+H_\tau d\tau$. Post localization, this induces a bilinear form $H_1(\C^\times,\C)\times (H_1(\C^\times,\C)\otimes\mathfrak{C}^\ast)\rightarrow\C\otimes\mathfrak{C}^\ast$ also denoted by $\langle\cdot,\cdot\rangle$.

For a fixed-energy process , the inverse complex power of $\mathrm{F}$ is
\begin{eqnarray}\label{fixed-energy process}
 \mathcal{M}_\lambda[\mathrm{F};\alpha]
&\stackrel{\lambda_{\C^\times}}{\rightarrow}&\int_{\C^\times} e^{-\frac{1}{2}\langle z,H\rangle}f(\langle r,H_r\rangle)z^\alpha\delta(\tau-\Delta t)\;d\log z\;\;\;\;\;\;\;\;\alpha\in\mathbb{S}_\lambda\notag\\
&=&\int_{\C^\times} e^{-\frac{1}{2}\langle\tau,H_\tau\rangle}F(\langle r,H_r\rangle)r^\alpha\tau^\alpha\delta(\tau-\Delta t)\;d\log r\,d\log\tau\;\;\;\;\;\;\;\;\alpha\in\mathbb{S}_\lambda\notag\\
&=&\chi_{\Delta t}(H_\tau)\int_{\R_\pm}e^{-\frac{1}{2}\langle r,H_r\rangle}f(\langle r,H_r\rangle){r}^\alpha\;d\log {r}
\end{eqnarray}
where $\chi_{\Delta t}(H_\tau):=e^{-\frac{1}{2}\langle\tau,H_\tau\rangle}$, $|\Delta t|\equiv |t_b-t_a|\geq0$, and the delta functional for the gamma integrator from \cite{LA1} was used which forces $\alpha=1$ in the $\tau$ integral. The interpretation of $\langle \tau,H_\tau\rangle|_{\tau=\Delta t}$ as an imaginary chemical potential is evident. Changing the scale of $\Delta t$ changes the chemical potential and hence the fixed energy of the process, and \S \ref{renormalization} quantifies how the process changes with the scale $\mathrm{s}$.

\begin{example}\label{Dirichlet L-function}
Take $\mathrm{H}\equiv \mathrm{N}$ the counting functional from \emph{Example \ref{zeta example}}, fix $\tau=\Delta t$, and put $f_0=1$, $f_{n>0}=0$. Then $\mathcal{M}_{\R_\pm\times i\Delta t}[\mathrm{Tr}\,\mathrm{F};\alpha]=\sum_{\varepsilon_i\in\mathbb{Z}_+}e^{-\varepsilon_i\Delta t/2}\varepsilon_i^{-\alpha}$. Scale $\Delta t\rightarrow \mathrm{s}\Delta t=:\Delta t_{\mathrm{s}}$ with $\mathrm{s}\in[1,\infty)$
 so that $2\pi/(-i\Delta t_{\mathrm{s}})=1$. Observe the phase $\chi(\varepsilon_i):=e^{-\varepsilon_i\Delta t_{\mathrm{s}}}$ is periodic $\chi(\varepsilon_i)=\chi(\varepsilon_i+P)$ with period $P=2k$ and $k\in\mathbb{Z}$. Consequently, $\chi:\mathbb{Z}/P\mathbb{Z}\rightarrow\C$ is a Dirichlet character that can be extended to a function on the real numbers $\tilde{\chi}:\R\rightarrow\C$ and then Fourier decomposed as $\tilde{\chi}(r)=\frac{1}{\sqrt{P}}\sum_{a=0}^{P-1}\hat{\chi}_ae^{2\pi i \frac{ra}{P}}$ where $\hat{\chi}_a=\frac{1}{\sqrt{P}}\sum_{n=0}^{P-1}{\chi}(n)e^{-2\pi i \frac{an}{P}}$ and $r\in\R$. Evidently, when $H\equiv Ndz= Ndr+Nd\tau$ is the positive integer counting operator, $\mathrm{tr}(e^{-\langle\Delta t_{\mathrm{s}},N\rangle}N^{-\alpha})$  determines a Dirichlet L-function $L(\alpha,\chi)
=\sum_{\varepsilon_i\in\mathbb{Z}_+}\frac{\tilde{\chi}(\varepsilon_i)}{\varepsilon_i^\alpha}$.
\end{example}

By analogy with the preceding counting functional example, we define and examine $L$-functionals for a fixed-energy process induced by a Hamiltonian $\mathrm{H}(g_\lambda)\in L_B(\mathcal{H})$  where Hilbert space $\mathcal{H}$ is a separable Lagrangian subspace (with a Hilbert structure) of a symplectic state-space $Z_a$ assumed to be a vector space for now.
\begin{definition}
Let $Z_a$ be a separable symplectic vector space with $\LL\subset Z_a$ a Lagrangian subspace, and suppose the Banach space of pointed maps $\mathcal{B}\equiv\Gamma((\C^\times,\tau_a),(\LL,l_a))$ furnishes a representation $\rho$ of $G^\C$.
 For functionals $\mathrm{F}\in\mathbf{F}_{\mathbb{S}}(G^\C)$ and localization $\lambda_{\C^\times}:G^\C\rightarrow \C^\times$ such that $\mathrm{F}(g_\lambda)\in L_B(\mathcal{B})$ models a fixed-energy process and is holomorphic, the $L$-functional  is
\begin{equation}
L^+_\lambda(\alpha,F,\phi)
:=\phi^{-1/2}L^+_\lambda(\alpha,F)
:=\phi^{-1/2}\mathcal{M}_\lambda[\mathrm{Tr}\,\mathrm{F};\alpha]
\end{equation}
where $\phi$ is a constant obeying $|\phi|=1$, the trace is taken in the $\rho$ representation, and $\mathcal{M}_\lambda[\mathrm{Tr}\,\mathrm{F};\alpha]$ is required to be meromorphic on $\C$. Remark there will be restrictions on $\mathrm{Tr}\,\mathrm{F}(g)$ to render the Mellin integral and hence the definition well-defined for all of $\C$. The $+$ superscript denotes an $L$-functional associated with a functional trace.
\end{definition}

Recall functional trace only converges for $\alpha\in\mathbb{S}_\lambda$, but an $L$-functional by definition is a meromorphic function on the entire complex plane. Hence, $L^+_\lambda(\alpha,F)$ needs to be analytically continued.

\begin{application}
In the prototypical case where $f(\langle r,H_r\rangle)=1$ and $\langle l|e^{-\frac{1}{2}\langle r,H_r\rangle}|l\rangle$ is quadratic in $l\in\LL$, the continuation is achieved by deriving a functional equation based on Poisson summation and the scaling property of Fourier transform \emph{(see e.g. \cite[\S1.1]{B})}. When $f(\langle r,H_r\rangle)\neq1$, the physical model of a symplectic state-space provides more flexibility: Let $\{\psi_{i}\}$ (which is countable) span a Lagrangian subspace $\LL\subset Z_a$. The cyclically invariant trace  allows a symplectic transformation to a different Lagrangian subspace without changing the Hamiltonian dynamics. In particular, one can transform to an orthogonal (with respect to the symplectic form on $Z_a$) Lagrangian subspace $\hat{\LL}\subset Z_a$. Because $\LL,\hat{\LL}$ have underlying locally compact abelian group structure and $\hat{\hat{\LL}}\cong \LL$, the two orthogonal Lagrangian subspaces are Pontryagin duals, and the symplectic transformation is essentially a Fourier transform. Since the symplectic transformation doesn't change the dynamics, the Mellin transform $\mathcal{M}_\lambda[\mathrm{Tr}\,\mathrm{F};\alpha]$ captures the same fixed-energy process for any Lagrangian subspace. The upshot is the $L$-functional can be formulated in $\LL$ or $\hat{\LL}$, and the two perspectives taken together allow for its analytic continuation.

To see this explicitly, recall the calculation of the time-evolution reproducing kernel in subsection \emph{\ref{point-to-point}} and note that $|\mathrm{tr}({K}_H^{\alpha}(\psi_{i},\psi_{j}))|
=|\mathrm{tr}(\hat{K}_{{H}}^{\alpha}(\hat{\psi}_i,\hat{\psi}_j))|$ where orthonormal bases $\{\psi_{i}\}$ and $\{\hat{\psi}_{i}\}$ span $\LL$ and $\hat{\LL}$ respectively, ${K}_H^{\alpha}(\psi_i,\psi_j):=\langle \psi_{i}|\mathcal{M}_\lambda[\mathrm{F};\alpha]|\psi_{j}\rangle$, and $\hat{|\psi_{i}}\rangle\equiv S\,|\psi_{i}\rangle$  for symplectic $S:\LL\rightarrow \hat{\LL}$.\footnote{The equality follows from $\langle \psi_{i}|\mathcal{M}_\lambda[\mathrm{F};\alpha]|\psi_{j}\rangle
=\langle \hat{\psi}_{i}|S\mathcal{M}_\lambda[\mathrm{F};\alpha]S^{-1}|\hat{\psi}_{j}\rangle=:\langle \hat{\psi}_{i}|\hat{\mathcal{M}}_\lambda[{\mathrm{F}};\alpha]|\hat{\psi}_{j}\rangle$.}
  Since $\langle l|\langle r,H_r\rangle|l\rangle$ is quadratic in $l$, expand $F(\langle r,H_r\rangle)$ in a basis of Hermite polynomials $F(\langle r,H_r\rangle)=e^{-\frac{1}{2}\langle r,H_r\rangle}\sum_n f_nH_n(\langle r,H_r\rangle)$. Note $\langle r,H_r\rangle$ is linear in $r$.

 Suppose $\{\psi_{i}\}$ diagonalizes $H$ in $\LL$ with eigenvalues $(\lambda^{(\tau)}_i,\lambda^{(r)}_i)$, and let $\mathit{\Pi}\in L_B(\mathcal{H})$ be a period operator such that $[\mathit{\Pi},H]=0$ and $\langle \psi_{i}|\langle\Delta t,\mathit{\Pi}\rangle|\psi_{i}\rangle=2\pi i$. Fourier expand the periodic function
\begin{equation}
\chi_{\Delta t}(\lambda^{(\tau)}_i):=\langle \psi_{i}|e^{-\frac{1}{2}\langle\Delta t,H_\tau\rangle}|\psi_{i}\rangle
=\langle \psi_{i}|e^{-\frac{1}{2}\langle\Delta t,H_\tau+\mathit{\Pi}\rangle}|\psi_{i}\rangle
=P^{-d/2}\sum_{\{a^i\}=-\infty}^{\infty}\hat{\chi}(a)e^{2\pi i \frac{a\cdot\lambda^{(\tau)}_i}{P}}
\end{equation}
 where $P=\langle \psi_{i}|\langle\Delta t,\mathit{\Pi}\rangle|\psi_{i}\rangle/\Delta t\in\mathbb{Z}_+$ and $d=\mathrm{dim}(Z_a)$.
 Then we have
\begin{eqnarray}
\mathcal{M}_\lambda[\mathrm{Tr}\,\mathrm{F};\alpha]
&\stackrel{\lambda_{\C^\times}}{\rightarrow}& \int_{\R_\pm} \mathrm{tr}\left( \chi_{\Delta t}(H_\tau)F(\langle r,H_r\rangle)\right) r^\alpha\;d\log r\;\;\;\;\;\;\;\;\alpha\in\mathbb{S}_\lambda\notag\\
&=&\pm P^{-d/2}\sum_{\{a^i\}}\hat{\chi}(a)\left(\int_{0}^{1/\sqrt{P}}+\int_{1/\sqrt{P}}^\infty\right) \mathrm{tr}\left( e^{2\pi i \frac{a H_r}{P}}F(\langle r,H_r\rangle)\right) r^\alpha\;d\log r\;.\notag\\
\end{eqnarray}
The $\int_{1/\sqrt{P}}^\infty$ integral is an entire function of $\alpha\in\C$. But the $\int_0^{1/\sqrt{P}}$ integral has a potential convergence problem as $r\rightarrow 0$ when $\alpha\notin\mathbb{S}_\lambda$, and it is not clear if $\mathcal{M}_\Gamma[\mathrm{Tr}\,\mathrm{F};\alpha]$ is well-defined on $\C$.

One remedy is to use the trace in $\hat{\LL}$ for the $\int_0^{1/\sqrt{P}}$ integral by applying a symplectic transformation $S$. It follows from the scaling property of the Fourier transform of quadratic $F(\langle r,H\rangle)$ and Hermite polynomials that $S\,F(\langle r,H\rangle)\,S^{-1}=\mathcal{F}(F(\langle r,H\rangle))=\phi\, r^{-n}{F}(\langle r^{-1},\hat{H}\rangle)$ where the phase $\phi=1$ or $\phi=i^{-n/2}$ for a definite or indefinite metric respectively in $\hat{\LL}$ and $n=d/2=\mathrm{dim}(\hat{\LL})$. Hence, recalculating \emph{(\ref{fixed-energy process})} for the trace of the Fourier transform gives
\begin{eqnarray}
\int_{0}^{1/\sqrt{P}}\mathrm{tr}\left( \chi_{\Delta t}(H_\tau)F(\langle r,H_r\rangle)\right) r^{\alpha}\;d\log r
&=&\phi\int_{0}^{1/\sqrt{P}}\mathrm{tr}\left({\chi}_{\Delta t}(\hat{H}_\tau)\ast{F}(r^{-1},\hat{H}_r)\right) r^{\alpha-n}\;d\log r\notag\\
&=&\phi\int_{1/\sqrt{P}}^\infty\mathrm{tr}\left({\chi}_{\Delta t}(\hat{H}_\tau)\ast{F}(r,\hat{H}_r)\right) r^{n-\alpha}\;d\log r\notag\\
\end{eqnarray}
where $(\cdot\ast\cdot)$ denotes the convolution of Fourier transforms, the restriction $\delta(\tau-\Delta t)$ forces $\alpha-n=1$ in the $\tau$ integral, and we used invariance of the Haar measure under $r\rightarrow r^{-1}$ in the second line.

So in the diagonal basis of $H$,
\begin{equation}
L^+_\lambda(\alpha,F)
=\phi^{1/2}\int_{1/\sqrt{P}}^\infty\sum_i \left(\phi^{1/2}{\chi}_{\Delta t}({\hat{\lambda}^{(\tau)}}_i)\ast{F}(r,{\hat{\lambda}^{(r)}}_i)r^{n-\alpha}
+\phi^{-1/2}\chi_{\Delta t}(\lambda^{(\tau)}_i)\cdot F(r,\lambda^{(r)}_i)r^\alpha\right)\;d\log r\;.
\end{equation}
Evidently $L^+_\lambda(\alpha,F,\phi):=\phi^{-1/2}L^+_\lambda(\alpha,F)=L^+_\lambda(n-\alpha,{\hat{F}},\phi^{-1})$ verifies a functional equation, and therefore $L^+_\lambda(\alpha,F,\phi)$ can be extended to the entire complex plane.\footnote{We have seen that $L^+_\lambda(\alpha,F,\phi)$ is periodic in $\Delta t$ and invariant under $r\rightarrow r^{-1}$ in the right-half complex plane. We can of course rotate this to the upper-half complex plane. This hints that the hyperbolic plane and $SL(2,\mathbb{Z})$ have some part to play in the story of fixed-energy processes and perhaps more generally when $\tau$ is a $2$-chain or a $2$-cycle in $\C^\times$.}
\end{application}

\begin{example}\label{momentum L-function}
Revisit the context in \emph{\S \ref{scattering}}. Scale $\Delta t\rightarrow \mathrm{s}\Delta t=2\pi i$ so that $P=2k$ with $k\in\mathbb{Z}$ and put $m=0$ to make the integrals manageable. The Hamiltonian is $H=2p_b^2$ and $f_0=1$, $f_{n>0}=0$. Away from the poles, we get
\begin{equation}
\mathcal{M}_\Gamma[\mathrm{Tr}\,\Box^{-1};\alpha]
=\pi^3i^{\alpha+1}(\alpha-1)\csc(\pi\alpha)
=-\mathcal{M}_\Gamma[\mathrm{Tr}\,\mathcal{F}(\Box^{-1});\alpha]\;.
\end{equation}
Observe that (since $d=4$ and $\phi=-1$)
\begin{eqnarray}
L^+_\Gamma(\alpha,F,-1)
&:=&i\pi^3 e^{\pi i(\frac{\alpha}{2}+1)}\mathcal{M}_\Gamma[\mathrm{Tr}\,\Box^{-1};\alpha]\notag\\
&=&i^{-1}\pi^3e^{-\pi i(\frac{\alpha}{2}+1)}\mathcal{M}_\Gamma[\mathrm{Tr}\,\mathcal{F}(\Box^{-1});\alpha]
=L^+_\Gamma(2-\alpha,\hat{F},-1)\;.
\end{eqnarray}
\end{example}

To implement analytic continuation more generally, expand $\mathrm{F}$ in a basis of algebraic elements $\Upsilon_n^{(s)}\in\mathbf{F}_{\mathbb{S}_{\mathcal{R}}}(G^\C)$  such that $\Upsilon_n^{(s)}:G^\C\rightarrow\mathfrak{C}^\ast$ by $g\mapsto e^{-\frac{1}{2}\langle\rho(g),H\rangle}L_n^{(s)}(\langle\rho(g),H\rangle)$ where $L_n^{(s)}(\langle\rho(g),H\rangle)$ are operator-valued, generalized Laguerre polynomials with $-1<s\in\R$. In the diagonal basis in $\LL$, this becomes $\langle \psi_i|\Upsilon_n^{(s)}(g)|\psi_i\rangle=e^{-\frac{1}{2}\langle\Delta t,\lambda^{(\tau)}_i\rangle}e^{-\frac{1}{2}\lambda^{(r)}_i\, r}L_n^{(s)}(\lambda^{(r)}_i\, r)$.

 Making use of the family of Laguerre integrators $\mathcal{L}_\lambda[\mathrm{F};\alpha,n,s]$ introduced in Appendix \ref{ast and star}, matrix elements (in the diagonal basis) of the inverse complex power of $\mathrm{F}$ for a fixed-energy process can be expressed as
\begin{eqnarray}
\langle\psi_i|\mathcal{M}_\Gamma[\mathrm{F};\alpha,s]|\psi_i\rangle
&\stackrel{\lambda_{\C^\times}}{\rightarrow}&
e^{-\frac{1}{2}\langle\Delta t,\lambda^{(\tau)}_i\rangle}{\lambda^{(r)}_i}^{-\alpha}\lim_{N\rightarrow\infty}\sum_{n=0}^N f_n^{(s)}
\frac{1}{2^\alpha\Gamma(\alpha)}\int_{\R_\pm}e^{-\frac{1}{2}\tilde{r}} L_n^{(s)}(\tilde{r})\tilde{r}^\alpha\;d\log \tilde{r}\notag\\
&=&e^{-\frac{1}{2}\langle\Delta t,\lambda^{(\tau)}_i\rangle}|2\lambda^{(r)}_i|^{-\alpha}\lim_{N\rightarrow\infty}\sum_{n=0}^N f_n^{(s)}\mathcal{L}_\Gamma[\mathrm{E}^{-\mathrm{Id}/2};\alpha,n,s]\notag\\
&=&e^{-\frac{1}{2}\langle\Delta t,\lambda^{(\tau)}_i\rangle}|2\lambda^{(r)}_i|^{-\alpha}\sum_{n=0}^\infty a_n^{(s)}\,\tensor[_2]{\mathrm{F}}{_1}(-n,\alpha;s+1;2)
 \;\;\;\;\;\;\;\;\alpha\in\langle0,\infty\rangle\notag\\
\end{eqnarray}
where $\tilde{r}={\lambda^{(r)}_i}^{-1}r$ (recall $H$ is Hermitian) and $a_n^{(s)}:=f_n^{(s)} \left(\begin{array}{c}
 s+n \\
 n
 \end{array}\right)$.
If $\lambda^{(r)}_i\gtrless 0$, only the $\R_\pm$ integral contributes yielding the absolute value term in the second and third lines. The associated $s$-dependent operator expressions of the Dirichlet Green's function and zeta function for initial and final states $\psi_a,\psi_b$ are then defined by

\begin{equation}
\langle\psi_b|\mathcal{M}_\Gamma[\mathrm{F};\alpha,s]|\psi_a\rangle
:=\langle\psi_b|e^{-\frac{1}{2}\langle\Delta t,H_\tau\rangle}|2H_r|^{-\alpha}|\psi_a\rangle\sum_{n=0}^\infty a_n^{(s)}\,\tensor[_2]{\mathrm{F}}{_1}(-n,\alpha;s+1;2)\;,
\end{equation}
and
\begin{equation}\label{L zeta}
\mathcal{M}_\Gamma[\mathrm{Tr}\,\mathrm{F};\alpha,s]
 :=\mathrm{tr}(e^{-\frac{1}{2}\langle\Delta t,H_\tau\rangle}|2H_r|^{-\alpha})\sum_{n=0}^\infty a_n^{(s)} \,\tensor[_2]{\mathrm{F}}{_1}(-n,\alpha;s+1;2)
 \;\;\;\;\;\;\;\;\alpha\in\langle0,\infty\rangle\;.
\end{equation}
Evidently, for special $H_r$ (e.g. $H_r=Id$) the zeta functions for Laguerre-integrable functionals can be analytically continued to meromorphic functions on $\C$; which is not too surprising. The drawback of the zeta function (\ref{L zeta}) is that it doesn't establish a functional equation, which is a desired feature as it allows for meromorphic continuation for more generic $H_r$.

To get a functional equation, suppose that $\mathcal{M}_\lambda[\mathrm{E}^{-\mathrm{H}_r}\ast\mathrm{E}^{-\mathrm{H}_r};\alpha]
=\mathcal{M}_\lambda[\mathrm{E}^{-\mathrm{H}_r}\star\mathrm{E}^{-\mathrm{H}_r};\alpha]$. Then
 Appendix E suggests a modification that will do the trick: apply the shift $\alpha\rightarrow \alpha+s/2$, choose Haar measure $\nu(g_\psi):=2^{-(\alpha+s/2)}\,\nu(g)/\Gamma(\alpha+s/2)$, and define an $n$-dependent and $s$-dependent $L$-functional $L^+_\lambda(\alpha,\Upsilon_n^{(s)},\phi)
:=\phi^{-n}\mathcal{M}_\lambda[\mathrm{Tr}\,\Upsilon_n^{(s)};\alpha+s/2, s]$. For $\phi=i$, using Lemma \ref{Psi lemma} and Theorem \ref{Psi theorem} we find
\begin{eqnarray}
L^+_\psi(\alpha,s,F,\phi)
&:=&\sum_{n=0}^\infty a_n^{(s)}L^+_\lambda(\alpha,\Upsilon_n^{(s)},\phi)\notag\\
&=&\mathrm{tr}(e^{-\frac{1}{2}\langle\Delta t,H_\tau\rangle}|H_r|^{-\alpha-s/2})\sum_{n=0}^\infty a_n^{(s)}\,i^{-n}\tensor[_2]{\mathrm{F}}{_1}(-n,\alpha+s/2;s+1;2)\notag\\
&=&\mathrm{tr}(e^{-\frac{1}{2}\langle\Delta t,H_\tau\rangle}|H_r|^{\alpha-1-s/2})\sum_{n=0}^\infty a_n^{(s)}\,i^n\tensor[_2]{\mathrm{F}}{_1}(-n,1-\alpha+s/2;s+1;2)\notag\\
&=&L^+_\psi(1-\alpha,s,F,\phi^{-1})
\end{eqnarray}
This defines a class of $L$-functions based on functionals of the form $\mathrm{F}^{(s)}=\sum f_n^{(s)}\Upsilon_n^{(s)}$ for Hamiltonians satisfying $\mathcal{M}_\lambda[\mathrm{E}^{-\mathrm{H}_r}\ast\mathrm{E}^{-\mathrm{H}_r};\alpha+{s}/2]
=\mathcal{M}_\lambda[\mathrm{E}^{-\mathrm{H}_r}\star\mathrm{E}^{-\mathrm{H}_r};\alpha+{s}/2]$. It would be interesting to understand the physical interpretation of $\mathrm{E}^{-\mathrm{H}_r}\ast\mathrm{E}^{-\mathrm{H}_r}
=\mathrm{E}^{-\mathrm{H}_r}\star\mathrm{E}^{-\mathrm{H}_r}$ in the $C^\ast$-algebra $\mathit{\Pi}^{(\alpha)}(\mathbf{F}_{\mathbb{S}_{\mathcal{R}}}(G^\C))$. For such $\mathrm{H}_r$, one can anticipate a connection between the Mellin transform of $\mathrm{tr}\,e^{-\langle z,H\rangle}$ and modular forms --- hence (by the modularity theorem) a connection between complex one-parameter subgroups of $G^\C$ representing $\mathrm{E}^{-\mathrm{H}}$ and elliptic curves when $\mathrm{E}^{-\mathrm{H}_r}\ast\mathrm{E}^{-\mathrm{H}_r}
=\mathrm{E}^{-\mathrm{H}_r}\star\mathrm{E}^{-\mathrm{H}_r}$.

There are some obvious extensions of the preceding considerations: i) The fixed energy and Hermitian $H$ conditions can be relaxed so that $\tau$ is a $1$- or $2$-chain in $\C^\times$ with suitably restricted $\langle\tau,H\rangle$ to render the Mellin transform well-defined. ii) The vector state-space (with symplectic structure) can be promoted to a symplectic vector bundle. iii) One can localize to an Abelian group other than $\C^\times$.

Regarding extension iii): A direct connection with number theory can be had by localizing onto the abelian group of ideles over the Gaussian rationals $\lambda_{\mathbb{A}^\times_{\mathbb{Q}[i]}}
:\phi_{\mathfrak{h}}(\C)\rightarrow\mathbb{A}^\times_{\mathbb{Q}[i]}$. For fixed-energy processes, functional Mellin then implements $p$-adic Mellin transforms on the locally compact topological group of ideles $\mathbb{A}^\times_{\mathbb{Q}}$.

\begin{example}For a quick illustration, return to \emph{Example \ref{Dirichlet L-function}} and take $\Delta t=0$. This yields $L_\Gamma(\alpha,N)=\sum_{n\in\mathbb{Z}_+}n^{-\alpha}=\zeta(\alpha)$ which is the tracial form of the Riemann zeta function.

On the other hand, if we localize to $\mathbb{A}^\times_{\mathbb{Q}}$ and choose $\mathrm{F}(g)=e^{-\frac{1}{2}g_{\infty}}\prod_{p}1_{\mathbb{Z}_p}(g_p)$ where $g\in\mathbb{A}^\times_{\mathbb{Q}}$ and $1_{\mathbb{Z}_p}(g_p)$ is the indicator function on the $p$-adic integers $\mathbb{Z}_p$, then idelic integration gives the Euler product form of Riemann zeta \emph{(see e.g.\cite[\S 2.2]{GH})}
\begin{eqnarray}
\mathcal{M}_\Gamma[\mathrm{F};\alpha]
\stackrel{\lambda_{\mathbb{A}^\times_{\mathbb{Q}}}}{\rightarrow}
\frac{1}{\Gamma(\alpha)}\int_{\mathbb{A}^\times_{\mathbb{Q}}}\mathrm{F}(g)
|g|^\alpha_{\mathbb{A}}\,d^\times g
&=&\frac{1}{\Gamma(\alpha)}\int_{\R_+}e^{-\frac{1}{2}x_\infty}|x|_\infty^{\alpha}\;d^\times x_\infty
\prod_p\int_{\mathbb{Z}_p\backslash\{0\}}|x_p|^\alpha_p\,d^\times x_p\notag\\
&=&\prod_p(1-p^{-\alpha})^{-1}\;\;\;\;\;\;\;\;\alpha\in\langle0,\infty\rangle\notag\\
&=&\zeta(\alpha)\;.
\end{eqnarray}
Since $\int_{\mathbb{A}^\times_{\mathbb{Q}}}=\prod_v \int_{\mathbb{Q}^\times_{v}}$ where $v$ denotes all primes including the prime at infinity, this can be viewed as a functional determinant $\zeta(\alpha)=\mathcal{M}_\Gamma[\mathrm{Det}\,\mathrm{F};\alpha]
\stackrel{\lambda_{\mathbb{Q}^\times_{p}}}{\rightarrow}
\prod\int_{\mathbb{Q}^\times_{p}}\mathrm{F}(g_p)
|g|^\alpha_{\mathbb{Q}_p}\,d^\times g_p$  over the finite places such that $\mathrm{F}(g_p)=1_{\mathbb{Z}_p}(g_p)$.
\end{example}
This example suggests a connection between functional traces over $\mathcal{H}$ and functional determinants over $\mathcal{H}'$ that becomes an equivalence given suitably chosen Hilbert spaces. That is, for special $\mathcal{H},\mathcal{H}'$ this implies the remarkable relation $\mathrm{tr}_{\mathcal{H}}e^{-\langle r,H\rangle}=e^{-\mathrm{tr}_{\mathcal{H}'}\langle r,H\rangle}$ for all $r\in\R_\pm$.

\subsection{Determinants}
In this subsection we define an $L$-functional based on (\ref{parametric Feynman}).

Recall for a general graph $\mathrm{G}$ with $V$ vertices associated with scalar particles and external momentum conservation, we have complex momentum states ${p}^v\in\C^4$ (which are sums of $n$ external momenta) incident at the $V$ vertices such that $\sum_{r=1}^V{p}^{v_r}=0$. Referring to \S\ref{graphs}, recall $M^2:=|\Bold{m}\rangle\langle\Bold{m}|$ is an $I\times I$ diagonal mass matrix in $L_B(\mathcal{H})$ with Hilbert space $\mathcal{H}\cong\C^{I}$ where $I=L+(V-1)$ and $L$ is the number of loops in $\mathrm{G}$. The rank $V-1$ incidence matrix $\Bold{\varepsilon}$ associated with $\mathrm{G}$ implies momentum conservation, and the quadratic form can be written $\mathrm{Q}_{\mathrm{G}}(\Bold{p},\Bold{\tau})=\mathrm{tr}\,P^2Q_\epsilon(\Bold{\tau})$ where $P^2:=|\Bold{p}\rangle\langle\Bold{p}|$ is a $(V-1)\times (V-1)$ diagonal matrix and $Q_\epsilon(\Bold{\tau}):=(\Bold{\varepsilon}\Delta(\Bold{\tau})\Bold{\varepsilon}^{\mathrm{T}})^{-1}$.

We want to express $\mathrm{Q}_{\mathrm{G}}(\Bold{p},\Bold{\tau})$ in terms of $\mathcal{H}\cong\C^{I}$: To that end, define $\tilde{Q}_\epsilon:=\Bold{\varepsilon}^{\mathrm{T}}Q_\epsilon\Bold{\varepsilon}$ and $|\tilde{\Bold{p}}\rangle:=\Bold{\varepsilon}_R|\Bold{p}\rangle\in\mathcal{H}$ where $\Bold{\varepsilon}_R$ is the right inverse of $\Bold{\varepsilon}$. Then the quadratic form  becomes $\mathrm{tr}(P^2Q_\epsilon(\Bold{\tau}))
=\mathrm{tr}(\tilde{P}^2\tilde{Q}_\epsilon(\Bold{\tau}))
=:\mathrm{tr}(\tilde{P}^2_{\mathrm{G}}\Delta^{-1}(\Bold{\tau}))$ where $\tilde{P}^2_{\mathrm{G}}:=\tilde{P}^2\tilde{Q}_\epsilon(\Bold{\tau})\Delta(\Bold{\tau})\in L_B(\mathcal{H})$ is an ``internal line'' momentum matrix. But observe that $\tilde{Q}_\epsilon(\Bold{\tau})\Delta(\Bold{\tau})=Id$ is the identity operator on $\mathcal{H}$, so the exponential term in (\ref{parametric Feynman}) can be expressed as
\begin{equation}
e^{-\mathrm{tr}[(\tilde{P}^2-{M}^2)\Delta^{-1}(\Bold{\tau})]}
=\mathrm{det}\,e^{-(\tilde{P}^2-{M}^2)\Delta^{-1}(\Bold{\tau})}
\end{equation}
where $\tilde{P}^2:=|\tilde{\Bold{p}}\rangle\langle\tilde{\Bold{p}}|
=\Bold{\varepsilon}_R|\Bold{p}\rangle\langle\Bold{p}|\Bold{\varepsilon}_R^\mathrm{T}\in L_B(\mathcal{H})$. Moreover, the first Symanzik polynomial in the $\tilde{\Bold{p}}$ representation reduces to $\mathrm{det}(\Delta(\Bold{\tau}))\prod_l\Bold{\tau}_l=1$. Hence the Feynman amplitude associated with $\mathrm{G}$ can be viewed as a functional determinant relative to $\mathcal{H}$;
\begin{equation}
I_{\mathrm{G}}({p}_1,\ldots,{p}_V;\Bold{\alpha})
=\langle\Bold{p}|\Bold{\varepsilon}_R^\mathrm{T}
(\mathrm{Det}\,(\tilde{\mathrm{P}}^2-{\mathrm{M}}^2))^{-\Bold{\alpha}}_\Gamma
\Bold{\varepsilon}_R|{\Bold{p}}\rangle
=\langle\tilde{\Bold{p}}|
\mathcal{M}_\Gamma[\mathrm{Det}\,\mathrm{E}^{-(\tilde{\mathrm{P}}^2-{\mathrm{M}}^2)};\Bold{\alpha}]
|\tilde{\Bold{p}}\rangle
\end{equation}
where $G^\C$ is the group of positive-definite diagonal matrices $\Delta^{i\R^I_+}$ with entries in $i\R_+$ and $\Bold{\alpha}=(\alpha_1,\ldots,\alpha_I)$.

Away from the poles, $(\tilde{P}^2-{M}^2)$ is a positive-definite diagonal matrix in which case the operator $\mathrm{det}(\tilde{P}^2-{M}^2)$ factors out of the Mellin transform by virtue of the invariant Haar measure. At the poles, the Mellin transform contributes a delta functional as we saw in \S\ref{point-to-point}.  Summing over all possible incidence matrices $\Bold{\varepsilon}_R$ for a given $I$ gives the  perturbative $n$-point correlation function at loop-level $L$ that can be renormalized according to \S\ref{renormalization}.

Following the arguments of the previous subsection, the correlation function satisfies a functional equation and hence can be extended to a meromorphic function of several complex variables $\Bold{\alpha}=(\alpha_1,\ldots,\alpha_I)\in\C^I$ if $\mathrm{E}^{-(\tilde{\mathrm{P}}^2-{\mathrm{M}}^2)}
\ast\mathrm{E}^{-(\tilde{\mathrm{P}}^2-{\mathrm{M}}^2)}
=\mathrm{E}^{-(\tilde{\mathrm{P}}^2-{\mathrm{M}}^2)}
\star\mathrm{E}^{-(\tilde{\mathrm{P}}^2-{\mathrm{M}}^2)}$ (since the Mellin integral over diagonal matrices is a product of one-dimensional Mellin integrals). In such a case, we will call $L_\Gamma^\times(\Bold{\alpha},\mathrm{det}\,(\tilde{{P}}^2-{{M}}^2))
:=\mathrm{Det}\,(\tilde{{P}}^2-{{M}}^2)^{-\Bold{\alpha}}_\Gamma$ a Feynman $L$-function where the determinant is with respect to $\mathcal{H}$. The $\times$ superscript denotes an $L$-functional associated with a functional determinant.

We propose to generalize the Feynman $L$-function to non-abelian fixed-energy processes by localizing according to $\lambda_{G^\C_{\mathbf{M}}} :\phi_{\mathfrak{h}}(\C)\rightarrow G^\C_{\mathbf{M}}\subseteq \C^{I\times I}$ where $G^\C_{\mathbf{M}}$ is the group of complex matrices $\mathbf{M}$. Polar decompose $\Bold{z}\in G^\C_{\mathbf{M}}$ as $\Bold{z}=\Bold{r}\Bold{\tau}$ where $\Bold{r}\in \mathbf{V}$ is positive-definite Hermitian and $\Bold{\tau}\in \mathbf{U}$ is unitary. For fixed-energy processes, $\delta(\Bold{\tau}-\Delta\Bold{t})$,  the domain of the Mellin integral reduces to the group $\mathbf{V}$. For Hamiltonian operator $H=H_{\Bold{\tau}}+H_{\Bold{r}}$ and standard Haar measure (denoted $\mathrm{sH}$), we have a matrix integral
\begin{equation}
\mathrm{Det}\,{H}^{-\Bold{\alpha}}_{\mathbf{V},\mathrm{sH}}
=\mathrm{det}\,\chi_{\Delta \Bold{t}}(H_{\Bold{\tau}})\int_{\mathbf{V}}\mathrm{det}\,(e^{-H_{\Bold{r}}\Bold{r}}\,\Bold{r}^{\Bold{\alpha}})
\,d\log\Bold{r_{\mathbf{V}}}
\end{equation}
where $\log\Bold{r}$ is defined in (\ref{log definition}) and $d\log\Bold{r_{\mathbf{V}}}=\mathrm{det}\,(\Bold{r}^{-\frac{I+1}{2}})
\,d\Bold{r}$. Notice the integral is just the matrix integral introduced in \cite{LA1} associated with a gamma integrator, and for positive-definite Hamiltonians and certain values of $\Bold{\alpha}$ it is the integral of a complex Wishart distribution. This connection between Wishart distributions and Feynman amplitudes is not so surprising as the Wishart eigenvalue distribution is related to Laguerre polynomials which we have seen are relevant to scattering processes.

If the spectrum of $H_{\Bold{r}}$ is positive definite, the operator can be extracted from the integral and we have $\mathrm{Det}\,{H}^{-\Bold{\alpha}}_{\mathbf{V},\mathrm{sH}}
=\mathrm{det}\,(\chi_{\Delta \Bold{t}}(H_{\Bold{\tau}})H_{\Bold{r}}^{-\Bold{\alpha}})\Gamma_I(\Bold{\alpha})$ where
\begin{equation}
\Gamma_I(\Bold{\alpha}):=\int_{\mathbf{V}}e^{-\mathrm{tr}\,\Bold{r}}
\,\mathrm{det}\,(\Bold{r}^{\Bold{\alpha}-\frac{I+1}{2}})
\,d\Bold{r}\;\;\;\;\;\;\;\;\Re(\alpha_l)>\frac{I+1}{2}\;.
\end{equation}
For $\alpha_l=\alpha_o$ for all $l\in I$ and some $\alpha_o>\frac{I+1}{2}$ (i.e. $\Bold{\alpha}=(\alpha_o,\ldots,\alpha_o)$) this is the multivariate gamma function. Hence, in this case at least, if $\mathrm{E}^{-\mathrm{H}_{\Bold{r}}}
\ast\mathrm{E}^{-\mathrm{H}_{\Bold{r}}}
=\mathrm{E}^{-\mathrm{H}_{\Bold{r}}}
\star\mathrm{E}^{-\mathrm{H}_{\Bold{r}}}$ then $\mathrm{Det}\,{H}^{-\Bold{\alpha}}_{\mathbf{V},\mathrm{sH}}$ satisfies a functional equation and so can be continued to a meromorphic function on $\C$ whose Feynman $L$-function is $L_{\Gamma_I}^\times(\Bold{\alpha},\mathrm{det}\,(H))
:=\mathrm{Det}\,H^{-\Bold{\alpha}}_{\Gamma_I}=\mathrm{det}\,(\chi_{\Delta \Bold{t}}(H_{\Bold{\tau}})H_{\Bold{r}}^{-\Bold{\alpha}})$.

\section{Conclusion}
Fourier transform has been a central theme in the discipline of functional integrals since their inception. The Fourier transform represents duality between locally compact abelian groups. We contend that the Mellin transform in the functional context goes beyond Fourier: It represents duality between Banach $\ast$-algebras. As such, functional Mellin is a useful addition to the toolbox  of mathematical physics. In particular, we used it to construct functional analogs of resolvent, trace, log, and determinant and presented several examples and applications. The primary applications probed the observable $\mathrm{E}^{-\mathrm{H}}$ whose associated evolution operator depends on a real or complex one-parameter subgroup $\phi_{\mathfrak{h}}$ and $1$-chains in $i\R_+\cup i\R_-$ or $\C^\times$. These one-parameter applications gave $\alpha$-dependent representations of QFT generating functionals, scattering amplitudes, renormalization group equations, density operators, and $L$-functions associated with fixed-energy processes. Besides offering new perspectives and interpretations of well-understood examples, these exercises support the claim that $\mathbf{F}_{\mathbb{S}}(G^\C)$ and its functional Mellin representations provide a unifying framework for QFT. It would be interesting to probe $\mathrm{E}^{-\mathrm{H}}$ with an evolution operator depending on oriented $2$-chains in $\C^\times$ utilizing Green's functions on fat graphs in the perturbative regime. One might anticipate some stringy connections.

Importantly, at a broader level functional Mellin is apposite in characterizing quantum systems in general: Given  some relevant representations of a topological group $G^\C$, functional Mellin defines a $C^\ast$-algebra for which the Mellin integrator acts as a $\ast$-homomorphism to the algebra of bounded linear operators on the Hilbert spaces carrying representations of $G^\C$. This means that, armed with functional Mellin and a starting topological group, one can construct and represent a non-commutative $C^\ast$-algebra --- without having to somehow deform a commutative algebra. Consequently, one can \emph{directly} model a system's quantum properties \emph{without first passing through the classical realm}. We exploit this idea in subsequent work.

\appendix

\section{Mellin transforms}\label{Mellin appx.}
\subsection{Basics}
Most of the following basic properties can
be found in \cite{FL}-\cite{ZEM}.
\begin{definition}
Let $f:(0,\infty)\rightarrow\C$ be a function such that $f\in
L^1(\R_+)$ with limits $\lim_{\x\rightarrow
0^+}f(\x)\rightarrow\mathcal{O}(\x^{-\mathrm{a}})$ and
$\lim_{\x\rightarrow\infty}f(\x)\rightarrow\mathcal{O}(\x^{-\mathrm{b}})$
for $\mathrm{a},\mathrm{b}\in\R$. Then the Mellin transform
$\widetilde{f}(\alpha)$ with $\alpha\in \langle
\mathrm{a},\mathrm{b}\rangle:=(\mathrm{a},\mathrm{b})\times
i\R\subset\C$ is defined by
\begin{equation}
\widetilde{f}(\alpha):=\mathcal{M}[f(\x);\alpha]:=\int_0^\infty
f(\x)\x^{\alpha-1}d\x\;.
\end{equation}

\end{definition}

The fundamental strip $\langle
\mathrm{a},\mathrm{b}\rangle\subset\C$ indicates the domain of
convergence. Since if
\begin{equation}
f(\x)|_{\x\rightarrow
0^+}=\mathcal{O}(\x^{-\mathrm{a}})\hspace{.15in}\mathrm{and}\hspace{.15in}
f(\x)|_{\x\rightarrow \infty}=\mathcal{O}(\x^{-\mathrm{b}})\,,
\end{equation}
then $\widetilde{f}(\alpha)$ exists in $\langle
\mathrm{a},\mathrm{b}\rangle$ where it is holomorphic and absolutely
convergent. More precisely,
\begin{eqnarray}
|\widetilde{f}(\alpha)|\leq\int_0^1|f(\x)|\,\x^{\Re(\alpha)-1}\;d\x
+\int_1^\infty|f(\x)|\,\x^{\Re(\alpha)-1}\;d\x\notag\\
\leq M_1\int_0^1\x^{\Re(\alpha)-1-\mathrm{a}}\;d\x
+M_2\int_1^\infty\x^{\Re(\alpha)-1-\mathrm{b}}\;d\x
\end{eqnarray}
for some finite constants $M_1,M_2$.

From the definition follows some important properties (for suitable
fundamental strips);
\begin{eqnarray}\label{properties}
c^{-\alpha}\widetilde{f}(\alpha)&=&\mathcal{M}[f(c\x);\alpha]
\hspace{1.1in}c>0\notag\\
\widetilde{f}(\alpha+d)&=&\mathcal{M}[\x^df(\x);\alpha]
\hspace{1.0in}d>0\notag\\
\frac{1}{|\mathrm{r}|}\widetilde{f}(\frac{\alpha}{\mathrm{r}})
&=&\mathcal{M}[f(\x^\mathrm{r});\alpha]
\hspace{1.1in}\mathrm{r}\in\R-\{0\}\;,\;\alpha\in\langle \mathrm{r}
\mathrm{a},\mathrm{r}
\mathrm{b}\rangle\notag\\
\frac{d^n}{d\alpha^n}\widetilde{f}(\alpha)
&=&\mathcal{M}[(\log\x)^nf(\x);\alpha]
\hspace{.65in}n\in\mathbb{N}\notag\\
-\alpha\widetilde{f}(\alpha)
&=&\mathcal{M}\left[\left(\x\frac{d}{d\x}\right) f(\x);\alpha\right]
\hspace{.48in}\notag\\
-\widetilde{f}(\alpha-1)
&=&\mathcal{M}\left[\frac{d}{d\x}f(\x);\alpha\right]
\hspace{.85in}\notag\\
-\frac{1}{\alpha}\widetilde{f}(\alpha+1)
&=&\mathcal{M}\left[\int_0^xf(\x')\;d\x';\alpha\right]\;.
\end{eqnarray}

The last three relations can be extended by iteration:
\begin{equation}\label{iteration}
(-1)^n\frac{\Gamma(\alpha+n)}{\Gamma(\alpha)}\widetilde{f}(\alpha)
=\mathcal{M}\left[\x^n\frac{d^n}{d\x} f(\x);\alpha\right]
\end{equation}\label{iteration 1}
for $n\in\mathbb{N}$ and
$\x^{\alpha+n-m}f^{(n-m)}(\x)|_0^\infty=0\;\forall
m\in\{1,\ldots,n\}$,
\begin{equation}
(-1)^n\frac{\Gamma(\alpha)}{\Gamma(\alpha-n)}\widetilde{f}(\alpha-n)
=\mathcal{M}\left[f^{(n)}(\x);\alpha\right]
\end{equation}
for $n\in\mathbb{N}$ and
$\x^{\alpha-n-1+m}f^{(n-m)}(\x)|_0^\infty=0\;\forall
m\in\{1,\ldots,n\}$, and
\begin{equation}
(-1)^n\frac{\Gamma(\alpha)}{\Gamma(\alpha+n)}\widetilde{f}(\alpha+n)
=\mathcal{M}\left[\left(\int_0^xf(\x')\;d\x'\right)^n;\alpha\right]
\end{equation}
where $\left(\int_0^xf(\x')\;d\x'\right)^n$ defines an iterated
integral
\begin{equation}
\left(\int_0^xf(\x')\;d\x'\right)^n:=\int_0^x\cdots\int_0^xf(\x_n)\ldots
f(\x_1)\;d\x_1\ldots d\x_n\;.
\end{equation}
The last two relations show that (for functions with appropriate
asymptotic conditions) the Mellin transforms of derivatives and
integrals are symmetrical under $n\rightarrow-n$. Indeed, this is
the basis of the definition of fractional derivatives. This suggests
an application to pseudo-differential symbols of the type
$A(\x,d/d\x)=\sum_{i=-\infty}^na_i(\x)d^i/d\x^i$.

The Mellin transform is directly related to the Fourier and
(two-sided) Laplace transforms by
\begin{equation}
\mathcal{M}[f(\x);\alpha]=\mathcal{F}[f(e^{\x});-i\alpha]
=\mathcal{L}[f(e^{-\x});\alpha]\;.
\end{equation}
From these relationships, the inverse Mellin transform can be
deduced;
\begin{equation}
f(\x)\stackrel{a.\,e.}{=}\frac{1}{2\pi
i}\int_{\mathrm{c}-i\infty}^{\mathrm{c}+i\infty}\x^{-\alpha}\widetilde{f}(\alpha)\;d\alpha
\end{equation}
where $\mathrm{c}\in(\mathrm{a},\mathrm{b})$ (provided
$\widetilde{f}(\alpha)$ is integrable along the path). The almost
everywhere (a.e.) designation can be dropped if $f(\x)$ is
continuous. Moreover, if $f(\x)$ is of bounded variation about
$\x_0$, then
\begin{equation}
\frac{f(\x_0^+)+f(\x_0^-)}{2}=\lim_{T\rightarrow\infty}\frac{1}{2\pi
i}\int_{\mathrm{c}-iT}^{\mathrm{c}+iT}
\x^{-\alpha}\widetilde{f}(\alpha)\;d\alpha\;.
\end{equation}

Using the inversion formula, the Parseval relation for the Mellin
transform follows from
\begin{equation}
\int_0^\infty g(\x)h(\x)\x^{\alpha-1}\;d\x =\frac{1}{2\pi
i}\int_{\mathrm{c}-i\infty}^{\mathrm{c}+i\infty}\widetilde{g}(\alpha')
\widetilde{h}(\alpha-\alpha')\;d\alpha'\;,
\end{equation}
assuming the necessary conditions on $g(\x)$ and $h(\x)$ to allow
for the interchange of integration order. In particular,
\begin{equation}
\mathcal{M}\left[g(\x)h(\x);1\right]=\int_0^\infty g(\x)h(\x)\;d\x
=\frac{1}{2\pi
i}\int_{\mathrm{c}-i\infty}^{\mathrm{c}+i\infty}\widetilde{g}(\alpha')
\widetilde{h}(1-\alpha')\;d\alpha'\;.
\end{equation}
Similarly,
\begin{equation}
\mathcal{M}\left[g(\x)\ast
h(\x);\alpha\right]:=\int_0^\infty\int_0^\infty
g(\x')h\left(\frac{\x}{\x'}\right)\,\x^{\alpha-1}\;\frac{d\x'}{\x'}d\x
=\widetilde{g}(\alpha)\widetilde{h}(\alpha)\;,
\end{equation}
and
\begin{equation}
\mathcal{M}\left[g(\x)\star
h(\x);\alpha\right]:=\int_0^\infty\int_0^\infty
g(\x\x')h(\x')\,\x'\x^{\alpha-1}\;\frac{d\x'}{\x'}d\x
=\widetilde{g}(\alpha)\widetilde{h}(1-\alpha)\;.
\end{equation}

\subsection{Expansions}
\begin{definition}The singular expansion of a meromorphic function
$f(\z)$ with a finite set $\mathcal{P}$ of poles  is defined to be the sum of its
Laurent expansions to order $\mathcal{O}(\z^0)$ about each pole,
i.e.
\begin{equation}
f(\z)\asymp\sum_{\varepsilon\in\mathcal{P}}
\mathrm{Laur}[f(\z),\varepsilon;\mathcal{O}(\z^0)]\;.
\end{equation}
\end{definition}
\begin{theorem}\emph{(\cite[th. 3]{FL})} Let $f(\x)$ have Mellin transform
$\widetilde{f}(\alpha)$ in $\langle \mathrm{a},\mathrm{b}\rangle$.
Assume
\begin{equation}
f(\x)|_{\x\rightarrow
0^+}=\sum_{\varepsilon,k}c_{\varepsilon,k}\x^{\varepsilon}(\log\x)^k
+\mathcal{O}(\x^{\mathrm{M}})
\end{equation}
where $-\mathrm{M}<-\varepsilon\leq \mathrm{a}$ and
$k\in\mathbb{N}$. Then $\widetilde{f}(\alpha)$ can be continued to a
meromorphic function in $\langle -\mathrm{M},\mathrm{b}\rangle$,
and it has the singular expansion
\begin{equation}
\widetilde{f}(\alpha)\asymp\sum_{\varepsilon,k}c_{\varepsilon,k}
\frac{(-1)^kk!}{(\alpha+\varepsilon)^{k+1}}\;.
\end{equation}
Likewise, if
\begin{equation}
f(\x)|_{\x\rightarrow
\infty}=\sum_{\varepsilon,k}c_{\varepsilon,k}\x^{-\varepsilon}(\log\x)^k
+\mathcal{O}(\x^{-\mathrm{M}})
\end{equation}
where $\mathrm{b}\leq\varepsilon< \mathrm{M}$ and $k\in\mathbb{N}$,
then $\widetilde{f}(\alpha)$ can be continued to a meromorphic
function in $\langle \mathrm{a},\mathrm{M}\rangle$, and it has the
singular expansion
\begin{equation}
\widetilde{f}(\alpha)\asymp\sum_{\varepsilon,k}c_{\varepsilon,k}
\frac{(-1)^{k+1}k!}{(\alpha-\varepsilon)^{k+1}}\;.
\end{equation}
\end{theorem}

Conversely, it can be shown that for $\widetilde{f}(\alpha)$
meromorphic in $\langle -\mathrm{M},\mathrm{b}\rangle$ (respectively
$\langle \mathrm{a},\mathrm{M}\rangle$) whose poles lie to the right
(respectively left) of the fundamental strip, then
\begin{eqnarray}
f(\x)|_{\x\rightarrow
0^+}&=&\sum_{\varepsilon_k\in\mathcal{P}}\mathrm{Res}
\left[\widetilde{f}(\alpha)\x^{\alpha},\alpha=\varepsilon_k\right]
+\mathcal{O}(\x^{\mathrm{M}})\notag\\
f(\x)|_{\x\rightarrow
\infty}&=&-\sum_{\varepsilon_k\in\mathcal{P}}\mathrm{Res}
\left[\widetilde{f}(\alpha)\x^{-\alpha},\alpha=\varepsilon_k\right]
+\mathcal{O}(\x^{-\mathrm{M}})
\end{eqnarray}
if $f(\x)$ is at least twice differentiable. More precisely,
\begin{theorem}\emph{(\cite[th. 4]{FL})}
Let $f(\x)$ have Mellin transform $\widetilde{f}(\alpha)$ in
$\langle \mathrm{a},\mathrm{b}\rangle$. Assume
$\widetilde{f}(\alpha)$ is meromorphic in $\langle
-\mathrm{M},\mathrm{b}\rangle$ such that
\begin{equation}
\widetilde{f}(\alpha)|_{|\alpha|\rightarrow\infty}
=\mathcal{O}(|\alpha|^{-\mathrm{r}})\;,\;\;\;\;\mathrm{r}>1
\end{equation}
and
\begin{equation}
\widetilde{f}(\alpha)\asymp\sum_{k,\varepsilon}
\frac{c_{k,\varepsilon}}{(\alpha-\varepsilon)^{k+1}}\;.
\end{equation}
Then
\begin{equation}
f(\x)|_{\x\rightarrow 0+}=\sum_{k,\varepsilon}
\frac{(-1)^k}{k!}c_{k,\varepsilon}
\x^{-\varepsilon}(\log\x)^k+\mathcal{O}(\x^\mathrm{M})\;.
\end{equation}
Likewise, if $\widetilde{f}(\alpha)$ is meromorphic in $\langle
\mathrm{a},\mathrm{M}\rangle$, then
\begin{equation}
f(\x)|_{\x\rightarrow \infty}=\sum_{k,\varepsilon}
\frac{(-1)^{k+1}}{k!}c_{k,\varepsilon}
\x^{-\varepsilon}(\log\x)^k+\mathcal{O}(\x^{-\mathrm{M}})\;.
\end{equation}
\end{theorem}

\subsection{Mellin distributions} The relationship between Mellin and
Fourier transforms allows the development of Mellin distributions.
Following \cite{SZ,ZEM};
\begin{definition}\emph{\cite[\S 7]{SZ}}
Let $f_I:\R^n_+:=\{\y\in\R^n:0<\y<\infty\}\rightarrow\C$ be a
function with support $I:=\{\x\in\R^n_+:0<\x\leq \x_0$ for some
$\x_0\in\R^n_+$. Take $f_I\in L^1(\R^n_+)$ with limits
$\lim_{\x\rightarrow
0^+}f_I(\x)\rightarrow\mathcal{O}(\x^{-\mathrm{a}})$ and
$\lim_{\x\rightarrow\infty}f_I(\x)\rightarrow\mathcal{O}(\x^{-\mathrm{b}})$
for $\mathrm{a},\mathrm{b}\in\R^n$. Then the Mellin transform
$\widetilde{f}(\alpha)$ with $\alpha\in \langle
\mathrm{a},\mathrm{b}\rangle:=(\mathrm{a},\mathrm{b})\times
i\R^n\subset\C^n$ is defined by (the analytic extension
of)\footnote{The substitution $\alpha\rightarrow -\alpha$ in the
exponent of $\x$ conforms with reference \cite{SZ}.}
\begin{equation}
\widetilde{f}(\alpha):=\mathcal{M}[f_I(\x);\alpha]:=\int_{\R^n_+}
f_I(\x)\x^{-\alpha-1}d\x\;.
\end{equation}
The notation $(\mathrm{a},\mathrm{b})$ denotes a poly-interval
$\{\y\in\R^n:\mathrm{a}<\y<\mathrm{b}\}$ and
$\x^{\alpha}:=\x_1^{\alpha_1}\cdot\ldots\cdot\x_n^{\alpha_n}$.
\end{definition}

Now, the close relationship with the Fourier transform motivates the
definition
\begin{definition}\emph{\cite[\S 5]{SZ}}
Let $\mathrm{v}\in\R^n$ and define
\begin{equation}
M_{\mathrm{v}}(I):=\{\psi\in C^\infty(I):\sup_{\x\in
I}\left|\x^{\mathrm{v}+1}
\left(\x\p_\x\right)^{\lambda}\psi(\x)\right|<\infty\}
\end{equation}
where $\lambda\in\mathbb{N}_0^n$ and $\mathbb{N}_0^n$ is the set of
non-negative multi-indices. Endow $M_{\mathrm{v}}(I)$ with the
topology defined by the sequence of seminorms
\begin{equation}
\varrho_{\mathrm{v},\lambda}(\psi)=\sup_{\x\in
I}\left|\x^{\mathrm{v}+1}
\left(\x\p_\x\right)^{\lambda}\psi(\x)\right|\;.
\end{equation}
Then $M_{(\mathrm{w})}(I)$ for
$\mathrm{w}\in\R^n_\infty:=(\R\cup\{\infty\})^n$ is defined to be
the inductive limit of $M_{\mathrm{v}}(I)$, i.e.
$M_{(\mathrm{w})}(I)=\lim_{\overrightarrow{\mathrm{v}<\mathrm{w}}}
M_{\mathrm{v}(I)}$. The dual space $M'_{(\mathrm{w})}(I)$ is
comprised of Mellin distributions and the total space of Mellin
distributions is
\begin{equation}
M'(I)=\bigcup_{\mathrm{w}\in\R^n_\infty}M'_{(\mathrm{w})}(I)\;.
\end{equation}
Finally, the Mellin transform of a distribution $\mathrm{T}\in
M'_{(\mathrm{w})}(I)$ is defined by
\begin{equation}
\widetilde{\mathrm{T}}(\alpha):=\mathcal{M}\left[\mathrm{T};\alpha\right]
:=\langle\mathrm{T},\x^{-\alpha-1}\rangle\;\;\;\;,\;\;\;\;\Re(\alpha)<\mathrm{w}\;.
\end{equation}
\end{definition}
Note the topological inclusions
\begin{equation}
D(I)\subset M_{\mathrm{v}}(I)\subset M'_{(\mathrm{w})}(I)\subset
D'(I)\;,
\end{equation}
and $\widetilde{\mathrm{T}}(\alpha)$ is well-defined on the set
\begin{equation}
\Omega_{\mathrm{T}}:=\bigcup_{\{\mathrm{v}:\mathrm{T}\in
M'_{(\mathrm{v})}(I)\}}\left[\Re(\alpha)<\mathrm{v}\right]\;.
\end{equation}

\begin{theorem}\emph{\cite[\S 8, th. 1]{SZ}}
$M'(I)$ coincides with the space of distribution on $\R^n$ supported on the closure of $I$ and restricted to $\R_+^n$.
\end{theorem}

Some of the important properties of the 1-dimensional Mellin
transform have their analogues for distributions:
The Mellin transform $\widetilde{\mathrm{T}}(\alpha)$ is holomorphic
on $\Omega_{\mathrm{T}}$ and
\begin{eqnarray}
\frac{\p}{\p\alpha_i}\widetilde{\mathrm{T}}(\alpha)
&=&\langle\mathrm{T},\x^{-\alpha-1}(-\log\x_i)\rangle\notag\\
\widetilde{\mathrm{T}}(\alpha-\beta)
&=&\mathcal{M}\left[\x^\beta\mathrm{T};\alpha\right]
\hspace{.65in}\Re(\alpha)-\Re(\beta)<\mathrm{w}\notag\\
\alpha^\gamma\widetilde{\mathrm{T}}(\alpha)
&=&\mathcal{M}\left[\left(\x\p_\x\right)^\gamma\mathrm{T};\alpha\right]
\hspace{.4in}\gamma\in\mathbb{N}_0^n
\;,\;\Re(\alpha)<\mathrm{w}\notag\\
(\alpha^\gamma+1)\widetilde{\mathrm{T}}(\alpha+\gamma)
&=&\mathcal{M}\left[\left(\p_\x\right)^\gamma\mathrm{T};\alpha\right]
\hspace{.5in}|\gamma|=1\;,\;\Re(\alpha)<\mathrm{w}-\gamma\;.\notag\\
\end{eqnarray}

\section{Exponential exercises}\label{exercises}
The exponential function plays a prominent role in ordinary Mellin
transforms, so we want to develop and characterize the functional
counterpart by looking at some specific cases of reduction to finite dimensional groups.

Let $\mathrm{E}:=\mathrm{exp}_{\mathbf{F}_{\mathbb{S}}(G^\C)}:=\sum_n\frac{1}{n!}(\cdot)^n$ stand for the
exponential on $\mathbf{F}_{\mathbb{S}}(G^\C)$ defined with the product given by the $\ast$-convolution.
Suppose $\mathfrak{C}^\ast\equiv\C$, $\lambda:G^\C\rightarrow \R_+$, and
$\mathrm{A}(g)=A g$ where
 $A\in\C_+:=\R_+\times i\R$. For the standard Haar measure on $\R_+$ this is just the
usual exponential Mellin transform
\begin{equation}
\mathcal{M}_{\R_+,H}\left[\mathrm{E}^{-\mathrm{A}};\alpha\right] :=\int_{\R_+}
e^{-A g}g^{\alpha-1}\;dg=\frac{\Gamma(\alpha)}{A^\alpha}
\;,\;\;\;\;\;\alpha\in\langle0,\infty\rangle_{H}\;.
\end{equation}
In particular,
\begin{equation}
\mathcal{M}_{\R_+,H}\left[\mathrm{E}^{-\mathrm{Id}};\alpha\right]
=\Gamma(\alpha) \;,\;\;\;\;\;\alpha\in\langle0,\infty\rangle_{H}\;.
\end{equation}
As a quick exercise, use Lemma \ref{commutative} with $\mathrm{F}_1(\tilde{g}g)=e^{-\tilde{g}g}$ and $\mathrm{F}_2(\tilde{g})\rho(\tilde{g})=e^{-\tilde{g}}\tilde{g}$ to deduce (for
$\mathrm{A}=\mathrm{Id}$ and a choice of $\lambda$ corresponding to
standard normalization)
\begin{eqnarray}
\Gamma(\alpha)\Gamma(1-\alpha)&=&\mathcal{M}_{\R_+,H}
\left[\int_{\R_+}e^{-\tilde{g}}\tilde{g}\,e^{-\tilde{g}g}\;d\log \tilde{g}
;\alpha\right]\notag\\
&=&\int_0^\infty\frac{t^{\alpha-1}}{1+t}\;dt\notag\\
&=&\pi\csc(\pi\alpha)\;,\;\;\;\;\;\alpha\in\langle0,1\rangle_{H}\;.
\end{eqnarray}
Notice the reduction in the fundamental strip $\mathbb{S}_\lambda$. Simple manipulations  yield the standard results
$\pi\alpha\csc(\pi\alpha)=\Gamma(1+\alpha)\Gamma(1-\alpha)$ and
$\Gamma(1+\alpha)/\Gamma(\alpha-1)=\alpha(\alpha-1)$.

However, the \emph{functional} Mellin transform provides a mechanism
to regularize; and with a suitable choice of $\lambda$,
\begin{equation}
\mathcal{M}_{\R_+,\Gamma}\left[\mathrm{E}^{-\mathrm{A}};\alpha\right]
:=\int_{\R_+}e^{-A g}g^{\alpha}
\;d\nu(g_\Gamma)=\frac{1}{A^\alpha}
\;,\;\;\;\;\;\alpha\in\langle0,\infty\rangle_\Gamma
\end{equation}
for $\nu(g_\Gamma):=\log g/\Gamma(\alpha)=\nu(g)/\Gamma(\alpha)$
where $\nu(g)$ is the normalized Haar measure on $\R_+$. To extend the
fundamental strip to the left of the imaginary axis, one can use
\begin{eqnarray}\label{extension}
\mathcal{M}_{\R_+,\Gamma^p}\left[\mathrm{E}^{-\mathrm{A}};\alpha\right]
&:=&\frac{\Gamma(\alpha)}{\Gamma(\alpha+p)}\int_{\R_+}\left(A
g\right)^{p}e^{-A g}g^{\alpha}
\;d\nu(g_\Gamma)\notag\\
&&\hspace{-.6in}=\frac{(-1)^p}{\Gamma(\alpha+p)}
\mathcal{M}_{\R_+,H}\left[g^p\frac{d^p}{dg}
\mathrm{E}^{-\mathrm{A}};\alpha\right]\;,
\;\;\;\;\alpha\in\langle-p,\infty\rangle_{\Gamma^p},\;\;p\geq 0\;.\notag\\
\end{eqnarray}

There are other ways to extend the
fundamental strip to the left of the imaginary axis (besides analytic continuation). For example,
defining $\not\!e^{-A g}:=e^{-A g}-e^{-g}$ yields
\begin{equation}
\mathcal{M}_{\R_+,H}\left[\not\!\mathrm{E}^{-\mathrm{A}};\alpha\right]
:=\int_{\R_+}\not\!e^{-A g}g^{\alpha}
\;d\nu(g_{H})=\Gamma(\alpha)\left(\frac{1}{A^\alpha}-1\right)\;,
\;\;\;\;\;\alpha\in\langle-1,\infty\rangle_{H}\;.
\end{equation}
For $\lim\alpha\rightarrow0^+$ this gives
$\int_{\R_+}\not\!e^{-A g} \;d\nu(g_H)=-\log A$,
and therefore (in this case)
\begin{equation}
\left.\mathcal{M}_{\R_+,H} \left[\not\!\mathrm{E}^{-\mathrm{A}};\alpha\right]\right|_{\alpha\rightarrow0^+}
=\left.\frac{d}{d\alpha}\mathcal{M}_{\Gamma}
\left[\mathrm{E}^{-\mathrm{A}};\alpha\right]\right|_{\alpha\rightarrow0^+}
\end{equation}
which suggests the definition
\begin{eqnarray}
\left.\widehat{\mathcal{M}}_{\lambda}
\left[\mathrm{F};\alpha\right]\right|_{\alpha\rightarrow0^+}
&=:&\left.\frac{d}{d\alpha}\mathcal{M}_{\lambda}
\left[\mathrm{F};\alpha\right]\right|_{\alpha\rightarrow0^+}\notag\\
&=:&\left.\int_{G^\C}
\mathrm{F}(g)g^\alpha\,\mathrm{log}_\lambda\,g \;\mathcal{D}_\lambda g\right|_{\alpha\rightarrow0^+}\notag\\
&=:&\left.\int_{G^\C}
\mathrm{F}(g)g^\alpha\;\widehat{\mathcal{D}}_\lambda g\right|_{\alpha\rightarrow0^+}
\end{eqnarray}
if the limit exists. In particular, for $\lambda:G^\C\rightarrow G^\C_\lambda$,
\begin{equation}\label{log measure}
g^\alpha\;\widehat{\mathcal{D}}_\lambda g\stackrel{\lambda}{\rightarrow}\frac{d}{d\alpha}\,g^{\alpha}d\nu(g_\lambda)=:g^\alpha d\widehat{\nu}(g_\lambda)\;.
\end{equation}
For example, choosing ${\nu}(g_\Gamma)$  yields $d\widehat{\nu}(g_\Gamma)=(\log g-\psi(\alpha))\; d\nu(g_\Gamma)$ where $\psi(\alpha)=\Gamma'(\alpha)/\Gamma(\alpha)$.
This motivates Definition \ref{functional log} for functional Log.

Moving on to the non-abelian case, suppose $\lambda:G^\C\rightarrow
GL(n,\C)_+:=SL(n,\C)\times\R_+$ and $\mathfrak{C}^\ast\equiv\C$. Define the
functional $\mathrm{E}^{-\mathrm{Tr}\,\mathrm{A}}:G^\C\rightarrow\C$ by
$g\mapsto e^{-\mathrm{tr}\,(A\cdot g)}$ with $A\in
GL(n,\C)_+$, and take
$\rho:G^\C\rightarrow\C_+$ by $g\mapsto\det g$. Then
\begin{eqnarray}
\mathcal{M}_{GL(n,\C)_+,\Gamma_n}\left[\mathrm{E}^{-\mathrm{Tr}\,\mathrm{A}};\alpha\right]
&=&\int_{GL(n,\C)_+}e^{-\mathrm{tr}(A
g)}\det g^{\alpha} d\nu(g_{\Gamma_n})\;,
\;\;\alpha\in\mathbb{S}_{\Gamma_n}\notag\\
&=&\int_{GL(n,\C)_+}e^{-\mathrm{tr}(A
g)}\left(\det{g}\right)^{\alpha}e^{i\varphi(\alpha)} d\nu(g_{\Gamma_n})\;,
\;\;\alpha\in\mathbb{S}_{\Gamma_n}\notag\\
&=&\int_{GL(n,\C)_+}e^{-\mathrm{tr}
(g)}\left(\det A^{-1}\det g\right)^{\alpha}e^{i\varphi(\alpha)}  d\nu(g_{\Gamma_n})\;,
\;\;\alpha\in\mathbb{S}_{\Gamma_n}\notag\\
&=&\det A^{-\alpha}\;,
\;\;\alpha\in\mathbb{S}_{\Gamma_n}
\end{eqnarray}
where $\varphi(\alpha)$ is a phase, $\nu(g_{\Gamma_n}):=\nu(g)/\Gamma_n(\alpha)$ with $\nu(g)$ the
Haar measure on $GL(n,\C)_+$, and $\Gamma_n(\alpha)$ a complex
multi-variate gamma function defined by
\begin{eqnarray}
\Gamma_n(\alpha)
:=\mathcal{M}_{GL(n,\C)_+,H}\left[\mathrm{E}^{-\mathrm{Tr}\,\mathrm{Id}};\alpha\right]
&:=&\int_{GL(n,\C)_+}e^{-\mathrm{tr}(g)}
\det g^{\alpha}\;d\nu(g),
\;\;\;\;\;\alpha\in\mathbb{S}_{H}\;.\notag\\
\end{eqnarray}
In particular, if $\alpha=1\in\mathbb{S}_{\Gamma_n}$, then
\begin{equation}
\mathcal{M}_{GL(n,\C)_+,\Gamma_n}\left[\mathrm{E}^{-\mathrm{Tr}\,\mathrm{A}};1\right]
=\int_{GL(n,\C)_+}e^{-\mathrm{tr}(A g)}
\det g\;d\nu(g_{\Gamma_n})=(\mathrm{det}A)^{-1}
\end{equation}

Remark that $\Gamma_n(\alpha)$ is not a well-defined object unless one restricts to a compact subgroup of $GL(n,\C)_+$. Otherwise, the price of extracting $\det A^{-\alpha}$ from the integral comes with the price of regularizing this possibly singular normalization.

Generalizing further, suppose $\lambda:G^\C\rightarrow GL(n,\C)_+$ but now
$\mathfrak{C}^\ast\equiv L_B(\C^n)$ the space of bounded linear maps on $\C^n$ and
$\mathrm{E}^{-\mathrm{A}}:G^\C\rightarrow L_B(\C^n)$ by
$g\mapsto e^{-a\cdot g}$ with $a\in GL(n,\C)_+$ and $\rho(a)=A\in L_B(\C^n)$. The Haar
normalized functional Mellin transform yields
\begin{eqnarray}
\mathcal{M}_{GL(n,\C)_+,H}\left[\mathrm{E}^{-\mathrm{A}};\alpha\right]
&:=&\int_{GL(n,\C)_+}e^{-A \rho(g)}\,\rho({g}^{\alpha})
\;d\nu(g)\;, \;\;\;\;\;\alpha\in\mathbb{S}_{H}\notag\\
&=&\int_{GL(n,\C)_+}e^{-\rho(g)}\,\rho((A^{-1}{g})^{\alpha})
\;d\nu(g)\;, \;\;\;\;\;\alpha\in\mathbb{S}_{H}\notag\\
&=:&{A}_{H}^{-\alpha}\;,
\;\;\;\;\;\alpha\in\mathbb{S}_{H}
\end{eqnarray}
which defines the element
$A_{H}^{-\alpha}\in M_s(\mathfrak{C}^\ast)$ for $\alpha\in\mathbb{S}_{H}$.

Unless $a$ is in the center of $GL(n,\C)_+$ or we restrict to a subgroup of $GL(n,\C)_+$, this can't be reduced further without explicit computation, i.e. ${A}_{H}^{-\alpha}\neq(A)^{-\alpha}$ in general. However, various restrictions allow for various degrees of simplification. For example, if $A$ is self-adjoint and $G^\C$ is restricted to a one-parameter subgroup generated by $\log a\in\mathfrak{gl}(n,\C)$, then more can be done. So let us take the subgroup $\phi_{\log a}(\R)\leq G^\C$. Then, since $a\in\phi_{\log a}(\R)$,
\begin{eqnarray}
\mathcal{M}_{\lambda}[\mathrm{E}^{-\mathrm{A}};\alpha]
&=&\int_{\phi_{\log a}(\R)}e^{-A \rho(g)}\rho({g}^{\alpha})
\;\mathcal{D}_\lambda g\;, \;\;\;\;\;\alpha\in\mathbb{S}_{\Gamma}\notag\\
&=&\int_{\phi_{\log a}(\R)}e^{-\rho(g)}\rho((a^{-1}{g})^{\alpha})
\;\mathcal{D}_\lambda g\;, \;\;\;\;\;\alpha\in\mathbb{S}_{\Gamma}\notag\\
&=&A^{-\alpha}\int_{\phi_{\log a}(\R)}e^{-\rho(g)}\rho({g}^{\alpha})
\;\mathcal{D}_\lambda g\,\;, \;\;\;\;\;\alpha\in\mathbb{S}_{\Gamma}\notag\\
&=:&A^{-\alpha}{\mathcal{N}}_\Gamma(\alpha)\;, \;\;\;\;\;\alpha\in\langle0,\infty\rangle_\Gamma
\end{eqnarray}
The second line follows from the left-invariance of the Haar measure, and the third line from the fact that $a$ and $g$ commute and $\rho$ is a representation. Most of the time we will absorb the normalization ${\mathcal{N}}_\Gamma(\alpha)$ into the measure.

\section{Commuting Mellin and the exponential}\label{sec. key theorem}

The fundamental relationship between exponentials, determinants, and
traces in finite dimensions, i.e. $\exp\mathrm{tr} M=\det \exp M$,
also characterizes the functional analogs. In the functional context, consider $\mathrm{F}=-\mathrm{Log}\,\mathrm{A}$. Then formally, for
$\mathrm{F}\in\mathbf{F}_{\mathbb{S}}(G^\C)$ and suitable $\lambda$,
\begin{equation}\label{important relation}
e^{\mathrm{tr}\,\mathrm{log}\,{A}^{-1}_\lambda} \sim
e^{\mathrm{tr}\int_{G^\C}e^{-\mathrm{A}(g)}\log_{\lambda} g\;\mathcal{D}_\lambda g}
 \,\sim\,
\int_{G^\C}  e^{-\mathrm{tr}(\mathrm{A}(g))}\det g
\;\mathcal{D}_\lambda g =\mathrm{det}\,A_\lambda^{-1}\;.
\end{equation}
This important relation represents a deep connection between Poisson processes and functional Mellin transforms/gamma
integrators (as indicated for example in \cite{LA1}). It is a particular case of the following theorem:
\begin{theorem}\label{key theorem}
Let $\mathrm{E}^{-\mathrm{E}^{-\mathrm{A}}}\in \mathrm{Mor}_C(G^\C,\mathfrak{C}^\ast)$  by  $g\mapsto e^{-{(e^{-\mathrm{A}(g)})}}$ and $\mathrm{E}^{-\mathrm{Tr}\,\mathrm{E}^{-\mathrm{A}}}\in \mathrm{Mor}_C(G^\C,\mathfrak{C}^\ast)$  by  $g\mapsto e^{-{\mathrm{tr}\,(e^{-\mathrm{A}(g)})}}$ with $e^{-\mathrm{A}(g)}:=\sum_n\frac{(-\mathrm{A}(g))^n}{n!}$ trace class.\footnote{One must prescribe ${(\mathrm{A}(g))}^n$. Three examples are $(\mathrm{A}(g))^n=A^n\rho(g)^n$ and $(\mathrm{A}(g))^n=\int_0^{\rho(g)}A^n(s)\,ds$ and $(\mathrm{A}(g))^n=\int_0^{\rho(g)}\int_0^{s_{n-1}}\cdots\int_0^{s_2}
\int_0^{s_1}\,A^n(s_1)A^n(s_2)\cdots A^n(s_n)\;ds_1\,ds_2\, \cdots ds_n$ where $A^n\in\mathfrak{C}^\ast$.} If $\mathrm{Tr}\,\mathrm{E}^{-\mathrm{A}}$ and $\mathrm{E}^{-\mathrm{Tr}\,\mathrm{E}^{-\mathrm{A}}}$ are Mellin integrable for a common domain
$\alpha\in\mathbb{S}_\lambda$, then
\begin{equation}
e^{-\mathrm{tr}\,A_\lambda^{-\alpha}} =\mathrm{det}\,
(\mathrm{E}^{-\mathrm{A}})_\lambda^{-\alpha}\;,\;\;\;\;\;\alpha\in\mathbb{S}_\lambda\;.
\end{equation}
\end{theorem}

\emph{Proof}: First, for suitable functionals $\mathrm{F}$, an
immediate consequence of the definitions and the relationship
between $(\mathrm{exp}\,\mathrm{tr})$ and
$(\det\,\mathrm{exp})$ in $\mathfrak{C}^\ast$ is
\begin{eqnarray}
\mathcal{M}_{\lambda}
\left[\mathrm{Det}\,\mathrm{E}^{-\mathrm{F}};\alpha\right]
&=&\int_{G^\C}\det\left(e^{-\mathrm{F}(g)
}g^\alpha\right)\,\mathcal{D}_\lambda g\notag\\
&=&\int_{G^\C} e^{-\mathrm{tr}\,\mathrm{F}(g)}{\det}\, g^\alpha\,\mathcal{D}_\lambda g\notag\\
&=&\mathcal{M}_{\lambda}
\left[\mathrm{E}^{-\mathrm{Tr}\,\mathrm{F}};\alpha\right]
\end{eqnarray}
where the second line follows as soon as $\mathrm{F}(g)$ is
trace class.

\begin{lemma}\label{key lemma}
Assume $\mathrm{Tr}\, \widetilde{\mathrm{F}}\in \mathrm{Mor}_C(G^\C,\C)$ with $\widetilde{\mathrm{F}}(g)$ analytic. Suppose the
functional Mellin transforms of $\,\mathrm{Tr}\, \widetilde{\mathrm{F}}$ and
$\mathrm{E}^{-\mathrm{Tr}\, \widetilde{\mathrm{F}}}$ exist for common $\alpha\in\mathbb{S}_\lambda$ for a given $\lambda$. Then
\begin{equation}
e^{-\mathcal{M}_{\lambda}\left[\, \mathrm{Tr}\,
\widetilde{\mathrm{F}};\alpha\right]}
=\mathcal{M}_{\lambda}\left[\mathrm{E}^{-\mathrm{Tr}\,
\widetilde{\mathrm{F}}};\alpha\right]\,,\;\;\;\;\;\alpha\in\mathbb{S}_\lambda\;.
\end{equation}
\end{lemma}
\emph{proof}: For $\alpha\in\mathbb{S}_\lambda$, the Mellin transform of $\mathrm{Tr}\,
\widetilde{\mathrm{F}}$ exists by assumption so the integral of $\mathrm{Tr}\,
\widetilde{\mathrm{F}}$ is holomorphic and converges absolutely. Hence,
$e^{-\mathcal{M}_{\lambda} \left[\mathrm{Tr}\, \widetilde{\mathrm{F}};\alpha\right]}$
represents an absolutely convergent series for $\alpha\in\mathbb{S}_\lambda$. Also, recall $\mathrm{E}^{-\mathrm{Tr}\,
\widetilde{\mathrm{F}}}=\sum_n\frac{(-1)^n}{n!}(\mathrm{Tr}\,
\widetilde{\mathrm{F}})^n$ where $(\cdot)^n$ denotes the $n$-fold $\ast$-product in $\mathbf{F}_{\mathbb{S}_{\mathcal{R}}}(G^\C)$. Therefore,
\begin{eqnarray}
e^{-\mathcal{M}_{\lambda} \left[\mathrm{Tr}\, \widetilde{\mathrm{F}};\alpha\right]}
=\sum_{n=0}^\infty\frac{(-1)^n}{n!}\mathcal{M}_{\lambda}
\left[\mathrm{Tr}\, \widetilde{\mathrm{F}};\alpha\right]^n
&=&\sum_{n=0}^\infty\frac{(-1)^n}{n!}\mathcal{M}_{\lambda}
\left[(\mathrm{Tr} \,\widetilde{\mathrm{F}})^n;\alpha\right]\notag\\
&=&\lim_{N\rightarrow\infty}\mathcal{M}_{\lambda}
\left[\sum_{n=0}^N\frac{(-1)^n}{n!}(\mathrm{Tr}\,
\widetilde{\mathrm{F}})^n;\alpha\right]\notag\\
&=&\mathcal{M}_{\lambda}\left[\mathrm{E}^{-\mathrm{Tr}\,
\widetilde{\mathrm{F}}};\alpha\right]\;.
\end{eqnarray}
Moving the power of $n$ into the Mellin transform in the first line follows from induction using Lemma \ref{commutative}.\footnote{Of course the nature of $G^\C$ and $\mathfrak{C}^\ast$ may severely restrict $\alpha\in\mathbb{S}_\lambda$. For example, $\alpha=1$ for non-abelian $G^\C$ and non-commutative $\mathfrak{C}^\ast$.} The second line is obvious from the linearity of functional Mellin. The last line follows from analytic $e^{-\mathrm{tr}\,\widetilde{\mathrm{F}}(g)}$ and the assumption that
$\mathrm{E}^{-\mathrm{Tr}\,
\widetilde{\mathrm{F}}}$ is Mellin integrable. Together these imply that $\lim_{N\rightarrow\infty}\sum_{n=0}^N\frac{(-1)^n}{n!}(\mathrm{tr}\,
\widetilde{\mathrm{F}}(g))^n\rightarrow e^{-\mathrm{tr}\,
\widetilde{\mathrm{F}}(g)}$ point-wise and $|\sum_{n=0}^N\frac{(-1)^n}{n!}(\mathrm{tr}\,
\widetilde{\mathrm{F}}(g))^n|\leq e^{-\Re(\mathrm{tr}\,
\widetilde{\mathrm{F}}(g))}$ such that $e^{-\Re(\mathrm{tr}\,
\widetilde{\mathrm{F}}(g))}$ is integrable. Hence, by Lebesgue Dominated Convergence, $\lim_{N\rightarrow\infty}||\sum_{n=0}^N\frac{(-1)^n}{n!}(\mathrm{tr}\,
\widetilde{\mathrm{F}}(g))^n-e^{-\mathrm{tr}\,
\widetilde{\mathrm{F}}(g)}||_{1,\lambda}\rightarrow0$ for all $\lambda\in\Lambda$. Therefore, $\lim_{N\rightarrow\infty}||\sum_{n=0}^N\frac{(-1)^n}{n!}(\mathrm{Tr}\,
\widetilde{\mathrm{F}})^n-\mathrm{E}^{-\mathrm{Tr}\,
\widetilde{\mathrm{F}}}||_{\mathbf{F}}\rightarrow0$. We stress the lemma holds
\emph{only} for $\alpha\in\mathbb{S}_\lambda$ and it can certainly happen that $\mathbb{S}_\lambda=\emptyset$. $\qED$

To finish the proof, put $\widetilde{\mathrm{F}}\equiv\mathrm{E}^{-\mathrm{A}}$ in the lemma to get;
\begin{equation}
\mathcal{M}_{\lambda}\left[\mathrm{E}^{-\mathrm{Tr}\,
{\mathrm{E}^{-\mathrm{A}}}};\alpha\right]=\int_{G^\C} e^{-\mathrm{tr}\left(e^{-\mathrm{A}(g)}\right)}\,\det g^\alpha\,\mathcal{D}_\lambda g
=\left(\mathrm{Det}\,{(\mathrm{E}^{-\mathrm{A}})}\right)^{-\alpha}_\lambda
=\mathrm{det}\,(\mathrm{E}^{-\mathrm{A}})_\lambda^{-\alpha}
\end{equation}
and
\begin{equation}
\mathcal{M}_{\lambda} \left[\mathrm{Tr}\,\mathrm{E}^{-\mathrm{A}};\alpha\right]
=\int_{G^\C}\mathrm{tr}\left(e^{-\mathrm{A}(g)}g^{\alpha}\right)\,\mathcal{D}_\lambda g
=\left(\mathrm{Tr}\,\mathrm{A}\right)^{-\alpha}_\lambda
=\mathrm{tr}\,A_\lambda^{-\alpha}\;.
\end{equation}
$\QED$

\begin{corollary}\label{key corollary}
Under the conditions of \emph{Lemma \ref{key lemma}}, replace
$\mathrm{Tr}\, \widetilde{\mathrm{F}}$ with
$\mathrm{V}\in \mathrm{Mor}_C(G^\C,\mathfrak{C}^\ast)$
where now $\mathfrak{C}^\ast=L_H(\mathcal{H})$ is the (commutative) algebra of Hermitian linear operators on some Hilbert space. Then \emph{Lemma \ref{commutative}} and \emph{Theorem \ref{prop 4.2}} imply
\begin{equation}
\mathcal{M}_{\lambda}\left[\mathrm{E}^{-\mathrm{V}};\alpha\right]
=e^{-\mathcal{M}_{\lambda}\left[\mathrm{V};\alpha\right]}
\,,\;\;\;\;\;\alpha\in\mathbb{S}_\lambda\;.
\end{equation}
\end{corollary}
In particular, if $\mathrm{A}(g)=Ag$ is self-adjoint, the corollary implies
\begin{equation}
 (\mathrm{E}^{-\mathrm{A}})^{-\alpha}_\lambda=\mathcal{M}_{\lambda}
[\mathrm{E}^{-\mathrm{E}^{-\mathrm{A}}};\alpha]=e^{-\mathcal{M}_{\lambda}
[\mathrm{E}^{-\mathrm{A}};\alpha]}=e^{-A_\lambda^{-\alpha}}\;,
\end{equation}
and then Proposition \ref{funtional determinant} leads to $e^{-\mathrm{tr}\,A_\lambda^{-\alpha}}
=\mathrm{det}\,e^{-A_\lambda^{-\alpha}}$ when $\alpha\in\mathbb{S}_\lambda$.

\begin{remark}
According to the corollary, when $\mathrm{V}$ is exponential-like we can interpret functional Mellin and the exponential map as commuting operations. To see this, observe that the exponential map $\mathrm{Exp}$ (in distinction to the functional $\mathrm{E}^{-\mathrm{A}}$) can be viewed as an algebra endomorphism $\mathrm{Exp}:\mathbf{F}_{\mathbb{S}}(G^\C)\rightarrow\mathbf{F}_{\mathbb{S}}(G^\C)$ defined by $\mathrm{Exp}(-\mathrm{A}):=\mathrm{E}^{-\mathrm{A}}=\sum_n\frac{(-1)^n}{n!}(\mathrm{A})^n$. Then the corollary, which is a consequence of the fact that functional Mellin is a $\ast$-homomorphism, gives  $(\mathrm{Exp}(-\mathrm{A}))_\lambda^{-\alpha}=\mathrm{exp}(-(\mathrm{A})^{-\alpha}_\lambda)
=\mathrm{exp}(-{A}^{-\alpha}_\lambda)$ where $\mathrm{exp}(-{A}^{-\alpha}_\lambda):=\sum_n\frac{(-1)^n}{n!}({A}_\lambda^{-\alpha}))^n$. In other words, $\mathcal{M}_{\lambda}\circ\mathrm{Exp}=\mathrm{exp}\circ\mathcal{M}_{\lambda}$.
\end{remark}

\begin{example}
Continuing with our prevailing example; consider the one-parameter subgroup $\phi_{\mathfrak{a}}(\R)$, localization $\lambda_{\R_\pm}:\phi_{\mathfrak{a}}(\R)\rightarrow\R_+\cup\R_-$, and $\mathrm{A}(g)=Ag$ with $A\in\mathfrak{C}^\ast$ self-adjoint. Let $M_\lambda=A_\lambda^{-1}$ and choose a suitable normalization so that $M_\lambda{\rightarrow}M=A^{-1}$. Then
\begin{equation}
e^{-\zeta_{{M}^{-1}}(1)}
\stackrel{\lambda_{\R_\pm}}{\longleftarrow}e^{-\mathrm{tr}\,M_\lambda}
=e^{-\mathcal{M}_\lambda[\mathrm{Tr}\,\mathrm{E}^{\,-(\mathrm{M}^{-1})};1]}
=\mathrm{det}\, e^{-\,M_\lambda}
\stackrel{\lambda_{\R_\pm}}{\longrightarrow}\mathrm{det}\,e^{-M}\;.
\end{equation}
Similarly, take $\mathrm{log}M^{-1}_\lambda=A^{-1}_\lambda$, choose a suitable normalization so that $M_\lambda{\rightarrow}M$ with $M$ positive definite,  and use \emph{(\ref{trace/log})} to get (if the limit exists)
\begin{equation}\label{standard expression}
e^{-{\zeta}'_{{M}}(0)}
\stackrel{\lambda_{\R_\pm}}{\longleftarrow}
e^{-\widehat{\mathcal{M}}_\lambda[\mathrm{Tr}\,\mathrm{E}^{-\mathrm{M}};0]}
=e^{-(\mathrm{Tr}\,\mathrm{Log}\,\mathrm{M})^{-1}_\lambda}
=e^{-{\mathcal{M}}_\lambda[\mathrm{Tr}\,\mathrm{E}^{-\mathrm{Log}\,\mathrm{M}};1]}
=\mathrm{det}\,e^{-\mathrm{log}\,M^{-1}_\lambda}
\stackrel{\lambda_{\R_\pm}}{\longrightarrow}
\det M\;.
\end{equation}

Evidently the theorem reproduces the standard expressions for this special case.\emph{\cite{RS,VOR}} Moreover, the \textbf{functional} relation $e^{-\widehat{\mathcal{M}}_\lambda[\mathrm{Tr}\,\mathrm{E}^{-\mathrm{M}};0]}
=\mathrm{det}\,e^{-\mathrm{log}\,M^{-1}_\lambda}$  and the subsequent `localized' \textbf{function} relation $\exp\mathrm{tr}(\log M)=\det M$ expressed in \emph{(\ref{standard expression})} are consistent with the conclusion in \emph{Example \ref{replica trick}} that the $\log$ of a Mellin transform can be represented as the topological localization of a functional $\mathrm{Log}$.

\end{example}

We emphasize that Theorem \ref{key theorem} and the above properties should be understood as a family of statements at
the functional level that \emph{may} be explicitly realized only for
appropriate choices of $\lambda$ leading to non-empty $\mathbb{S}_\lambda$.

\section{Relation to crossed products}\label{relation to crossed products}
The ingredients necessary to define crossed products of $C^\ast$-algebras\cite{W} are: i) a ``dynamical system'' $(A,G,\varepsilon)$ where $A$ is a $C^\ast$-algebra, $G$ is a locally compact group, and $\varepsilon:G\rightarrow Aut(A)$ is a continuous homomorphism; ii) some Hilbert space $\mathcal{H}$; iii) an algebra representation $\varpi:A\rightarrow L_B(\mathcal{H})$; and iv) a unitary, group representation $U:G\rightarrow U(\mathcal{H})$. The two representations are required to satisfy the ``covariance condition''
\begin{equation}\label{covariance condition}
\varpi(\varepsilon_g(a))=U_g\varpi(a)U_g^\ast\;,\;\;\;\;g\in G\;,\;\;a\in A\;.
\end{equation}
With these objects, a $\ast$-representation on $\mathcal{H}$ of $C_c(G,A)$ (continuous compact morphisms $\mathrm{f}:G\rightarrow A$) is supplied by the integral
\begin{equation}\label{crossed product}
\varpi\rtimes U(\mathrm{f}):=\int_G\varpi(\mathrm{f}(g))U_g\;d\mu(g)
\end{equation}
where $\mathrm{f}\in C_c(G,A)$ and $\mu$ is a Haar measure on $G$.

A product and involution are introduced on $C_c(G,A)$ according to
\begin{equation}
(\mathrm{f}_1\ast \mathrm{f}_2)(g):=\int_G \mathrm{f}_1(\tilde{g})\varepsilon_{\tilde{g}}(\mathrm{f}_2(\tilde{g}^{-1}g))\;d\mu(\tilde{g})
\end{equation}
and
\begin{equation}
\mathrm{f}^\ast(g):=\Delta(g^{-1})\varepsilon_g(\mathrm{f}(g^{-1})^\ast)
\end{equation}
where $\Delta$ is the modular function on $G$. Completion of $C_c(G,A)$ with respect to the norm $\|\mathrm{f}\|:=\mathrm{sup}\|\varpi\rtimes U(\mathrm{f})\|$
is a $C^\ast$-algebra called the crossed product denoted by $A\rtimes_\varepsilon G$.

The crucial property of this construction is a one-to-one correspondence between non-degenerate covariant representations associated with $(\varpi,U)$ and non-degenerate representations of  $A\rtimes_\varepsilon G$ which preserves direct sums, irreducibility, and equivalence. So the $C^\ast$-algebra $A\rtimes_\varepsilon G$ can be used to model the $C^\ast$-algebra encoded in the system $(A,G,\varepsilon)$ endowed with a covariant representation $(\varpi,U)$. We recognize the covariant condition as an algebra automorphism by a group element; which, in particular, for the evolution operator in quantum mechanics becomes the integrated Heisenberg equation.

Let's compare with functional Mellin. Suppose $\lambda:G^\C\rightarrow G_\lambda$. Identify  $\pi\circ\rho\equiv U$ (with suitable restrictions if necessary) and choose $\mathcal{D}_\lambda g\equiv d\mu(g)$ with $g\in G_\lambda$, then
\begin{equation}
\pi\left(\mathcal{M}_\lambda[\mathrm{F},1]\right)=\int_{G_\lambda} \pi(f(g))U(g)\;d\mu(g)
\end{equation}
 where $\pi:\mathfrak{C}^\ast\rightarrow L_B(\mathcal{H})$. As soon as $\pi(f(g))$ is Mellin integrable w.r.t. $G_\lambda$, this integral and the integral in  (\ref{crossed product}) represent the same object in $L_B(\mathcal{H})$ iff $\varpi\circ\mathrm{f}\equiv\pi\circ f$. Keep in mind that the nature of $\mathrm{f}\in C_C(G,A)$ versus $f\in C_C(G_\lambda,\mathfrak{C}^\ast)$ is quite dependent on the nature of $A$ versus $\mathfrak{C}^\ast$: If they are both simultaneously commutative or non-commutative, then $\mathrm{f}$ and $f$ at least have the chance of representing the same object if $A$ and $\mathfrak{C}^\ast$ are isomorphic. Otherwise, they are distinctly different. Mathematically, we can always choose $A\equiv\mathfrak{C}^\ast$ and $\varpi\equiv\pi$. In this case, the difference between crossed products and functional Mellin is that $A\rtimes_\varepsilon G$ is the $C^\ast$-algebra of $\mathrm{f}\in C_C(G,A)$ satisfying the \emph{covariance condition} (\ref{covariance condition}) while $\mathbf{F}_{\mathbb{S}}(G^\C)$  is the $C^\ast$-algebra of \emph{equivariant} $f\in C_C(G_\lambda,A)$.

For application to quantum physics, the pivotal point in this difference comes down to $\varepsilon:G\rightarrow Aut(A)$ and dynamics. Suppose $A\equiv\mathfrak{C}^\ast$ is commutative. By Gelfand duality, there is some topological space $X$ such that $A\equiv C_0(X)$ (the algebra of complex valued continuous morphisms on $X$ vanishing at infinity). Non-trivial $\varepsilon$ reflects a basic assumption about the dynamical system; that $G$ acts on $X$ and this is accounted for by $\varepsilon_h(\mathrm{f}(g))(x)=\mathrm{f}(g)(h^{-1}\cdot x)$ for $x\in X$. But then the covariance condition is required to encode dynamics through the adjoint action on $L_B(\mathcal{H})$. Insofar as crossed-product quantization (virtually always) starts with a classical ``dynamical system'' $(C_0(X),G,\varepsilon)$ with covariant representation $(\varpi,U)$, the crossed product $A\rtimes_\varepsilon G$ realizes a concrete quantization of the commutative algebra $C_0(X)$. On the other hand, for functional Mellin the group is already contained in $\mathfrak{A}$ by construction, and so it acts by inner automorphisms which automatically incorporates the covariance condition. Moreover, if $\mathfrak{C}^\ast\equiv A$ is assumed to act on some $X$, then by equivariance $f(gh)(x)=f(g)\rho(h)(x)=f(g)(h^{-1}\cdot x)$. However, the involution and product that are defined for functional Mellin do not depend on $\varepsilon$ --- unlike $A\rtimes_\varepsilon G$.  Evidently, even though $\mathbf{F}_{\mathbb{S}}(G^\C)$ is a $C^\ast$-algebra it is not isomorphic to $A\rtimes_\varepsilon G$ in general, and it would be difficult to attach a classical interpretation to the functions $f\in C_C(G_\lambda,A)$ in relation to some dynamical system.

Now suppose $A\equiv\mathfrak{C}^\ast$ is \emph{non-commutative} and $G$ is its group of units for some dynamical system. Then $G$ acts on $A$ by inner automorphisms which means the covariance condition is automatic and $\varepsilon$ is unneeded. Setting $\varepsilon\equiv Id$ brings the product and involution of crossed products $A\rtimes_{Id} G$ into agreement with functional Mellin for $\alpha=1$. But then, the only way (it seems) to save the non-commutative $C^\ast$-algebra structure of $A\rtimes_{Id} G$ is to insist that the morphisms $\mathrm{f}$ be equivariant. In this situation $\mathbf{F}_{\mathbb{S}}(G^\C)\cong A\rtimes_{Id} G$ and representations furnished by $\pi\left(\mathcal{M}_\lambda[\mathrm{F},1]\right)$ are in one-to-one relation to $A\rtimes_{Id} G$ and therefore in one-to-one relation to the system $(A,G,Id)$ with covariant representation $(\varpi\equiv\pi,U)$ and equivariant $\mathrm{f}$. But note this dynamical system in not classical --- until expectations are taken. Whereas the previous paragraph described the quantization process $\mathrm{classical}\rightarrow\mathrm{quantum}$; this paragraph describes $\mathrm{quantum}\rightarrow\mathrm{classical}$.

But for $\mathrm{quantum}\rightarrow\mathrm{classical}$ how can we know anything about the functions $f\equiv\mathrm{f}$ without $C_0(X)$? Happily, spectral theory allows to represent $(\pi\circ f)(g)$ in terms of an operator valued function $\hat{f}(\rho(g))$ and Mellin integrators supply the resolvent of $\rho(g)$. If we don't venture outside of $A$ to find evolution operators and we use functional calculus to represent $\pi(f(g))\equiv \hat{f}(\rho(g))$, then functional Mellin and a choice of $G^\C$ fully determine a quantum system. That is, once we settle on $G^\C$ and find relevant representations and their furnishing Hilbert spaces, functional Mellin defines a $C^\ast$-algebra $\mathbf{F}_{\mathbb{S}}(G^\C)$ that contains quantum dynamics, and Mellin integrators furnish representations of this algebra in $L_B(\mathcal{H})$.

To further highlight the similarity between crossed products and functional Mellin, extend the integrated form of $(\varpi, U)$ to $G^\C$ according to
\begin{equation}
\varpi\rtimes U^{(\alpha)}(\mathrm{f}):=\int_{G^\C}\varpi(\mathrm{f}(g))U_{g^\alpha}\;d\mu(g)\;.
\end{equation}
Likewise, extend the involution $\mathrm{f}^\ast(g^{1+\alpha}):=\Delta(g^{-1})\varepsilon_{g^\alpha}(\mathrm{f}(g^{-(1+\alpha)})^\ast)$ and define the $\ast$-product
\begin{equation}
(\mathrm{f}_1\ast \mathrm{f}_2)(g^\alpha):=\int_{G^\C} \mathrm{f}_1(\tilde{g})\varepsilon_{\tilde{g}}(\mathrm{f}_2(\tilde{g}^{-1}g^\alpha))\;d\mu(\tilde{g})\;.
\end{equation}
Then we claim $A\rtimes_{\varepsilon}G^\C$ (after completion w.r.t. to a suitable norm) is a $C^\ast$-algebra and $\varpi\rtimes U^{(\alpha)}$ is a $\ast$-homomorphism because:
\begin{itemize}
\item
\begin{equation}
(\mathrm{f}_1\ast\mathrm{f}_2)\ast\mathrm{f}_3=\mathrm{f}_1(\ast\mathrm{f}_2\ast\mathrm{f}_3)
\end{equation}
follows immediately from the star product using the invariance of the Haar measure.
  \item\begin{equation}
  \left(\mathrm{f}^\ast(g^{1+\alpha})\right)^\ast
  =(\Delta(g^{-1})^\ast\varepsilon_{g^\alpha}(\mathrm{f}(g^{-(1+\alpha)}))
  =\mathrm{f}(g^{1+\alpha})
\end{equation}
where the first equality follows from the covariance conditions and the second from the invariance of the Haar measure which implies $\mathrm{f}(g^{(1+\alpha)})=\Delta(g^{-1})^\ast\varepsilon_{g^\alpha}(\mathrm{f}(g^{-(1+\alpha)}))$.
\item
\begin{eqnarray}
\|\mathrm{f}^\ast\|_{\alpha}&:=&\int_{G^\C}\|\mathrm{f}^\ast(g^{1+\alpha})\| \;d\mu(g)\notag\\
&=&\int_{G}\|\Delta(g^{-1})\varepsilon_{g^\alpha}(\mathrm{f}(g^{-(1+\alpha)})^\ast)\| \;d\mu(g)\notag\\
&=&\int_{G^\C}\|\Delta(g^{-1})^\ast
\left(\varepsilon_{g^\alpha}(\mathrm{f}(g^{-(1+\alpha)})^\ast)\right)^\ast\| \;d\mu(g)\notag\\
&=&\int_{G^\C}\|\Delta(g^{-1})^\ast
\varepsilon_{g^\alpha}(\mathrm{f}(g^{-(1+\alpha)})\| \;d\mu(g)\notag\\
&=&\int_{G^\C}\|\mathrm{f}(g^{-(1+\alpha)})\| \;d\mu(g)\notag\\
&=&\|\mathrm{f}\|_{\alpha}\;.
\end{eqnarray}
\item
  \begin{eqnarray}
\left(\mathrm{f}_1^\ast\ast \mathrm{f}_2^\ast\right)(g^{1+\alpha})
&=&\int_{G^\C}\mathrm{f}^\ast_1(\tilde{g})
\varepsilon_{\tilde{g}}(\mathrm{f}_2^\ast(\tilde{g}^{-1}g^{1+\alpha}))
\;d\mu(\tilde{g})\notag\\
&=&\int_{G^\C}\Delta(\tilde{g}^{-1})\varepsilon_{\tilde{g}}(\mathrm{f}_1(\tilde{g}^{-1})^\ast)
\varepsilon_{\tilde{g}}(\Delta(g^{-1}\tilde{g}))
\varepsilon_{\tilde{g}^{-1}g^{1+\alpha}}(\mathrm{f}_2(g^{-(1+\alpha)}\tilde{g})^\ast)
\;d\mu(\tilde{g})\notag\\
&=&\Delta({g}^{-1})\int_{G^\C}\varepsilon_{\tilde{g}}(\mathrm{f}_1(\tilde{g}^{-1})^\ast)
\varepsilon_{\tilde{g}}
\left(\varepsilon_{\tilde{g}^{-1}g^{1+\alpha}}(\mathrm{f}_2(g^{-(1+\alpha)}\tilde{g})^\ast)\right)
\;d\mu(\tilde{g})\notag\\
&=&\Delta({g}^{-1})\int_{G^\C}\varepsilon_{\tilde{g}}(\mathrm{f}_1(\tilde{g}^{-1})^\ast)
\varepsilon_{g^{1+\alpha}}(\mathrm{f}_2(g^{-(1+\alpha)}\tilde{g})^\ast)
\;d\mu(\tilde{g})\notag\\
&=&\Delta({g}^{-1})\int_{G^\C}
\varepsilon_{g^{1+\alpha}\tilde{g}}(\mathrm{f}_1(\tilde{g}^{-1}g^{-(1+\alpha)})^\ast)
\varepsilon_{g^{1+\alpha}}(\mathrm{f}_2(\tilde{g})^\ast)
\;d\mu(\tilde{g})\notag\\
&=&\Delta({g}^{-1})\varepsilon_{g^{1+\alpha}}\int_{G^\C}\left(\mathrm{f}_2(\tilde{g})
\mathrm{f}_1(\tilde{g}^{-1}g^{-(1+\alpha)})\right)^\ast
\;d\mu(\tilde{g})\notag\\
&=&\Delta({g}^{-1})\varepsilon_{g^{1+\alpha}}
\left(\mathrm{f}_2\ast\mathrm{f}_1(g^{(1+\alpha)}\right)^\ast
=\left(\mathrm{f}_2\ast \mathrm{f}_1\right)^\ast(g^{1+\alpha})
\end{eqnarray}
\item
\begin{eqnarray}
\varpi\rtimes U^{(\alpha)}(\mathrm{f})^\ast
&=&\int_{G^\C}\left(\varpi(\mathrm{f}(g))U_{g^\alpha}\right)^\ast\;d\mu(g)\notag\\
&=&\int_{G^\C}U_{g^{-\alpha}}\varpi(\mathrm{f}(g)^\ast)\;d\mu(g)\notag\\
&=&\int_{G^\C}U_{g^{\alpha}}\varpi(\mathrm{f}(g^{-1})^\ast)\Delta(g^{-1})\;d\mu(g)\notag\\
&=&\int_{G^\C}\varpi\left(\varepsilon_g(\mathrm{f}(g^{-1})^\ast\Delta(g^{-1}))\right)U_{g^\alpha}\;d\mu(g)\notag\\
&=&\int_{G^\C}\varpi\left(\mathrm{f}^\ast(g^{-1})\right)U_{g^\alpha}\;d\mu(g)\notag\\
&=&\varpi\rtimes U^{(\alpha)}(\mathrm{f}^\ast)
\end{eqnarray}
\item
\begin{eqnarray}
\varpi\rtimes U^{(\alpha)}(\mathrm{f}_1\ast\mathrm{f}_2)
&=&\int_{G^\C\times G^\C}
\varpi\left(\mathrm{f}_1(g))\varepsilon_g(\mathrm{f}_2(g^{-1}\tilde{g}))\right)U_{\tilde{g}}
\;d\mu(g,\tilde{g})\notag\\
&=&\int_{G^\C\times G^\C}
\varpi\left(\mathrm{f}_1(g))U_g\varpi(\mathrm{f}_2(g^{-1}\tilde{g}))\right)U_{g^{-1}\tilde{g}}
\;d\mu(g,\tilde{g})\notag\\
&=&\int_{G^\C\times G^\C}
\varpi\left(\mathrm{f}_1(g))U_g\,\varpi(\mathrm{f}_2(\tilde{g}))\right)U_{\tilde{g}}
\;d\mu(g,\tilde{g})\notag\\
&=&\varpi\rtimes U^{(\alpha)}(\mathrm{f}_1)\cdot \varpi\rtimes U^{(\alpha)}(\mathrm{f}_2)
\end{eqnarray}
where the last equality follows from Fubini.
\end{itemize}

\section{Some comments on $\ast$ v.s. $\star$}\label{ast and star}
Recall lemma \ref{commutative} for the case of commutative $\mathfrak{C}^\ast$. When $\mathrm{F}_2=\mathrm{F}^\ast_1$ we have,
\begin{eqnarray}
\mathcal{M}_\lambda\left[\left(\mathrm{F}\ast
\mathrm{F}^\ast\right);\alpha\right]&=& \mathcal{M}_\lambda\left[
\mathrm{F};\alpha\right]\mathcal{M}_\lambda\left[ \mathrm{F}^\ast;\alpha\right]
= \mathcal{M}_\lambda\left[
\mathrm{F};\alpha\right]\mathcal{M}_\lambda\left[ \mathrm{F};\alpha\right]^\ast\\
\mathcal{M}_\lambda\left[\left(\mathrm{F}\star
\mathrm{F}^\ast\right);\alpha\right]&=& \mathcal{M}_\lambda\left[
\mathrm{F};\alpha\right]\mathcal{M}_\lambda\left[ {\mathrm{F}}^\ast;1-\alpha\right]
=\mathcal{M}_\lambda\left[
\mathrm{F};\alpha\right]\mathcal{M}_\lambda\left[ {\mathrm{F}};1-\alpha\right]^\ast\;.
\end{eqnarray}
Accordingly, for the two algebras distinguished by their $\ast$ v.s. $\star$ product, at $\alpha=1/2$ we get norm equality
$\|\mathcal{M}_\lambda\left[\left(\mathrm{F}\ast
\mathrm{F}^\ast\right);1/2\right]\|=\|\mathcal{M}_\lambda\left[\left(\mathrm{F}\star
\mathrm{F}^\ast\right);1/2\right]\|$ and simultaneous representations. But generically we get neither norm equality nor simultaneous representations.

However, if there exists a class of holomorphic transforms $\psi_\lambda(\alpha):=\mathcal{M}_\lambda\left[{\Psi};\alpha\right]\in\mathfrak{C}^\ast$ that satisfy a functional equation of the form $\psi_\lambda(\alpha)=\varepsilon_\lambda\psi_\lambda(1-\alpha)$ with $\varepsilon_\lambda\in\C$ such that $|\varepsilon_\lambda|=1$, then (for this class)  the $\ast$-convolution and $\star$-convolution yield norm equivalence  and simultaneous projective representations for all $\alpha\in\mathbb{S}_\lambda$!

\begin{lemma}\label{Psi lemma}
Suppose $\mathfrak{C}^\ast$ is commutative and there exists a family of holomorphic functions $\psi_\lambda(\alpha):=\mathcal{M}_\lambda\left[{\Psi};\alpha\right]\in\mathfrak{C}^\ast$ such that $\psi_\lambda(\alpha)=\varepsilon_\lambda\psi_\lambda(1-\alpha)$ with $\varepsilon_\lambda\in\C$ and $|\varepsilon_\lambda|=1$
\begin{eqnarray}\label{curious}
&&\hspace{-.75in}\|\mathcal{M}_\lambda\left[\left({\Psi\ast\Psi^\ast}\right);\alpha\right]\|
=\|\mathcal{M}_\lambda\left[\left({\Psi\star\Psi^\ast}\right);\alpha\right]\|
=\|\psi_\lambda(\alpha)\|^2
\;\;\;\;\;\;\forall \alpha\in\mathbb{S}\;.
\end{eqnarray}
\end{lemma}

Before exploring this lemma, we introduce a new integrator. It happens that Laguerre polynomials are germane in this context, so define algebraic elements $\mathrm{A}_n^{(s)}\in\mathbf{F}_{\mathbb{S}_{\mathcal{R}}}(G^\C)$  where $\mathrm{A}_n^{(s)}:G^\C\rightarrow\mathfrak{C}^\ast$ by $g\mapsto e^{-\rho(g)}L_n^{(s)}(\rho(g))$ with $L_n^{(s)}(\rho(g))$ being operator-valued generalized Laguerre polynomials such that (for unitary $\rho$ and $-1<s\in\R$). Then
\begin{equation}
\mathcal{M}_\lambda\left[\mathrm{Det}\,(\mathrm{A}_n^{(s)});\alpha\right]
=\int_{G^\C}e^{-\mathrm{tr}(g)}\mathrm{det}(L_n^{(s)}(g))\,\det g^\alpha
\,\mathcal{D}_\lambda g\;.
\end{equation}
Let $\Ta$ be the space of continuous pointed maps
$\tau:(\mathbb{T},\ti_a)\rightarrow(\C^\times,1)$ where $\mathbb{T}=[\ti_a,\ti_b]\subset\R$ and $\C^\times:=\C\backslash\{0\}$. For $G^\C\cong T_a$ and localization $\lambda_{\R_+}:T_a\rightarrow\R_+$ and choosing the gamma Haar measure, this reduces to
\begin{eqnarray}
\mathcal{M}_{\R_+,\Gamma}\left[\mathrm{Det}\,(\mathrm{A}_n^{(s)});\alpha\right]
&=&\frac{1}{\Gamma(\alpha)}\int_0^\infty e^{-t}L_n^{(s)}(t)t^{\alpha-1}\,dt\notag\\
&=&\left(\begin{array}{c}
                                 s-\alpha+n \\
                                 n
                               \end{array}\right)
                               \;\;\;\;\;\;0<\alpha.
\end{eqnarray}

This motivates to introduce a Laguerre-type functional integral defined by
\begin{equation}\label{Laguerre}
\mathcal{L}_\lambda\left[\mathrm{F};\alpha,n,s\right]
:=\int_{G^\C}F(g)\,L_n^{(s)}(g)g^{\alpha}\,\mathcal{D}_\lambda g
:=\int_{G^\C}F(g)\,\mathcal{D}_\lambda l_{n,\alpha}^{(s)}(g)
\end{equation}
where $\mathcal{D}_\Lambda l_{n,\alpha}^{(s)}(g)$ denotes a family of Laguerre integrators. We can add $\mathcal{D}_\lambda l_{n,\alpha}^{(s)}(g)$ to the list of non-Gaussian integrators introduced in \cite{LA1}. Remark that Hermite polynomials (relevant for normal ordered operators) are a special class of Laguerre polynomials on $(G^\C\times G^\C)_\Delta$.

For suitable $\mathrm{F}$, one can series expand $\mathrm{F}=\sum_n\,a_n^{(s)}\,\mathrm{A}_n^{(s)}$  with the  help of Laguerre integrators. Observe that $\mathcal{L}_\lambda\left[\mathrm{E}^{-\mathrm{Id}};\alpha,n,s\right]
=\mathcal{M}_\lambda\left[\mathrm{A}_n^{(s)};\alpha\right]$ and $\mathcal{L}_\lambda\left[\mathrm{E}^{-\mathrm{Id}};\alpha,0,s\right]
=\mathcal{M}_\lambda\left[\mathrm{E}^{-\mathrm{Id}};\alpha\right]$. Explicitly then,
\begin{equation}
\mathcal{M}_\lambda\left[\mathrm{F};\alpha,s\right]
:=\lim_{N\rightarrow\infty}\sum_{n=0}^N a_n^{(s)}\mathcal{M}_\lambda\left[\mathrm{A}_n^{(s)};\alpha\right]
=\lim_{N\rightarrow\infty}\sum_{n=0}^N a_n^{(s)}\mathcal{L}_\lambda\left[\mathrm{E}^{-\mathrm{Id}};\alpha,n,s\right]
\end{equation}
where
\begin{equation}
a_n^{(s)}
=\frac{1}{{c_{m,n}^{(s)}}}\int_{G^\C}F(g)L_m^{(s)}(g)g^{s+1}
\,\mathcal{D}_\lambda g
=\frac{1}{{c_{m,n}^{(s)}}}\mathcal{L}_\lambda\left[\mathrm{F};s+1,m,s\right]
\end{equation}
and the constants $c_{m,n}^{(s)}=\mathcal{L}_\lambda\big[\mathrm{A}_m^{(s)};s+1,n,s\big]$.

For example, with the localization $\lambda_{\R_+}:T_a\rightarrow\R_+$ and $\rho(g)=g Id$ where $Id$ is the identity in $\mathfrak{C}^\ast$, we have $\mathcal{L}_{\R_+,\Gamma}\big[\mathrm{E}^{-\mathrm{Id}};\alpha,n,s\big]=\big(\begin{array}{c}
 s-\alpha+n \\
 n
 \end{array}\big)\,Id$ and
\begin{eqnarray}
c_{m,n}^{(s)}=\mathcal{L}_{\R_+,\Gamma}\left[\mathrm{A}_m^{(s)};s+1,n,s\right]
&=&\int_0^\infty e^{-g}L_m^{(s)}(g)L_n^{(s)}(g)\,g^{s+1}\,d\nu(g_\Gamma)=\delta_{m,n}
\left(\begin{array}{c}
 n+s \\
 n
 \end{array}\right)\,Id\;.\notag\\
\end{eqnarray}
So calculating $\mathcal{M}_\lambda\left[\mathrm{F};\alpha,s\right]$ boils down to calculating $\mathcal{L}_\lambda\left[\mathrm{F};s+1,n,s\right]$ (for suitable $\mathrm{F}(g)$). In particular, suppose we are given $\mathrm{F}(g)=\sum_{m=0}^\infty F_m^{(s)}\mathrm{A}_m^{(s)}(g)$ with $F_m^{(s)}\in\mathfrak{C}^\ast$. Then
\begin{eqnarray}
\mathcal{L}_{\R_+,\Gamma}\left[\mathrm{F};s+1,n,s\right]
&=&\lim_{N\rightarrow\infty}\sum_{m=0}^N\int_0^\infty
F_m^{(s)}e^{-g}L_m^{(s)}(g)L_n^{(s)}(g)\,g^{s+1}\,d\nu(g_\Gamma)\notag\\
&=&\lim_{N\rightarrow\infty}\sum_{m=0}^N F_m^{(s)}\,c_{m,n}^{(s)}\notag\\
&=&\left(\begin{array}{c}
 n+s \\
 n
 \end{array}\right)F_n^{(s)}\;,
\end{eqnarray}
and
\begin{equation}
\mathcal{M}_\lambda\left[\mathrm{F};\alpha,s\right]
=\sum_{n=0}^{\infty}F_n^{(s)}\left(\begin{array}{c}
                                 s-\alpha+n \\
                                 n
                               \end{array}\right)\;.
\end{equation}

With this diversion in mind, return to the lemma. Following \cite{GIL,BCKV,COF}, define
\begin{definition}
For $-1<s\in\R$ and $\Psi\in\mathbf{F}_{\mathbb{S}_{\mathcal{R}}}(G^\C)$ such that $\Psi(g)=\rho(g)^{s/2}e^{-\rho(g)/2}$,
\begin{equation}
\mathcal{L}_\lambda\left[\Psi;\alpha,n,s\right]
:=\int_{G^\C}g^{s/2}e^{-g/2}\,L_n^{(s)}(g)g^{\alpha}
\,\mathcal{D}_\lambda g\;.
\end{equation}
\end{definition}
Suppose again $\lambda_{\R_+}:T_a\rightarrow \R_+$ and choose Haar measure $\nu(g_\psi):=2^{-(\alpha+s/2)}\,\nu(g)/\Gamma(\alpha+s/2)$. This yields a family of the $\psi_\lambda(\alpha)$ of Lemma \ref{Psi lemma};
\begin{eqnarray}
\psi_n^{(s)}(\alpha)
&:=&\mathcal{L}_{\R_+,\psi}\left[\Psi;\alpha,n,s\right]\notag\\
&=&\frac{1}{2^{\alpha+s/2}\Gamma(\alpha+s/2)}
\int_0^\infty t^{s/2}e^{-t/2}\,L_n^{(s)}(t)\,t^{\alpha-1}\,dt\;\;\;\;\;\;\;\;0<\Re(\alpha+s/2)\notag\\
&=:&\frac{1}{2^{\alpha+s/2}}\mathcal{M}_{\R_+,\Gamma}[\Psi_n^{(s)};\alpha+s/2,s]\notag\\
&=:&\mathcal{M}_{\R_+,\psi}[\Psi_n^{(s)};\alpha+s/2,s]
\end{eqnarray}
where we have defined $\Psi_n^{(s)}(g):=e^{-\rho(g)/2}L_n^{(s)}(\rho(g))=e^{\rho(g)/2}{A}_n^{(s)}(\rho(g))$.

Using the series representation of Laguerre, this can be expressed as (\cite[eq. 7.414(7.)]{GR})
\begin{equation}\label{hypergeometric representation}
\psi_n^{(s)}(\alpha)=\frac{(s+1)^{(n)}}{n!}\,\tensor[_2]{\mathrm{F}}{_1}(-n,\alpha+s/2;s+1;2)\;,
\end{equation}
and thereafter analytically continued to all $\alpha\in\C$. This family of functions satisfies
\begin{theorem}\label{Psi theorem}\hfill
\begin{itemize}
  \item $\psi_n^{(s)}(\alpha)=(-1)^n\psi_n^{(s)}(1-\alpha)\;\forall\,\alpha\in\C$
  \item All zeros $\alpha_0$ of $\psi_n^{(s)}(\alpha)$ are simple and lie on the critical line $\Re(\alpha)=1/2$.
\end{itemize}
\end{theorem}

\emph{proof}: The first point follows readily from (\ref{hypergeometric representation}) and the identity
\begin{eqnarray}
\tensor[_2]{\mathrm{F}}{_1}(a,b;c;z)&=&(1-z)^{-a}\tensor[_2]
{\mathrm{F}}{_1}\left(a,c-b;c;\frac{z}{z-1}\right)\;.\notag
\end{eqnarray}
For proof of the second point we refer to \cite[th. 4]{BCKV}. Note that, in the course of the proof, it is shown that $\psi_n^{(s)}(1/2+i\sigma)$ are orthogonal on $\R$ with respect to the measure $2^{s+1}|\Gamma(1/2+i\sigma+ s/2|^2\,d\sigma$ where $\sigma\in\R$.
$\QED$

We have learned that there exist elements $\Psi_n^{(s)}
\in\mathbf{F}_{\mathbb{S}_{\mathcal{R}}}(T_a)$ such that functional Mellin of both $\ast$ and $\star$ products act by multiplication up to a phase; also
\begin{equation}\label{curious1}
\mathcal{M}_\psi\left[\left(\Psi_n^{(s)}\ast
\Psi_m^{(r)}\right);\alpha+s/2,s\right]=(-1)^n\mathcal{M}_\psi\left[\left(\Psi_n^{(s)}\star
\Psi_m^{(r)}\right);\alpha+s/2,s\right]\;,
\end{equation}
and
\begin{equation}\label{curious2}
\|\mathcal{M}_\psi\left[\left({\Psi_n^{(s)}\ast\Psi_n^{(s)}}\right);\alpha+s/2,s\right]\|
=\|\mathcal{M}_\psi\left[\left({\Psi_n^{(s)}\star\Psi_n^{(s)}}\right);\alpha+s/2,s\right]\|
=\|\psi_n^{(s)}(\alpha)^2\|\;.
\end{equation}
Further, $\psi_n^{(s)}(\alpha_0)=0$ iff $\alpha_0=1/2+i\sigma_0$ with $\sigma_0\in\R$ and $-1<s$. Evidently, functionals of $\tau\in T_a$ degrees of freedom can be series expanded \emph{along the critical line} in terms of the class of functions $\psi_n^{(s)}(\alpha)=\mathcal{M}_{\psi}[({\Psi_n^{(s)}});\alpha+s/2,s]$. In light of this, it is curious and perhaps significant that (\ref{curious}) holds for $\psi_n^{(s)}(\alpha)$. We don't fully understand it's physical implications.\footnote{However, observe that in a quantum physics setting $\mathrm{Log}\,\Psi^{(s)}$ corresponds to a complex effective action, and the counting/evolution along a contour $\mathit{\Gamma}$ is associated with entropy/action. The fact that functional Mellin of both $\ast$ and $\star$ products yield (projective)representations supports the idea that, in this context, ${\tau}(\ti)$  represents a complex evolution parameter along a contour in $\C^\times$. And the critical line $\Re(\alpha)=1/2$ seems to indicate an isentropic process characterized by a duality or equivalence between the notions of real-valued action and information entropy. From this perspective, $\psi_n^{(s)}(\alpha)$ has the earmarks of a Loschmidt amplitude, which would suggest the isentropic process represents non-equilibrium, unitary evolution (see e.g. \cite{HEYL}).}


\begin{thebibliography}{99}

\bibitem{DY/YU} E.B. Dynkin and A.A. Yushkevich, \emph{Markov
Processes}, Plenum Press, New York (1969)

\bibitem{FR} M. Freidlin, \emph{Functional Integration and
Partial Differential Equations}, Princeton University Press,
Princton, New Jersey (1985)

\bibitem{FRI1}A. Friedman, \emph{Stochastic Differential
Equations and Applications, Vols. 1 and 2}, Academic Press, New
York (1969)

\bibitem{FE/HI}R.P. Feynman and A.R. Hibbs, \emph{Quantum
Mechanics and Path Integrals}, McGraw-Hill, New York (1965)

\bibitem{GROS}
C. Grosche, How to solve path integrals in quantum mechanics, \emph{J. Math. Phys.} \textbf{36}(5), 2354 (1995)

\bibitem{KLEIN}
H. Kleinert, \emph{Path Integrals in Quantum Mechanics, Statistics, Polymer Physics, and Financial Markets}, World Scientific, New Jersey (2006)

\bibitem{KL}
J.R. Klauder, \emph{A Modern Approach to Functional Integration}, Springer, New York (2010)

\bibitem{CA/D-W3}
P. Cartier  and C. DeWitt-Morette, \emph{Functional Integration:
Action and Symmetries}. Cambridge University Press, Cambridge (2006)

\bibitem{AL}
S. Albeverio, R. H{\o}egh-Krohn, S. Mazzucchi, \emph{Mathematical theory of Feynman
path integrals. An Introduction. 2nd and enlarged edition}. Lecture Notes in
Mathematics, Vol. 523. Springer-Verlag, New York (2008)

\bibitem{LA1}
J. LaChapelle, A Framework for Non-Gaussian Functional Integrals with Applications to Quantum Field Theory and Number Theory, submitted

\bibitem{POL}
J. Polchinski, String Theory I. and II., Cambridge Univ. Press, Cambridge, (1998)

\bibitem{KLT}
Z. Bern, Perturbative Quantum Gravity and its Relation to Gauge Theory, \emph{Living Reviews in Relativity}, \textbf{5}(5) (2002)

\bibitem{LANDS1}
N.P. Landsman, Lecture Notes on $C^\ast$-Algebras, Hilbert $C^\ast$-modules,
and Quantum Mechanics, arXiv:math-ph/9807030, (1998)

\bibitem{LANDS2}
N.P. Landsman, Quantization and Superselection sectors I and II, \emph{Rev. Math. Phys.} \textbf{2}(1), 45--104  (1990)

\bibitem{LANDS3}
N.P. Landsman, Rieffel induction as generalized quantum
Marsden-Weinstein reduction, \emph{J. Geo. and Phys.} \textbf{15}, 285--319
(1995)

\bibitem{STR}
M. Strassler, Field Theory Without Feynman Diagrams: One-Loop Effective Actions, \emph{Nucl. Phys. B} \textbf{385}(1-2), 145--184 (1992)

\bibitem{SCH}
C. Schubert, Perturbative quantum field theory in the string inspired formalism, \emph{Phys, Rept.} \textbf{355}, 73--234 (2001)

\bibitem{LD}
L. Donnay, Celestial holography: An asymptotic symmetry perspective, \emph{Phys, Rept.} \textbf{1073}, 1--41 (2024)

\bibitem{FL}
P. Flajolet, X. Gourdon, D. Phillipe, Mellin Transforms and
Asymptotics: Harmonic Sums, \emph{Theo. Comp. Sci.} \textbf{144}, 3--58
(1995)

\bibitem{SZ}
Z. Szmydt, B. Ziemian, {The Mellin transformation and Fuchsian
type partial differential equations}, Kluwer Academic Publishers,
Boston (1992)

\bibitem{BBO}
J. Bertrand, P. Bertrand, J. Ovarlez, \emph{The Mellin Transform}(ch. 11), The Transforms and Applications Handbook: Seoncd Edition, ed. A.D. Poularikas,  CRC Press, Boca Raton (2000)

\bibitem{ZEM}
A.H. Zemanian, Generalized Integral Transformations, Dover Pub., New York (1987)


\bibitem{LA3}
J. LaChapelle, Functional Integration on Constrained Function
Spaces I and II, arXiv:math-ph/1212.0502 (2012) and arXiv:math-ph/1405.0461 (2014)

\bibitem{GGR}
M. Gisonni, T. Grava, G. Ruzza, Laguerre Ensemble: Correlators, Hurwitz Numbers and Hodge Integrals, \emph{Ann. Henri Poincar\'{e}} \textbf{21}, 3285--3339 (2020)

\bibitem{HM}
K.H. Hofmann, S.A. Morris,  {The Structure of Compact Groups},
Walter de Gruyter, Berlin (1998)

\bibitem{W}
D.P. Williams,  {Crossed Products of $C^\ast$-algebras}, American
Mathematical Society, Providence, Rhode Island (2007)


\bibitem{GLO}
H. Gl$\ddot{o}$ckner, Algebras whose groups of units are Lie groups, \emph{Studia Mathematica } \textbf{153}(2), 147--177 (2002)

\bibitem{LEE}
Dong, H.L., The structure of complex Lie groups, Research Notes in Mathematics 429,
Chapman \& Hall/CRC, London (2002)

\bibitem{NAK}
M. Nakahara, Geometry, Topology and Physics, Adam Hilger, New York, (1990)

\bibitem{TAO}
T. Tao, The standard branch of the matrix logaritm, What's new. https://terrytao.wordpress.com/2015/05/03/the-standard-branch-of-the-matrix-logarithm/ (2015)

\bibitem{HIG}
N.J. Higham, {Functions of Matrices}, SIAM, Philadelphia, Pennsylvania, (2008)

\bibitem{NEEB}
 K.-H. Neeb, Infinite-Dimensional Lie Groups, 3rd. cycle. Monastir (Tunisie), 1--77 (2005)

\bibitem{BN}
D. Belti\c{t}$\breve{a}$, M. Nicolae, On Universal Enveloping Algebras in a Topological Setting, \emph{Studia Mathematica } \textbf{230}(1), 1--29 (2015)

\bibitem{BNEEB}
D. Belti\c{t}$\breve{a}$, K.-H. Neeb, Nonlinear Completely Positive Maps and Dilation Theory for Real Involutive Algebras, \emph{Integr. Equ. Oper. Theory} \textbf{83}, 517--562 (2015)


\bibitem{RS}
D.B. Ray, I.M. Singer, R-Torsion and the Laplacian on Riemannian manifolds, \emph{Adv. Math.} \textbf{7}(2), 145-210
(1971)

\bibitem{VOR}
A. Voros, Spectral Functions, Special Functions and the Selberg Zeta Function, \emph{Commun. Math. Phys.} \textbf{110}, 439--465 (1987)

\bibitem{LAM}
W. Lamb, Fractional Powers of Operators Defined on a Fr\'{e}chet Space, \emph{Proc. Edinburgh Math. Soc.} \textbf{27},165--180
(1984)

\bibitem{BAL}
A.V. Balakrishnan, Fractional Powers of Closed Operators and the Semigroups Generated by Them, \emph{Pacific J. Math.} \textbf{10}(2), 419--437
(1960)

\bibitem{KOM}
H. Komatsu, Fractional Powers of Operators, \emph{Pacific J. Math.} \textbf{19}(2), 285--346
(1966)

\bibitem{GR}
I.S. Gradshteyn, I.M. Ryzhik, Table of Integrals, Series, and Products, Elsevier AP, USA (2007)


\bibitem{BN}
D. Bump, E.K.-S. Ng, On Riemann's zeta function, \emph{Math. Z.} \textbf{192}, 195--204 (1986)

\bibitem{GIL}
G. Gilbert, Zeros of Symmetric, Quasi-finite, Orthogonal Polynomials, \emph{J. Math. Anal. and App.} \textbf{157}, 346--350 (1991)

\bibitem{BCKV}
D. Bump, K.-K. Choi, P. Kurlberg, J. Vaaler, A Local Riemann Hypoethesis, I, \emph{Math. Z.} \textbf{233}(1), 1--18 (2000)

\bibitem{COF}
M.W. Coffey, Special functions and the Mellin transforms of Laguerre and Hermite functions, \emph{Analysis}, \textbf{27}(1) (2009)

\bibitem{HEYL}
M. Heyl, Dynamical quantum phase transitions: a review, \emph{Rep. Prog. Phys.} \textbf{81}(5) (2018)

\bibitem{FOF}
J.M. Figueroao-O'Ferrill, The Theory of Induced Representation in Field Theory, QMW-PH-95-?

\bibitem{IZ}
C. Itzykson, J.-B. Zuber, {Quantum Field Theory},
McGraw-Hill, USA (1980)


\bibitem{BK}
Z. Bern, D. A. Kosower, The Computation of Loop Amplitudes in Gauge Theories, Fermilab preprint Fermilab-Pub-91/111-T (1991)


\bibitem{SP}
E.R. Speer, Generalized Feynman Amplitudes, Ann. Math. Stud. No. 62, Princeton University Press, USA (1969)

\bibitem{LE}
G. Leibbrandt, Introduction to the technique of dimensional regularization, Rev. Mod. Phys. \textbf{47}(4), 849 (1975)

\bibitem{PDM}
P. Cartier, C. DeWitt-Morette, A new perspective on functional integration, \emph{J. Math. Phys.} \textbf{36}(5), 2237--2312 (1995)

\bibitem{LA2}
J. LaChapelle,  Path integral solution of linear second order
partial differential equations: I and II. \emph{Ann. Phys.}
\textbf{314}, 362--424 (2004)

\bibitem{DS}
P. Dai, W. Siegel, Worldline Green functions for arbitrary Feynman Diagrams, \emph{Nucl. Phys. B} \textbf{770}(1-2), 107--122 (2007)

\bibitem{AHL}
N. Arkani-Hamed, S. He, T. Lam, Stringy canonical forms, \emph{JHEP} \textbf{69} (2021)

\bibitem{AFSPT}
N. Arkani-Hamed, H. Frost, G. Salvatori, P-G. Plamondon, H. Thomas, All Loop Scattering as a Counting Problem, arXiv:2309.15913 (2023)

\bibitem{VAS}
D.V. Vassilevich, Heat kernel expansion: user's manual, \emph{Phys. Reports}, \textbf{388}(5-6), 279-360 (2003)

\bibitem{DGS}
I. Dubovyk, J. Gluza, G. Somogyi, {Mellin-Barnes Integrals: A Primer on Particle Physics Applications}, Lecture Notes in Physics, Springer Nature (2022)

\bibitem{WE}
S. Weinzierl, Feynman Integrals, arXiv:hep-th/2201.03593v2 (2022)

\bibitem{LLY}
S.-H. Lai, J.-C. Lee, Y. Yang, The Lauricella functions and exact string scattering amplitudes,  \emph{JHEP}, \textbf{062} (2016)

\bibitem{BDH}
D.C. Brydges, J. Dimock, T.R. Hurd, Estimates on renormalization group transformations, \emph{Can. J. Math.}, \textbf{50}, 756-793 (1998)

\bibitem{BBS}
R. Bauerschmidt, D.C. Brydges, G. Slade, Introduction to a Renormalization Group Method, Lecture Notes in Mathematics, Springer (2019)

\bibitem{AKW}
M. Alford, A. Kapustin, F. Wilczek, Imaginary chemical potential and finite fermion density on the lattice, \emph{Phys. Rev. D.}, \textbf{59}, 054502 (1999)

\bibitem{DL}
M. D'Elia, M.-P. Lombardo, Finite density QCD via an imaginary chemical potential, \emph{Phys. Rev. D.}, \textbf{67}, 014505 (2003)

\bibitem{B}
D. Bump, Automorphic Forms and Representations, Cambridge Univ. Press, (1998)

\bibitem{GH}
D. Goldfeld, J. Hundley, Automorphic Representation and $L$-Functions for the General Linear Group vol 1., Cambridge Univ. Press, (2011)

\end{thebibliography}
\end{document}